\documentclass[11pt]{article}
\usepackage{jheppub}

\input{epsf}
\usepackage{epsfig}
\usepackage{amssymb}
\usepackage{amsfonts}
\usepackage{amsbsy}
\usepackage[all,cmtip]{xy}
\usepackage{amsmath}
\usepackage{dsfont}
\usepackage{bbm}
\usepackage{upgreek}
\usepackage{amssymb,amscd}
\usepackage{mathrsfs}
\usepackage{amsmath,amsthm}
\usepackage{slashed}
\usepackage{ulem}
\usepackage{dsfont}
\usepackage{tikz}
\usepackage[utf8]{inputenc}
\usetikzlibrary{positioning}
\usetikzlibrary{shapes.geometric, arrows}
\tikzstyle{rect} = [rectangle,rounded corners,minimum width=3cm, minimum height=1cm, text centered, draw=black]
\tikzstyle{arrow} = [thick,->,>=stealth]
\usepackage{stackrel}
\tikzstyle{oct} = [regular polygon,regular polygon sides=8, draw,
    text width=2em, text centered]
\tikzstyle{line} = [draw, -latex']

\tikzstyle{hex} = [regular polygon,regular polygon sides=6, draw,
    text width=2em, text centered]
\tikzstyle{hexs}= [regular polygon,regular polygon sides=5, draw,
    text width=1.7em, text centered]

\DeclareRobustCommand{\coprod}{\mathop{\text{\fakecoprod}}}
\newcommand{\fakecoprod}{%
  \sbox0{$\prod$}%
  \smash{\raisebox{\dimexpr.9625\depth-\dp0}{\scalebox{1}[-1]{$\prod$}}}%
  \vphantom{$\prod$}%
}

\tikzset{
  font={\fontsize{8pt}{12}\selectfont}}

\makeatletter
\DeclareFontFamily{OMX}{MnSymbolE}{}
\DeclareSymbolFont{MnLargeSymbols}{OMX}{MnSymbolE}{m}{n}
\SetSymbolFont{MnLargeSymbols}{bold}{OMX}{MnSymbolE}{b}{n}
\DeclareFontShape{OMX}{MnSymbolE}{m}{n}{
    <-6>  MnSymbolE5
   <6-7>  MnSymbolE6
   <7-8>  MnSymbolE7
   <8-9>  MnSymbolE8
   <9-10> MnSymbolE9
  <10-12> MnSymbolE10
  <12->   MnSymbolE12
}{}
\DeclareFontShape{OMX}{MnSymbolE}{b}{n}{
    <-6>  MnSymbolE-Bold5
   <6-7>  MnSymbolE-Bold6
   <7-8>  MnSymbolE-Bold7
   <8-9>  MnSymbolE-Bold8
   <9-10> MnSymbolE-Bold9
  <10-12> MnSymbolE-Bold10
  <12->   MnSymbolE-Bold12
}{}

\let\llangle\@undefined
\let\rrangle\@undefined
\DeclareMathDelimiter{\llangle}{\mathopen}%
                     {MnLargeSymbols}{'164}{MnLargeSymbols}{'164}
\DeclareMathDelimiter{\rrangle}{\mathclose}%
                     {MnLargeSymbols}{'171}{MnLargeSymbols}{'171}
\makeatother

\def\be{ \begin{equation} }
\def\ee{ \end{equation}}

\def\exp{{\rm exp}}

\def\I{{\rm i}}

\def\half{\frac{1}{2}}
\def\ihalf{\frac{i}{2}}

\def\galf{\frac{1}{2e^2}}

\def\shalf{\frac{1}{\sqrt{2}}}
\def\p{\partial}

\def\one{{\hbox{ 1\kern-.8mm l}}}

\def\vx{{\vec{x}}}

\def\vv{{\rm v}}
\def\uu{{\rm u}}

\def\CC {{\cal C}}
\def\CD {{\cal D}}

\def\CG {{\cal G}}
\def\CH {{\cal H}}

\def\CL {{\cal L}}
\def\CM {{\cal M}}
\def\CN {{\cal N}}

\def\CR {{\cal R}}
\def\CV {{\cal V}}

\def\CG {{\cal G}}
\def\CH {{\cal H}}

\def\CQ {{\cal Q}}
\def\CS {{\cal S}}

\def\IC{\mathbb{C}}

\def\IR{{\mathbb{R}}}

\def\IZ{{\mathbb{Z}}}

\def\fa{\mathfrak{a}}
\def\fb{\mathfrak{b}}

\def\fg{\mathfrak{g}}

\def\ft{\mathfrak{t}}

\def\ft{\mathfrak{t}}
\def\fu{\mathfrak{u}}

\def\rmk#1{\bigskip\noindent{\bf Remark} }
\def\cnj#1{\bigskip\noindent{\bf Conjecture:} }

\def\hk{{hyperk\"ahler~}}

\def\fMM{\overline{\underline{\mathcal{M}}}}

\DeclareMathAlphabet{\mathpzc}{OT1}{pzc}{m}{it}

\def\hfMM{\widehat{\underline{\overline{\mathcal{M}}}}}

\def\ua{\underline{a}}

\def\um{\underline{m}}
\def\un{\underline{n}}

\def\utn{\underline{\tilde{n}}}

\def\vx{{\vec{x}}}

\def\barphi{{\bar\phi}}
\def\bara{{\bar{a}}}
\def\barrho{{\bar\rho}}
\def\tildem{{\tilde{m}}}
\def\tildef{{\tilde{f}}}
\def\tilderho{{\tilde{\rho}}}
\def\barn{{\bar{n}}}
\def\tildea{{\tilde{a}}}
\def\tildeua{{\tilde{\underline{a}}}}

\def\tilden{{\tilde{n}}}

\def\tildepsi{{\tilde{\psi}}}
\def\tildep{{\tilde{p}}}
\def\uphi{{\underline{\phi}}}

\def\tildephi{{\tilde{\phi}}}
\def\upi{{\underline{\pi}}}
\def\utphi{{\tilde{\underline{\phi}}}}
\def\upsi{{\underline{\psi}}}
\def\utpsi{{\tilde{\underline{\psi}}}}

\def\umega{{\underline{\omega}}}

\def\utn{{\underline{\tilde{n}}}}
\def\un{{\underline{n}}}

\title{'t Hooft Defects and Wall Crossing in SQM}

\author[a]{T.~Daniel Brennan}\author[a]{,~Anindya Dey}\author[a]{,~ and Gregory W. Moore}

\affiliation[a]{NHETC and
Department of Physics and Astronomy, Rutgers University \\
126 Frelinghuysen Rd., Piscataway NJ 08855, USA}

\emailAdd{tdanielbrennan@physics.rutgers.edu, anindya.hepth@gmail.com, gwmoore@physics.rutgers.edu}

\abstract{ In this paper we study the contribution of monopole bubbling to the expectation value of supersymmetric 't Hooft defects in Lagrangian theories of class $\CS$ on $\IR^3\times S^1$. This can be understood as the Witten index of an SQM living on the world volume of the 't Hooft defect that couples to the bulk 4D theory. The computation of this Witten index has many subtleties originating from a continuous spectrum of scattering states along the non-compact vacuum branches. 
We find that even after properly dealing with the spectral asymmetry, the standard localization result for the 't Hooft defect does not agree with the result obtained from the AGT correspondence. In this paper we will explicitly show that one must correct the localization result by   adding an extra term to the standard Jeffrey-Kirwan residue formula. This extra term accounts for the contribution of ground states localized along the non-compact branches. This extra term 
restores both the expected symmetry properties of the line defect expectation value and reproduces the results derived using the AGT correspondence. 
}

\begin{document}

\maketitle

\section{Introduction}

't Hooft defects are an important tool for understanding non-perturbative phenomena in quantum field theory. They are one of the simplest non-local operators that are not  defined in terms of local fields. 't Hooft defects can be used as a handle to study a wide variety of topics relating to non-perturbative physics such as wall crossing \cite{Gaiotto:2008cd},  S-duality \cite{Kapustin:2006pk}, and quark confinement \cite{tHooft:1977nqb}. Thus, it is of interest to compute their expectation value.

In supersymmetric theories, one can compute the expectation value of (supersymmetric) ’t Hooft defects by using the method of localization. In recent years, this technique has been used with great success to compute line defects in many 4D $\CN=2$ theories on various manifolds \cite{Pestun:2007rz,Ito:2011ea,Brennan:2018yuj,Gomis:2011pf}. However, it has been known since \cite{Ito:2011ea} that the naive localization computation in $SU(N)$ gauge theories with N$_f =2N$ leads to incorrect results. The focus of this paper is to address this discrepancy. 


We will be primarily concerned with Lagrangian theories of class $\CS$ of type $A_1$ that have a gauge group $G=SU(2)$ on $\IR^3\times S^1$. \footnote{Though we focus on the case of $G=SU(2)$, our analysis will be generally applicable to theories with $G=\prod_i SU(2)_i$. See Section \ref{sec:quiverGT} for more details. } Here, the expectation value of an 't Hooft defect  of charge $P$ can be computed as a supersymmetric index \cite{Ito:2011ea,Brennan:2018yuj}:
\be
\langle L_{[P,0]}\rangle={\rm Tr}_{\CH_P} (-1)^F e^{- \beta H+ \epsilon_{_+}J_{_+} + m\cdot F+i \Theta\cdot Q}~,
\ee
where $\CH_P$ is the Hilbert space of the 4D theory 
in the presence of the 't Hooft defect $L_{[P,0]}$, $F$ is the fermion number, $\beta$ is the radius of the thermal circle, $H$ is the Hamiltonian, $\Theta=(\theta_e,\theta_m)$ is the vector of electric and magnetic  theta angles and $Q=(\gamma_e,\gamma_m)$ is the vector of asymptotic electric and magnetic charges. \footnote{Additionally, $J_+=J_3+J_R$ where $J_3$ generates rotations in the $1$-$2$ plane of $\IR^3$, $J_R$ is the Cartan generator for the R-symmetry group $SU(2)_R$, and additionally $F$ generates flavor symmetries. These symmetries have fugacities $\epsilon_+$ and $m$ respectively.} This index can be computed as the path integral of the 4D theory on $\IR^3\times S^1$ with appropriate boundary conditions by using localization \cite{Ito:2011ea}.

In these theories, the expectation values of 't Hooft defects can also be computed using the AGT correspondence \cite{Drukker:2009id,Alday:2009aq,Alday:2009fs,Drukker:2009tz}. 
This method relies on the fact that theories of class $\CS$ are constructed by compactifying the 6D $\CN=(0,2)$ theory along a Riemann surface with a topological twist \cite{Gaiotto:2009hg}. Due to the topological twist, we can compute the expectation value of 4D line defects exactly via loop operators in an associated CFT on the Riemann surface \cite{Alday:2009fs,Drukker:2009tz,Verlinde:1988sn,Drukker:2009id}. This provides a powerful check of  localization results for the expectation value of 't Hooft defects.

In theories of class $\CS$, the expectation value of 't Hooft defects are holomorhpic functions on the Seiberg-Witten moduli space \cite{Gaiotto:2010be,Dimofte:2011jd,Nekrasov:2011bc}.  In a weak coupling domain, which is implicitly chosen in specifying a Lagrangian, the expectation value can be written in terms of  a certain set of holomorphic Darboux coordinates known as (complexified) Fenchel-Nielson coordinates. These are denoted $\fa,\fb$ where $\fa$ is canonically defined and $\fb$ is a symplectic dual which, though not canonically defined, can be uniquely fixed  via the weak coupling expansion. \footnote{See Section \ref{sec:AGT} for more details on the definition of $\fa,\fb$. }

 In terms of these coordinates, the expectation value takes the general form of a finite Fourier expansion in $\fb$. It is convenient to write the Fourier expansion in the form 
\cite{Ito:2011ea,Alday:2009fs,Drukker:2009id}
\be\label{IntroL}
\langle L_{P,0}\rangle=\sum_{\substack{|\vv|\leq |P|\\\vv\in \Lambda_{cr}+P}} e^{  (\vv,\fb)} Z_{1-loop}(\fa,m,\epsilon_+;\vv) Z_{mono}(\fa,m,\epsilon_+;P,\vv)~,
\ee
where $P\in \Lambda_{mw}$ is the 't Hooft charge, \footnote{Here we we will generally use the notation $\Lambda_{mw}$ to denote the lattice of allowed 't Hooft charges. See Section \ref{sec:sec2} for more details.}  and  $(~,~)$ denotes the Killing form on $\fg$. 
 In this expansion, the coefficients $Z_{1-loop}(\fa,m,\epsilon_+;\vv)$, which are independent of $P$,  are computed from the one-loop determinant of the 4D path integral as in \cite{Ito:2011ea}. However, the coefficients $Z_{mono}(\fa,m,\epsilon_+;P,\vv)$, which are the focus of this paper, are much more subtle. \footnote{We will usually abbreviate these as $Z_{1-loop}(\vv)$ and $Z_{mono}(P,\vv)$. In the case of $G=SU(2)$ we will often use the notation $Z_{mono}(p,v)$ for $P=p h^1$, $\vv=v h^1$ where $h^1$ is the simple, positive magnetic weight $\Lambda_{mw}$. }
 
The Fourier decomposition above can naturally be interpreted as a sum of contributions from different monopole bubbling configurations. Monopole bubbling is the phenomenon in which smooth monopoles are absorbed by an 't Hooft defect \cite{Kapustin:2006pk}. This screens the 't Hooft charge and traps quantum degrees of freedom on the world volume of the defect, giving rise to a ``bubbling super quantum mechanics (SQM).'' Singular monopole configurations  decompose into ``bubbling sectors" labeled by the screened (effective) charge of the 't Hooft defect denoted $\vv$. Thus, each term of \eqref{IntroL} describes the contribution from monopole bubbling configurations with effective charge $\vv$. In each summand, $Z_{mono}(P,\vv)$ is the contribution from the bubbling SQM living on the 't Hooft defect. It can be computed as the twisted partition function of the bubbling SQM, which is formally equivalent to its Witten index.

As pointed out in \cite{Ito:2011ea}, the localization expression for the expectation value of 't Hooft defects does not match the same quantity computed using the AGT correspondence.
By examining the Fourier decomposition \eqref{IntroL}, the discrepancy can be distilled to a difference between the values of $Z_{mono}(P,\vv)$ predicted by AGT and localization. 
Thus, understanding the contribution of monopole bubbling is of critical importance to correctly computing the expectation value of 't Hooft defects. 

As we will see, the localization procedure requires the introduction of a regulator $\xi\in \ft^\ast$ and the localization answer for the monopole bubbling depends on the direction along which we send $\xi\to 0$. The regulator $\xi$ can be identified with an FI parameter in the SQM and the dependence on how we take 
$\xi\to0$ just reflects wall crossing in the SQM 
 \cite{JK,Hori:2014tda,Lee:2016dbm}. 
Further, the localization computation is not invariant under the Weyl symmetry of the flavor group \cite{Ito:2011ea}.  
In contrast, the AGT result requires no regulator and is  invariant under the flavor Weyl symmetry. 
This indicates that the localization expression for $Z_{mono}(P,\vv)$ is incorrect.

In this paper, we will explain the origin of these problems with the standard localization procedure applied to the computation of $Z_{mono}(P,\vv)$ and describe how to correct it by studying the example of type $A_1$ theories of class $\CS$ with $G=SU(2)$. The subtleties of this computation will come from trying to 
apply localization to a theory with a continuum of states coming from non-compact directions in field space with finite potential energy at infinity. We will show that the localization computation for the Witten index misses the contribution of BPS states along these non-compact directions and that by adding their contribution to the Witten index, we reproduce the AGT results. 


\subsection{Outline and Summary}

The outline of this paper is as follows. 
In Section \ref{sec:sec2} we will review 't Hooft defects and monopole bubbling on $\IR^3\times S^1$. Here we discuss how a string theory construction of monopole bubbling can be used to determine the bubbling SQM localized on the world volume of 't Hooft defects \cite{Brennan:2018yuj,Brennan:2018moe}.  Additionally, we review the AGT correspondence and present its results for the expectation value of the minimal 't Hooft defect in the $SU(2)$ ${\rm N}_f=4$ theory. This will be our key example throughout the paper.

In the following section we describe the localization computation of the  Witten index 
\be
I_W
:=
{\rm Tr}_{\CH}~(-1)^F e^{-\frac{\beta}{2} \{\CQ,\CQ\}+\fa Q_\fa+\epsilon_{_+} Q_{_+}+ m\cdot F}~,
\ee
of bubbling SQMs. \footnote{See \eqref{WittenInd} for a precise definition. } 
The bubbling SQMs for the theories we will be considering are  
$\CN=(0,4)$ quiver SQMs whose quivers $\Gamma(P,\vv)$ have the form
\begin{center}
\begin{tikzpicture}[
cnode/.style={circle,draw,thick,minimum size=9mm},snode/.style={rectangle,draw,thick,minimum size=9mm}]
\node[cnode] (1) {1};
\node[cnode] (2) [right=.4cm  of 1]{2};
\node[cnode] (3) [right=.4cm of 2]{3};
\node[cnode] (5) [right=0.8cm of 3]{\tiny{$k-1$}};
\node[cnode] (6) [right=0.4cm of 5]{$k$};
\node[cnode] (7) [right=0.8cm of 6]{$k$};
\node[cnode] (9) [right=0.8cm of 7]{$k$};
\node[cnode] (10) [right=0.4cm of 9]{\tiny{$k-1$}};
\node[cnode] (13) [right=0.8cm of 10]{{$3$}};
\node[cnode] (14) [right=0.4cm of 13]{$2$};
\node[cnode] (17) [right=0.4cm of 14]{1};
\node[snode] (18) [below=0.5cm of 6]{1};
\node[snode] (19) [below=0.5cm of 9]{1};
\node[snode] (20) [above=0.5cm of 7]{${\rm N}_f$};
\draw[-] (1) -- (2);
\draw[-] (2)-- (3);
\draw[dotted] (3) -- (5);
\draw[-] (5) --(6);
\draw[dotted] (7) -- (9);
\draw[dotted] (6) -- (7);
\draw[-] (9) -- (10);
\draw[dotted] (10) -- (13);
\draw[-] (13) -- (14);
\draw[-] (14) -- (17);
\draw[-] (6) -- (18);
\draw[-] (9) -- (19);
\draw[dashed] (20) -- (7);
\end{tikzpicture}
\end{center}
with $k$ repeated $n-2k+1$ times when $\vv\neq0$ and 
\begin{center}
\begin{tikzpicture}[
cnode/.style={circle,draw,thick,minimum size=9mm},snode/.style={rectangle,draw,thick,minimum size=9mm}]
\node[cnode] (1) {1};
\node[cnode] (2) [right=.4cm  of 1]{2};
\node[cnode] (3) [right=.4cm of 2]{3};
\node[cnode] (5) [right=0.8cm of 3]{\tiny{$k-1$}};
\node[cnode] (6) [right=0.4cm of 5]{$k$};
\node[cnode] (10) [right=0.4cm of 6]{\tiny{$k-1$}};
\node[cnode] (13) [right=0.8cm of 10]{{$3$}};
\node[cnode] (14) [right=0.4cm of 13]{$2$};
\node[cnode] (17) [right=0.4cm of 14]{1};
\node[snode] (18) [below=0.5cm of 6]{2};
\node[snode] (20) [above=0.5cm of 6]{${\rm N}_f$};
\draw[-] (1) -- (2);
\draw[-] (2)-- (3);
\draw[dotted] (3) -- (5);
\draw[-] (5) --(6);
\draw[-] (6) -- (10);
\draw[dotted] (10) -- (13);
\draw[-] (13) -- (14);
\draw[-] (14) -- (17);
\draw[-] (6) -- (18);
\draw[dashed] (20) -- (6);
\end{tikzpicture}
\end{center}
when $\vv=0$. Here $n$ and $k$ are defined by
\be
P=n\,\hat{h}^1\quad, \quad P-\vv=k H_1\quad,\quad  H_1\in \Lambda_{cr}~,\quad \hat{h}^1\in \Lambda_{cochar}~,
\ee
where $\Lambda_{cr}$ is the co-root lattice and $\Lambda_{cochar}$ is the cocharacter lattice which contains the 
weight lattice of allowed magnetic charges $\Lambda_{mw}\subset \Lambda_{cochar}$ and $\hat{h}^1, H_1$ are minimal magnetic weights and coroots respectively. \footnote{Here $\Lambda_{cochar}=\{P\in \ft~|~{\rm exp}\{2\pi P\}=\mathds{1}_G\}$. See Footnote \ref{foot:6} for a discussion of 't Hooft charge quantization in $\CN=2$ gauge theories. }

The Witten index of these theories can formally be computed as the path integral of the SQM on a circle of radius $\beta$. Applying localization to this path integral reduces the computation to an integral over the moduli space of BPS equations. By analyzing the singularities of the integrand, we find that it is not integrable and requires regularization. After regularizing the localized path integral, we find that it evaluates to a standard sum over 
residues according to the JK prescription \cite{JK} plus an additional boundary term  which is generically dependent on the gauge coupling $(e)$, thermal radius $(\beta)$, and FI-parameter $(\xi)$.  This dependence of the boundary term on $e,\beta,\xi$ is indicative of  continuous spectra of excited states in these theories that we identify as the scattering states along the non-compact Coulomb and mixed branches.

However, from AGT results, we know that $Z_{mono}(P,\vv)$ has no $\beta$ dependence.  The reason is that we are computing the expectation value of a reducible 't Hooft defect, which is defined as the product of minimally charged irreducible 't Hooft defects. \footnote{See Section \ref{sec:sec2} for the definition of irreducible and reducible singular monopoles. } In this setting, the FI-parameter $\xi$ can be interpreted as a UV regulator that introduces a spacing between the constituent 't Hooft defects making up the reducible 't Hooft defect. Thus, to compute properly  the expectation value of the line defect, we want heuristically to take the limit as ``$\xi\to 0$". 
This can be accomplished by taking the limit off the dimensionless combination $\xi/\beta\to 0$ with $\xi$ as our fixed length scale, which we can then identify as being equivalent to taking ``$\beta\to \infty$.''

Thus, we will identify $Z_{mono}(P,\vv)$ as the Witten index in the  limit as $\beta\to \infty$ \cite{Sethi:1997pa,Yi:1997eg}
\be
Z_{mono}(P,\vv)= I_{\CH_0}(\Gamma(P,\vv))=
\lim_{\beta\to \infty}I_W(\Gamma(P,\vv))~.
\ee
This effectively restricts the trace over the Hilbert space to the subspace of BPS ground states due to the suppression by $e^{-\frac{\beta}{2} \{\CQ,\CQ\}}$ of all non-BPS states. 
We will refer to this quantity as the \textit{ground state index}.

We then attempt to compute the ground state index by using localization. We find that  all boundary contributions vanish, reducing the localization result to a sum of residues according to the 
JK prescription. 
However, as we will demonstrate in Section \ref{sec:NullEx}, this sum over residues does not in general match the AGT calculations for $Z_{mono}(P,\vv)$ \cite{Ito:2011ea}. 

In Section \ref{sec:Coulomb}, we then propose a resolution to the problem of matching localization and AGT results. In particular, we show that \textit{while the localization computation is sensitive to the continuous spectrum of states on non-compact branches, it omits the BPS states localized there}. We explicitly demonstrate this in the several examples by directly computing the contribution of the BPS states on the non-compact branches to the ground state index in the Born-Oppenheimer approximation. This procedure is labor-intensive. It would be highly desirable to have a more efficient technique for computing the contribution of these states.

Thus, since the ground state index is the BPS state contribution to the Witten index, the fact that the localization computation of the ground state index omits states implies that the localization computation of the Witten index also omits states. 
Therefore, \textit{the standard localization computation of the Witten index in this class of SQMs is incorrect}.

Thus, we propose that 
\be
Z_{mono}=I_{\CH_0}=\lim_{\beta\to \infty}I_W=I_{\CH_0}^{(Loc)}+I_{asymp}~,
\ee
where $I_{asymp}=I_C+I_{mix}$ is the weighted trace over the BPS states localized along the Coulomb and mixed branches respectively and $I_{\CH_0}^{(Loc)}=I_{Higgs}$ 
 is the localization computation of $I_{\CH_0}$ which counts the BPS states localized along the Higgs branch states and is given by a sum over residues following the Jeffrey-Kirwan prescription ($Z^{JK})$. 
Note that $I_{asymp}$ is distinct from what is called the ``defect term'' or ``secondary term'' in the literature \cite{Sethi:1997pa,Yi:1997eg,Lee:2016dbm}. See Section \ref{sec:defect} for further discussion.

In Section \ref{sec:examples}, we then provide multiple non-trivial examples to check our conjecture. We additionally comment on the generalization to 4D $\CN=2$ quiver gauge theories with gauge group $G=\prod_i SU(2)_i$ and the relation to deconstruction of the 6D $\CN=(0,2)$ theory \cite{ArkaniHamed:2001ie}. 

In the final section we provide several more examples that illustrate that our proposal is on the right track. 
In these examples, there is the additional subtlety of Chern-Simons interactions in the SQM coming from integrating out 4D fundamental hypermultiplets. In total, we provide explicit examples for the two simplest bubbling SQMs in $SU(2)$ $\CN=2$ gauge theory with ${\rm N}_f=0,2,3,4$ fundamental hypermultiplets and in the $SU(2)$ $\CN=2^\ast$ theory.

\section{'t Hooft Defects  in 4D $\CN=2$ Theories}
\label{sec:sec2}

In Lagrangian 4D $\CN=2$ supersymmetric  gauge theories with compact semisimple gauge group $G$, supersymmetric 't Hooft defects are non-local disorder operators that stretch in the time direction at a fixed position in space. They are defined by imposing Dirac monopole boundary conditions on the fields in the path integral specified by the data of a spatial insertion point $\vx_n\in \IR^3$, 't Hooft charge\footnote{\label{foot:6}$\Lambda_{mw}$ is the lattice of allowed 't Hooft charges defined as the intersection of the cocharacter lattice $\Lambda_{cochar}(G)=\{P\in \ft~|~{\rm exp}(2\pi P)=\mathds{1}_G\}$ with the representation lattices $\Lambda_\mu=\{P\in \ft~|~\langle \mu,P\rangle\in \IZ\}$ where $\mu$ are the highest weights of the  representations of the matter Fermions.} $P_n\in \Lambda_{mw}$, and a choice of a preserved supercharges described by a phase\footnote{Note that $\zeta\in U(1)$ admits an analytic continuation to $\IC^\ast$.} $\zeta\in U(1)$. An 't Hooft defect specified by the data $(\vx_n,P_n,\zeta)$ imposes the boundary conditions
\be\label{tHooftBC}
\vec{B}=\frac{P_n}{2r_n^2}\hat{r}_n+O(r_n^{-3/2})\quad, \qquad X=-\frac{P_n}{2r_n}+O(r_n^{-1/2})~, 
\ee
where $\vec{B}$ is the magnetic field, $\vec{r}_n=\vx-\vx_n$, and $X$ is a real, adjoint scalar field which is related to the $\CN=2$ vector multiplet scalar field $\Phi$ by:
\be
\zeta^{-1}\Phi=Y+i X~. 
\ee
This defines an \textit{irreducible 't Hooft defect.}

In this paper, we will consider \textit{reducible 't Hooft defects} in $SU(N)$ gauge theories which are defined as the product of irreducible 't Hooft defects with minimal 't Hooft charge (minimal 't Hooft defects). We will introduce the notation 
\be
L_{\vec{p},0}=\prod_{I=1}^{N-1}(L_{h^I,0})^{p_I}\quad, \quad \vec{p}=(p_1,p_2,...,p_{N-1})~,
\ee
for a reducible 't Hooft defect of charge $P=\sum_I p_I h^I$ where $\{h^I\}_{I=1}^{N-1}\in \Lambda_{mw}$ is a basis of simple magnetic weights. The classical boundary condition for these operators is a coincident limit of minimal boundary conditions \footnote{As we discuss later, in the quantum theory, the coincident limit should only be taken after computing the expectation value.}
\begin{align}\begin{split}
&\vec{B}=\lim_{\vx_n^{\,(i)}\to \vx_n}\sum_{i}\frac{h^{I(i)}}{2|\vec{r}_n^{\,(i)}|^2}\hat{r}^{\,(i)}_n+O\left(|r_n|^{-3/2}\right)~,\\
&X=-\lim_{\vx_n^{\,(i)}\to \vx_n}\sum_{i}\frac{h^{I(i)}}{2|\vec{r}_n^{\,(i)}|}+O\left(|r_n|^{-1/2}\right)~, 
\end{split}\end{align}
near $\vx=\vx_n$ 
where $i=1,...,p$ indexes the constituent minimal 't Hooft defects, each of which has charge $h^{I(i)}\in \Lambda_{mw}$ and position $\vx_n^{(i)}$. Here we also use the notation where $\vec{r}_n^{\,(i)}=\vx-\vx_n^{\,(i)}$ and $\vx_n\in \IR^3$ is the insertion point of the reducible 't Hooft defect.

These boundary conditions locally satisfy the Bogomolny equation
\be\label{Bogomolny}
\vec{B}=\vec{D} X~,
\ee
giving us a natural identification of $P_n=\sum_i h^{I(i)}$ with the magnetic charge sourced by  the reducible 't Hooft defect.

\subsection{Singular Monopole Moduli Space and Monopole Bubbling}
\label{sec:singmono}

In this paper, we will be interested in using localization to compute the expectation value of reducible 't Hooft defects in 4D $G=SU(2)$ $\CN=2$  gauge theories with ${\rm N}_f\leq 4$ fundamental hypermultiplets or one adjoint hypermultiplet on $\IR^3\times S^1$. This will require an understanding of the BPS moduli space which we will now review. 

As shown in \cite{Kapustin:2005py,Ito:2011ea,Brennan:2018yuj,Gomis:2011pf,Brennan:2016znk}, the BPS moduli space in the presence of irreducible 't Hooft line defects on $\IR^3\times S^1$ is given by the moduli space of singular monopoles on $\IR^3$. Singular monopole moduli space describes the space of solutions to the Bogomolny equation \eqref{Bogomolny} subject to the local  boundary conditions \eqref{tHooftBC} at insertion points with asymptotic boundary conditions:
\be\label{AsymptoticBC}
\vec{B}=\frac{\gamma_m}{2r^2}\hat{r}+O(r^{-5/2})\quad, \qquad X=X_\infty-\frac{\gamma_m}{2r}+O(r^{-3/2})~, 
\ee
where $\gamma_m\in \Lambda_{cr}+\sum_n P_n$ is the asymptotic magnetic charge and $X_\infty\in \ft$ is the Higgs vev. 

The moduli space of solutions to the Bogomolny equation with respect to these boundary conditions is denoted $\fMM(\{P_n\},\gamma_m;X_\infty)$ and when non-empty, is a singular, non-compact \hk manifold of dimension
\be
{\rm dim}_\IR\, \fMM=4\sum_I \tilde{m}^I\quad, \quad \tilde\gamma_m=\gamma_m-\sum_n P^-_n=\sum_I \tilde{m}^I H_I\quad,\quad \tilde{m}^I\geq0~,
\ee
where $P_n^-$ is the Weyl image of $P_n$ in the totally negative chamber defined relative to $X_\infty$ \cite{Moore:2014jfa}. \footnote{This space is conjecturally non-empty only when all of the $\tilde{m}^I\geq0$ \cite{Moore:2014jfa}. } Roughly speaking, the dimension formula can be understood physically by attributing each minimally charged smooth monopole with 4 moduli/zero-modes. 

In this paper, we will only be concerned with the case of singular monopole moduli space associated to a single reducible 't Hooft defect. We will use the notation $\hfMM(P,\gamma_m;X_\infty)$ to denote the singular monopole moduli space with a single reducible 't Hooft defect of charge $P$ inserted at the origin. Heuristically, this can be thought of as 
\be
\hfMM(P, \gamma_m;X_\infty)=\lim_{\vx^{(i)}\to 0} \fMM(\{h^{I(i)}\},\gamma_m;X_\infty)\quad, \quad P=\sum_{i=1}^p h^{I(i)}~.
\ee

In the quantum theory we are considering the product of line defects centered around the origin whose charges add up to $P$. This requires first inserting the constituent minimal line defects and then taking their coincident limit. Consequently, the expectation value $\langle L_{\vec{p},0}\rangle$ is really given by 
\be
\langle L_{\vec{p},0}\rangle=\lim_{\vx^{(i)}\to 0}\left\langle L_{[h^{I(1)},0]}(\vx^{\,(1)})... L_{[h^{I(p)},0]}(\vx^{\,(p)})\right\rangle~.
\ee
Thus, in the upcoming discussion, when we say that the path integral localizes to $\hfMM(P,\gamma_m;X_\infty)$, we truly mean that it localizes to $\fMM(\{h^{I(i)}\},\gamma_m;X_\infty)$. Similarly, the ``properties'' of $\hfMM(P,\gamma_m;X_\infty)$ are those of $\fMM(\{h^{I(i)}\},\gamma_m;X_\infty)$ that survive in the limit as $m_W\times|\vx^{\,(i)}-\vx^{\,(j)}|\to 0$ , $\forall i\neq j$ where $m_W$ is the mass of the lightest W-boson. 

An important feature of singular monopole configurations is monopole bubbling. This describes the process in which 
smooth monopoles are absorbed by the 't Hooft defect, thereby screening the 't Hooft charge. These configurations are described by the singular locus of singular monopole moduli space, which is itself 
a set of nested singular reducible monopole moduli spaces. 
When a minimally charged smooth monopole is absorbed via bubbling, the number of bulk moduli decreases by 4 and the magnetic charge of the 't Hooft defect decreases by a simple coroot. Consequently, the reducible singular monopole moduli space has a stratification structure 
\be
\hfMM(P,\gamma_m;X_\infty)=\coprod_{|\vv|\leq |P|}\hfMM^{(s)}(\vv,\gamma_m;X_\infty)~, 
\ee
where $\vv$ is the effective (screened) 't Hooft charge and $\hfMM^{(s)}(\vv,\gamma_m;X_\infty)$ is the smooth component of $\hfMM(\vv,\gamma_m;X_\infty)$ \cite{Nakajima:2016guo}. Each factor $\hfMM^{(s)}(\vv,\gamma_m;X_\infty)$ describes the moduli associated to the unbubbled smooth monopoles in the presence of the screened reducible 't Hooft defect of effective charge $\vv$. 

The full geometry of singular monopole moduli space is then specified by the 
geometry of the transversal slices of the $\hfMM^{(s)}(\vv,\gamma_m;X_\infty)$ inside the full moduli space $\hfMM(P,\gamma_m;X_\infty)$, which we will denote $\CM(P,\vv)$. Physically, $\CM(P,\vv)$ encodes the moduli of the bubbled monopoles confined to the world line of the 't Hooft defect. 

As shown in \cite{Nakajima:2016guo,Brennan:2018yuj}, these transversal slices $\CM(P,\vv)$ are quiver varieties defined by a quiver $\Gamma(P,\vv)$ that 
takes the general form

\tikzstyle{hex} = [regular polygon,regular polygon sides=6, draw,
    text width=2em, text centered]
\tikzstyle{hexs}= [regular polygon,regular polygon sides=5, draw,
    text width=1.7em, text centered]
\tikzstyle{line} = [draw, -latex']

\begin{center}
\begin{tikzpicture}[node distance = 2.2cm, auto]
\node [hex] (1) {$\Gamma_{0,1}$};
\node (2) [hexs,right of = 1] {$\Sigma_1$};
\node (3) [hex,right of =2] {$\Gamma_{1,2}$};
\node (4) [hexs,right of= 3] {$\Sigma_2$};
\node (8) [hexs,right of=4,xshift=+0.5cm] {$\Sigma_{N-1}$};
\node (9) [hex,right of=8] {$\Gamma_{N-1,N}$};
\draw[-] (1) --  (2);
\draw[-] (2) --  (3);
\draw[-] (3) --  (4);
\draw[dotted] (4) -- (8);
\draw[-] (8) --  (9);
\end{tikzpicture}
\end{center}

\noindent Here the sub-quivers $\Sigma_I$ are of the form

\begin{center}
\begin{tikzpicture}[node distance=1.3cm,
cnode/.style={circle,draw,thick,text centered},snode/.style={rectangle,draw,thick,minimum size=10mm,text centered}]
\node [cnode] (1) {$k_I$};
\node (2) [cnode, right of=1] {$k_I$};
\node (3) [right of=2,xshift=-0.1cm] {};
\node (4) [right of=3,xshift=-0.8] {};
\node (5) [cnode, right of=4,xshift=-0.1cm] {$k_I$};
\node (6) [cnode, right of=5] {$k_I$};
\node (7) [snode, below of=1] {$\omega_{I,I-1}$};
\node (8) [snode, below of =6] {$\omega_{I,I+1}$};
\node (9) [left of=1,xshift=.1cm] {=};
\node (10) [hexs,left of=9,xshift=.1cm] {$\Sigma_I$};
\draw[-] (1) -- (2);
\draw[-] (3) -- (2);
\draw[dotted] (3) -- (4);
\draw[-] (4) -- (5);
\draw[-] (5) -- (6);
\draw[-] (6) -- (8);
\draw[-] (1) -- (7);
\end{tikzpicture}
\end{center}
where 
\be
P-\vv=\sum_I k_I H_I\quad, \quad \omega_{I,J}=\begin{cases}0&k_I \leq k_J\\1&k_I >  k_J \end{cases}\quad, \quad P=\sum_I n_I \hat{h}^I~~,~~ \hat{h}^I\in \Lambda_{cochar}
\ee
with $k_0=0$ and $k_N=0$ and  $k_I$ is repeated  in $\Sigma_I$ 
\be
\ell_I=n_I + 1 -|k_{I+1}-k_I|\omega_{I,I+1}-|k_{I-1}-k_I|\omega_{I,I-1}~,
\ee
times. Additionally, the sub-quiver $\Gamma_{I,I+1}$ is given by 

\begin{center}
\begin{tikzpicture}[node distance=2.4cm,
cnode/.style={circle,draw,thick,text centered,minimum size=14mm},snode/.style={rectangle,draw,thick,minimum size=8mm,text centered}]
\node [cnode] (1) {$k_I+1$};
\node (2) [cnode,right of=1] {$k_I+2$};
\node (3) [right of=2,xshift=-1cm] {};
\node (4) [right of=3,xshift=-2cm] {};
\node (5) [cnode, right of=4,xshift=-1cm] {$k_{I+1}-2$};
\node (6) [cnode, right of=5] {$k_{I+1}-1$};
\node (7) [left of=1,xshift=1cm] {=};
\node (8) [hex,left of=7,xshift=1cm] {$\Gamma_{I,I+1}$};
\draw[-] (1) -- (2);
\draw[-] (2) -- (3);
\draw[dotted] (3) -- (4);
\draw[-] (4) -- (5);
\draw[-] (5) -- (6);
\end{tikzpicture}
\end{center}
when $k_I<k_{I+1}$ and 
\begin{center}
\begin{tikzpicture}[node distance=2.4cm,
cnode/.style={circle,draw,thick,text centered,minimum size=14mm},snode/.style={rectangle,draw,thick,minimum size=8mm,text centered}]
\node [cnode] (1) {$k_I-1$};
\node (2) [cnode,right of=1] {$k_I-2$};
\node (3) [right of=2,xshift=-1cm] {};
\node (4) [right of=3,xshift=-2cm] {};
\node (5) [cnode, right of=4,xshift=-1cm] {$k_{I+1}+2$};
\node (6) [cnode, right of=5] {$k_{I+1}+1$};
\node (7) [left of=1,xshift=1cm] {=};
\node (8) [hex,left of=7,xshift=1cm] {$\Gamma_{I,I+1}$};
\draw[-] (1) -- (2);
\draw[-] (2) -- (3);
\draw[dotted] (3) -- (4);
\draw[-] (4) -- (5);
\draw[-] (5) -- (6);
\end{tikzpicture}
\end{center}
when $k_I>k_{I+1}$. \footnote{Note that in the degenerate cases where $\ell_I=0$ or $k_I=k_{I+1}$, we omit the degenerate sub-quiver and identify the end nodes of the newly connecting components. See \cite{Brennan:2018yuj} for full details.}

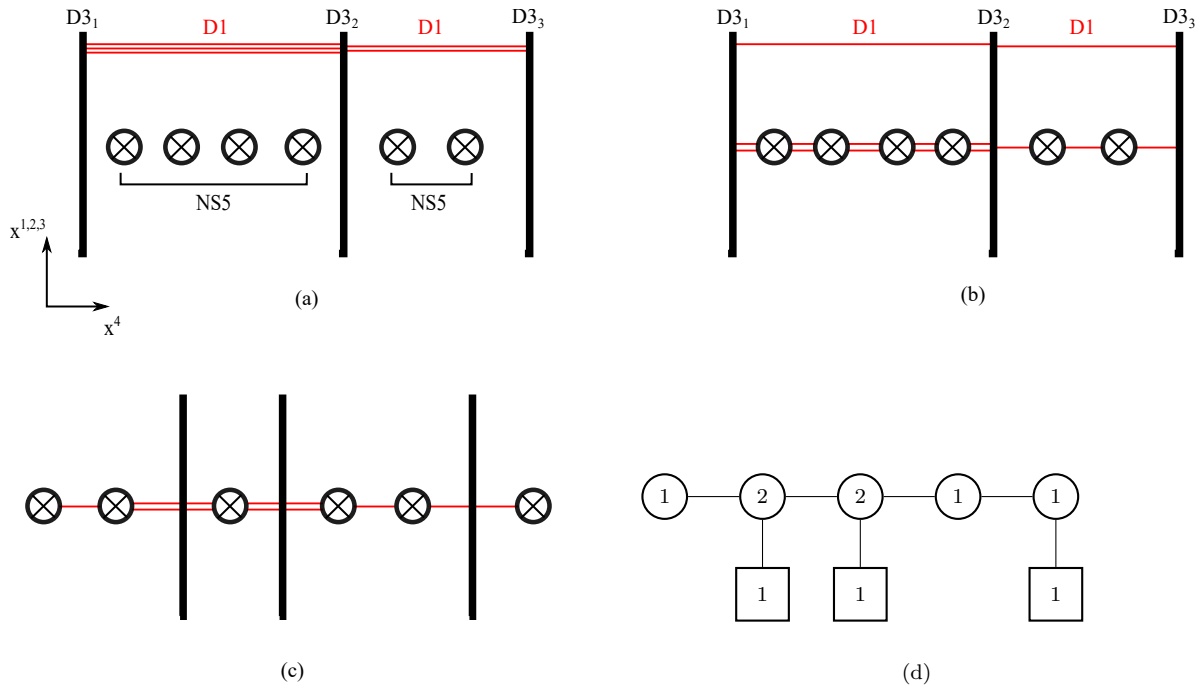
\begin{figure}
\includegraphics[scale=1,trim=0.5cm 24.5cm 4cm 0.6cm,clip]{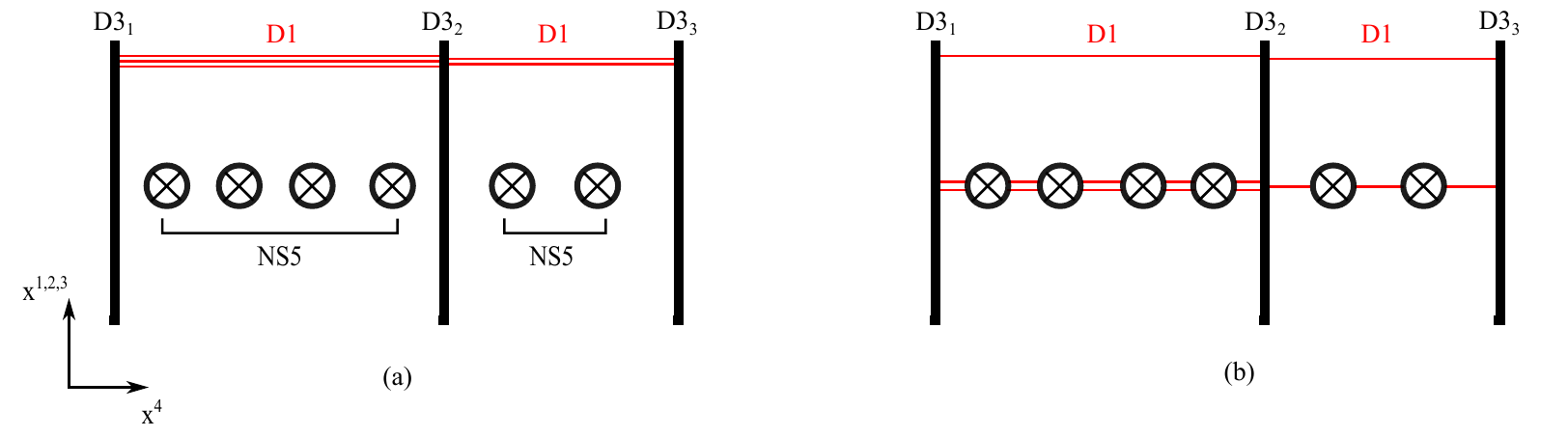}\\
\includegraphics[scale=1,trim=2.5cm 22cm 10cm 2.5cm,clip]{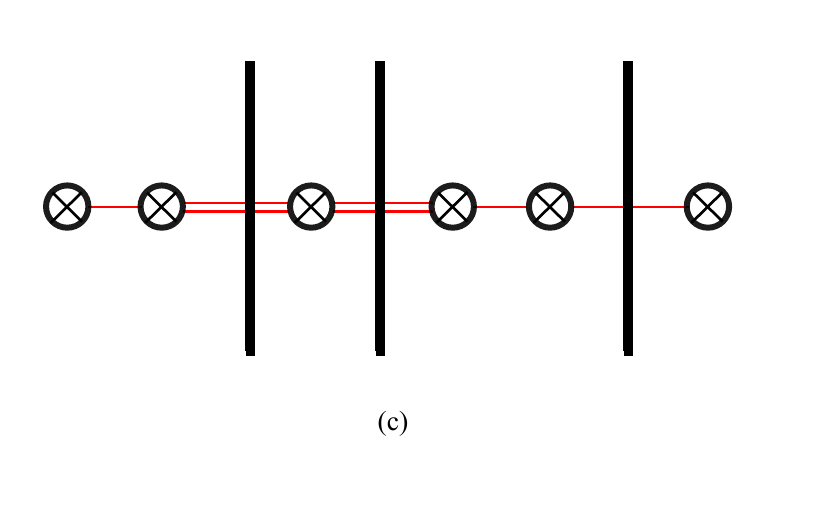}
\begin{tikzpicture}[node distance=1.3cm,
cnode/.style={circle,draw,thick,text centered},snode/.style={rectangle,draw,thick,minimum size=7mm,text centered}]
\node [cnode] (1) {1};
\node (2) [cnode, right of=1] {2};
\node (3) [cnode, right of=2] {2};
\node (4) [cnode, right of=3] {1};
\node (6) [cnode, right of=4] {1};
\node (7) [snode, below of=2] {1};
\node (8) [snode, below of =3] {1};
\node (9) [snode, below of =6] {1};
\node (10) [below of =8,yshift=-0.5cm,xshift=0.5cm] {};
\node (11) [below of=8,xshift=0.75cm,yshift=0.25cm]{(d)};
\draw[-] (1) -- (2);
\draw[-] (3) -- (2);
\draw[-] (3) -- (4);
\draw[-] (4) -- (6);
\draw[-] (7) -- (2);
\draw[-] (3) -- (8);
\draw[-] (6) -- (9);
\end{tikzpicture}
\caption{This figure shows many facets of the brane configuration describing singular monopoles and monopole bubbling in 4D $\CN=2$ SYM gauge theory for the example of $SU(3)$ gauge theory with $\gamma_m=3H_1+2H_2$ and $P=4 h^1+2h^2$ (a). (b)  displays an example of monopole bubbling where 3 monopoles have bubbled, screening the defect. By performing Hanany-Witten transformations (c), we can see that the effective SQM living on the D1-branes is given by a quiver SQM (d).}
\label{fig:HWMoves}
\end{figure}

\subsection{Brane Description and Fundamental Hypermultiplets}

\label{sec:branes}

In the quantum theory, the moduli of the bubbled monopoles gives rise to a $\CN=(0,4)$ quiver SQM on the world volume of the 't Hooft defect. This is specified by the same quivers that  define the transversal slices $\CM(P,\vv)$, namely $\Gamma(P,\vv)$. The $\Gamma(P,\vv)$ can be  derived via the following string theory construction \cite{Cherkis:1997aa,Brennan:2018yuj,Brennan:2018moe}.

Consider a 4D $\CN=2$ $SU(N)$ gauge theory on a stack of $N$ D3-branes\footnote{This requires projecting out the $U(1)$ center of mass degree of freedom and adding a mass deformation so that we integrate out the adjoint hypermultiplet of the $\CN=4$ vector multiplet. This can be implemented as an $\Omega$ deformation in the $x^{6,7,8,9}$-directions with parameter $m\to \infty$ similar to \cite{Hellerman:2011mv}. } localized at $x^{4,5,6,7,8,9}=0$. In a semiclassical chamber of the Coulomb branch, these branes separate along the $x^4$-direction such that they are localized at positions $\{x^4_I\}_{I=1}^N$ where $\sum_I x^4_I=0$. In the effective D3-brane world volume theory this seperation endows one of the real adjoint-valued Higgs fields, $X$, with the expectation value $X_\infty=\sum_I v_I H_I$ where  $v_I=x^4_{I+1}-x^4_I$.  
Here, smooth monopoles of charge $H_I$ are described by D1-branes localized at $x^{5,6,7,8,9}=0$ and $(x^1,x^2,x^3)=\vx_i$ that run between the D3$_I$- and D3$_{I+1}$-branes in the $x^4$-direction \cite{Diaconescu:1996rk}. 

In this construction of 4D $\CN=2$ $SU(N)$ SYM theory, a reducible 't Hooft defect at $\vx_\ast\in \IR^3$ of charge $P=\sum_I n_I \hat{h}^I$ are constructed from a collection of $\sum_I n_I$ NS5-branes that are localized at $(x^1,x^2,x^3)=\vx_\ast$, at distinct positions $x^4_\sigma$ in the $x^4$ direction with $n_I$ NS5-branes in between D3$_I$- and D3$_{I+1}$-brane where $\sigma=1,...,n=\sum_I n_I$ indexes the NS5-branes.  We can see that the collection of NS5-branes sources magnetic charge in the world volume of the D3-brane by performing a sequence of Hanany-Witten transformations. In this construction, Hanany-Witten transformations describe the action of pulling an NS5-brane through a D3-brane thereby creating a D1-brane that connects them \cite{Hanany:1996ie,Brennan:2018yuj}. Thus, by pulling all of the NS5-branes past the leftmost or rightmost D3-brane, we can go to a Hanany-Witten frame where all NS5-branes connect to the D3-branes via D1-branes. Then by the fact that the D1-branes end on codimension 3 submanifolds in the D3-branes,  we see that they do indeed source magnetic point charges in the world volume of the D3-branes \cite{Diaconescu:1996rk,Hanany:1996ie}.

Monopole bubbling then occurs when finite D1-branes stretched between the D3-branes become coincident with a collection of NS5-branes in the $(x^1,x^2,x^3)$-directions. The SQM describing bubbled degrees of freedom is given by the low energy effective theory of the bubbled D1-branes interacting with the NS5- and D3-branes \cite{Brennan:2018yuj}. This theory is espeically simple in a dual Hanany-Witten frame in which D1-branes only end on NS5-branes. In this frame, the effective D1-brane theory is  given by a quiver SQM described by the quivers $\Gamma(P,\vv)$ defined in Section \ref{sec:singmono}. In the D1-brane world volume theory, the $\Sigma_I$ sub-quivers correspond to the effective theory of the D1-branes stretched between NS5-branes in between the D3$_I$- and D3$_{I+1}$-branes that did not undergo a Hanany-Witten transformation while the quivers $\Gamma_{I,I+1}$ correspond to D1-branes ending on NS5-branes that were pulled past the D3$_I$-brane. See Figure \ref{fig:HWMoves}.

Now we will specialize to the case of an $SU(2)$ gauge theory. In this case we are considering the world volume theory of a stack of two D3-branes with finite D1-branes running between them. This theory can be coupled to fundamental hypermultiplets with complex masses\footnote{Note that here we have made a choice of $\zeta$ by choosing the D3-branes to be separated in the $x^4$-direction. This is also realized in the identification of directions for the real and complex parts of the mass of the fundamental hypermultiplets.} $\zeta^{-1}m^{(f)}=m_R^{(f)}+i m_I^{(f)}$ by introducing D7-branes that are localized at $x^4=m_I^{(f)}$, $x^5=m_R^{(f)}$.  As shown in \cite{Tong:2014yna}, this couples the quiver SQM describing the low energy effective theory of the D1-branes to a short $\CN=(0,4)$ fundamental Fermi-multiplet. Additionally, the inclusion of 4D fundamental matter mandates that the charge quantization of these theories satisfy $n_I\in 2\IZ_+$. \footnote{This is a consequence of the fact that in a theory with fundamental matter $\Lambda_{mw}/\Lambda_{cochar}=\IZ_2$. Hence, we will have $2p_I=n_I$ in these theories where the 't Hooft charge can be written as $P=\sum_I p_I h^I$ or $P=\sum_I n_I \hat{h}^I$ where $h^I\in \Lambda_{mw}$ and $\hat{h}^I\in \Lambda_{cochar}$. }

\begin{figure}
\includegraphics[scale=0.9,trim=0.5cm 21.5cm 4cm 0cm,clip]{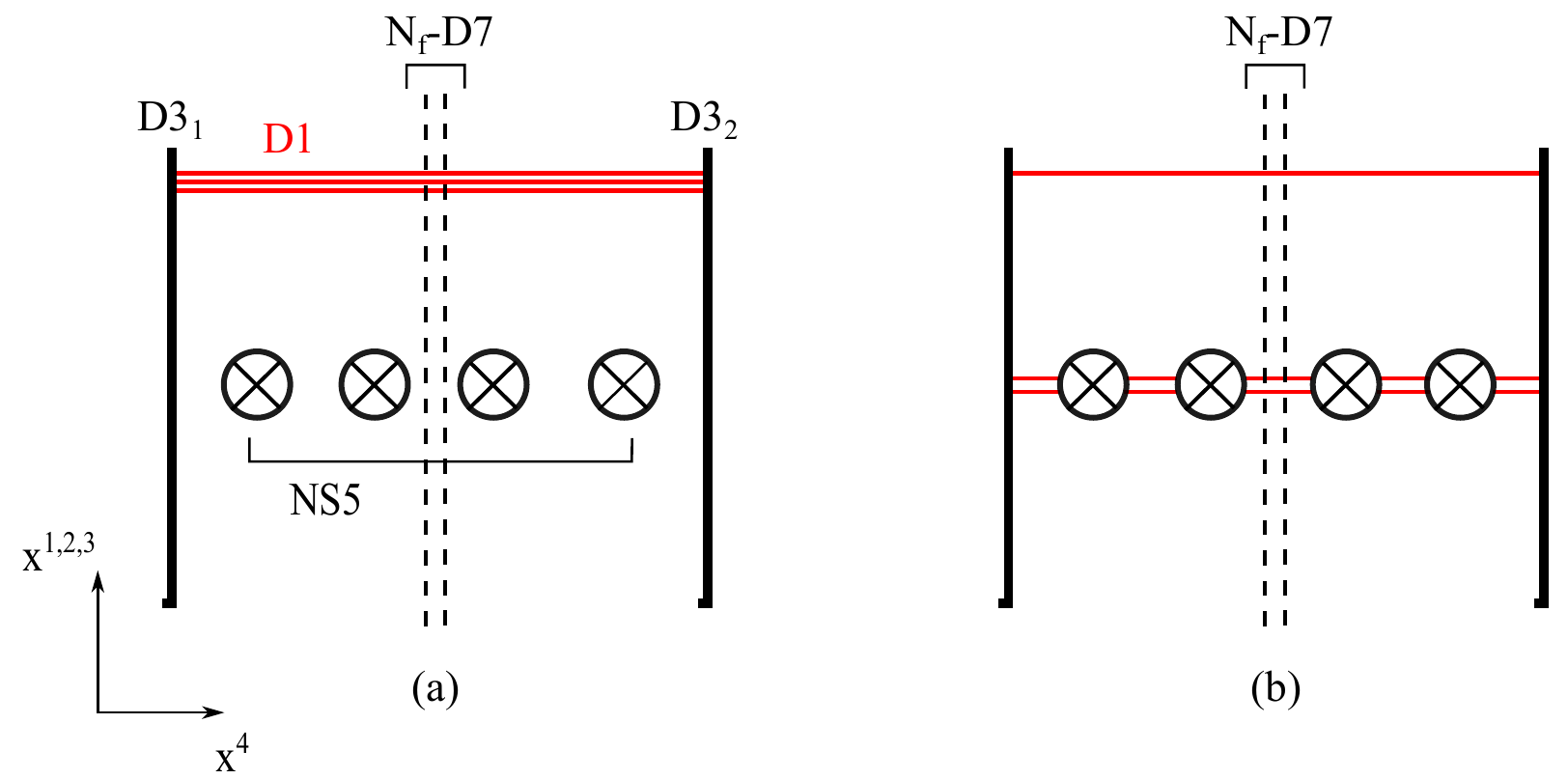}\\
\includegraphics[scale=0.9,trim=2.3cm 20cm 9cm 2cm,clip]{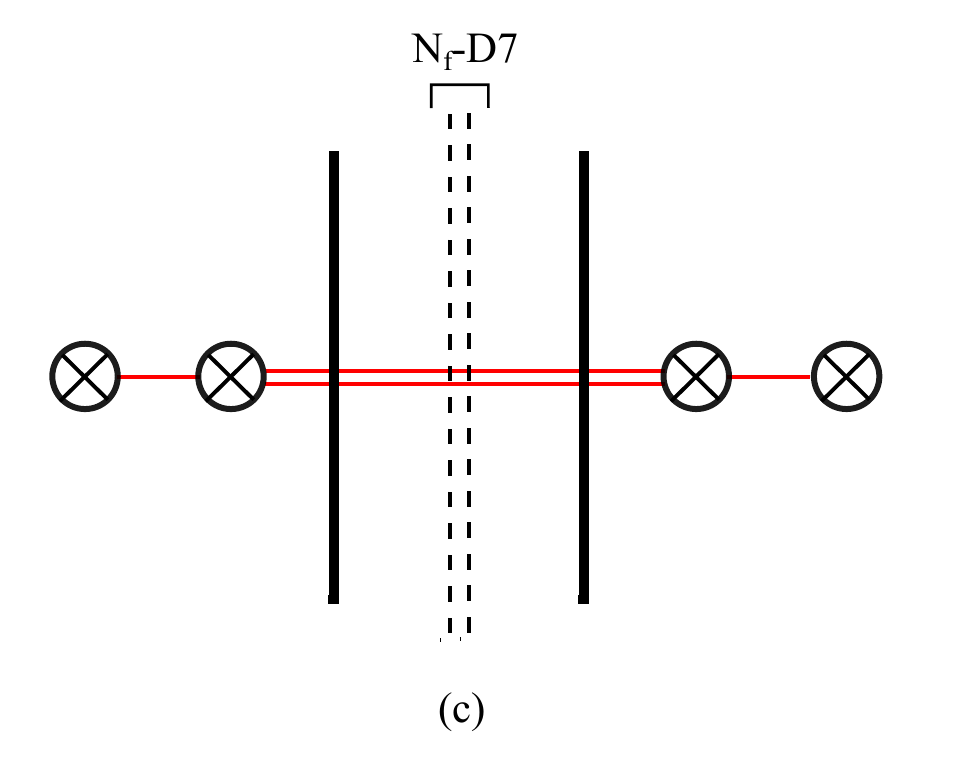}
\begin{tikzpicture}[node distance=1.3cm,
cnode/.style={circle,draw,thick,text centered},snode/.style={rectangle,draw,thick,minimum size=7mm,text centered}]
\node [cnode] (1) {1};
\node (2) [cnode, right of=1] {2};
\node (3) [cnode, right of=2] {1};
\node (4) [snode, below of=2] {2};
\node (5) [snode, above of=2] {N$_f$};
\node (10) [below of =4,yshift=-0.6cm] {};
\node (11) [below of=4,yshift=-0.5cm]{~\large{(d)}};
\node (12) [left of=1,xshift=-0.6cm]{};
\draw[-] (1) -- (2);
\draw[-] (3) -- (2);
\draw[-] (2) -- (4);
\draw[dashed] (2) -- (5);
\end{tikzpicture}
\caption{This figure shows many facets of the brane configuration describing singular monopoles and monopole bubbling in 4D $\CN=2$  gauge theory with N$_f$ fundamental hypermultiplets for the example of $SU(2)$ gauge theory with  $\gamma_m=3H_1$ 
and $P=4 h^1$ 
 (a). (b)  displays an example of monopole bubbling where 2 monopoles have bubbled,  screening the defect. By performing the Hanany-Witten transformations (c), we can see that the SQM living on the D1-branes is given by a quiver SQM (d).}
\label{fig:HWMovesD7}
\end{figure}
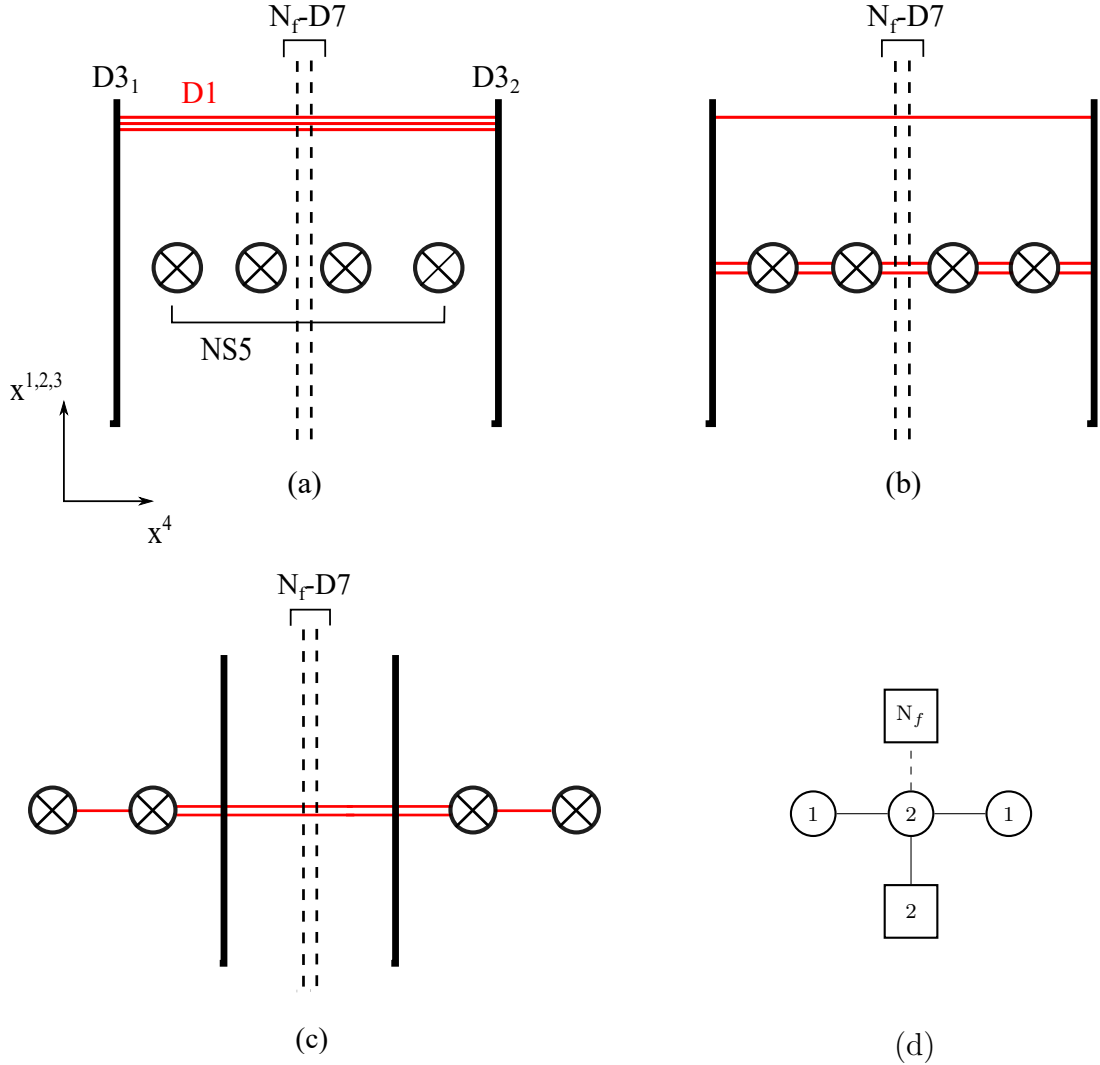

This brane configuration is summarized in the following table:

\begin{center}
\begin{tabular}{l|ccccc}
&$x^0$&$(x^1,x^2,x^3)$&$x^4$&$x^5$&$(x^6,x^7,x^8,x^9)$\\
\hline
D3$_I$&$-$&$-$&$x^4_I$&0&$\vec{0}$\\
D1$_i$&$-$&$\vx^{(i)}$&$[x^4_{I(i)},x^4_{I(i+1)}]$&0&$\vec{0}$\\
NS5$_\sigma$&$-$&$\vx_\ast$&$x^4_\sigma$&$-$&$-$\\
D7$_f$&$-$&$-$&$m_I^{(f)}$&$m_R^{(f)}$&$-$\\
\end{tabular}
\end{center}

However, since the D7-branes couple to the center of mass mode of the D3-branes, it is unclear how they depend on the initial choice of position of the NS5-branes and how they behave under Hanany-Witten transformations. Thus, it is unclear from this picture how the fundamental Fermi-multiplets corresponding to the 4D fundamental hypermultiplets couple to the bubbling SQM.  Rather, it is more clear if we use the T-dual brane configuration which is studied more carefully in \cite{Brennan:2018moe,Witten:2009xu}. 

Consider wrapping the D1/D3/NS5-brane configuration on a periodic $x^4$-direction and T-dualizing. As shown in \cite{Brennan:2018moe,Witten:2009xu}, the dual brane configuration is given by a stack of D4-branes wrapping charge $n$ Taub-NUT $(TN_n)$ where $n=\sum_I n_I$ whose gauge bundle is specified via Kronheimer's correspondence of the singular monopole configuration of the D3-branes \cite{KronCorr}. Notably, the gauge bundle is $U(1)_K$-invariant where $U(1)_K$ is the group of translations along the $S^1$-fiber in the $x^4$-direction whose lift to the gauge bundle is determined by the NS5-branes configuration \cite{Brennan:2018moe,Witten:2009xu}. 
In this description, the bubbling SQM is given by the theory of the bubbled $U(1)_K$-invariant D0-branes which is again a  quiver SQM of the form described above where now each node corresponds to a  weight under the $U(1)_K$ action \cite{Witten:2009xu,Brennan:2018yuj}. 

Adding hypermultiplets to the 4D theory corresponds to adding D8-branes to the T-dual theory. \footnote{Since we are studying the supersymmetric gauge theory in the weak coupling domain, it is reasonable to neglect the back reaction from the D8-branes as we do here. } Each D8-brane is localized at $x^5=m_R^{(f)}$  and wraps the Taub-NUT space. \footnote{This only corresponds to a real mass deformation that can be analytically continued to a complex mass.} 
 Each D8-brane is T-dual to a D7-brane that gives rise to a fundamental hypermultiplet in the 4D gauge theory on the D3-branes.   In this setting, global flavor symmetry of the 4D theory implies that the gauge field on the stack of D8-branes, 
 which must be $U(1)_K$-invariant, is flat. 
Here the $U(1)_K$-invariance is necessary so that there is a 1-1 mapping between the effective 5D and 4D theory under T-duality. 

However, due to the non-trivial action of $U(1)_K$ on Taub-NUT, flat gauge connections on Taub-NUT are graded by the first Chern class which we can identify with the weight of the $U(1)_K$ action \cite{Brennan:2018yuj,Brennan:2018moe}. Thus, since we have global flavor symmetry in 4D, the gauge field of the D8-branes must have trivial first Chern class and hence zero weight under the action of $U(1)_K$. 
%
%
Then, since the Fermi-multiplets come from D0-D8 strings, the only $U(1)_K$ invariant modes are those that couple the D0-branes with trivial $U(1)_K$ weight to the D8-branes: all other modes are projected out under T-dualizing back to the 4D theory. Thus, 
the fundamental Fermi-multiplets coming from 4D fundamental hypermultiplets couple to the gauge node with  vanishing $U(1)_K$-weight, which corresponds to the unique, central node of $\Gamma(P,\vv)$.  

In summary, the bubbling SQM quiver, $\Gamma(P,\vv)$,  for $SU(2)$ gauge theories with ${\rm N}_f$-flavors is generically of the form

\begin{center}
\begin{tikzpicture}[
cnode/.style={circle,draw,thick,minimum size=9mm},snode/.style={rectangle,draw,thick,minimum size=9mm}]
\node[cnode] (1) {1};
\node[cnode] (2) [right=.4cm  of 1]{2};
\node[cnode] (3) [right=.4cm of 2]{3};
\node[cnode] (5) [right=0.8cm of 3]{\tiny{$k-1$}};
\node[cnode] (6) [right=0.4cm of 5]{$k$};
\node[cnode] (7) [right=0.8cm of 6]{$k$};
\node[cnode] (9) [right=0.8cm of 7]{$k$};
\node[cnode] (10) [right=0.4cm of 9]{\tiny{$k-1$}};
\node[cnode] (13) [right=0.8cm of 10]{{$3$}};
\node[cnode] (14) [right=0.4cm of 13]{$2$};
\node[cnode] (17) [right=0.4cm of 14]{1};
\node[snode] (18) [below=0.5cm of 6]{1};
\node[snode] (19) [below=0.5cm of 9]{1};
\node[snode] (20) [above=0.5cm of 7]{${\rm N}_f$};
\draw[-] (1) -- (2);
\draw[-] (2)-- (3);
\draw[dotted] (3) -- (5);
\draw[-] (5) --(6);
\draw[dotted] (7) -- (9);
\draw[dotted] (6) -- (7);
\draw[-] (9) -- (10);
\draw[dotted] (10) -- (13);
\draw[-] (13) -- (14);
\draw[-] (14) -- (17);
\draw[-] (6) -- (18);
\draw[-] (9) -- (19);
\draw[dashed] (20) -- (7);
\end{tikzpicture}
\end{center}
\noindent where the length of the quiver is $n-1$ with $k$ occuring $n-2k+1$ times where
\be
P=n\,\hat{h}^1\quad, \quad P-\vv=k H_1\quad, \quad n\in 2\IZ_+~~,~~\hat{h}^1\in \Lambda_{cochar}~. 
\ee
and  the ${\rm N}_f$ fundamental Fermi-multiplets are coupled to the $(n/2)^{th}$ gauge node. 
Additionally, when $n=2k$, $\Gamma(P,\vv)$ takes the special form

\begin{center}
\begin{tikzpicture}[
cnode/.style={circle,draw,thick,minimum size=9mm},snode/.style={rectangle,draw,thick,minimum size=9mm}]
\node[cnode] (1) {1};
\node[cnode] (2) [right=.4cm  of 1]{2};
\node[cnode] (3) [right=.4cm of 2]{3};
\node[cnode] (5) [right=0.8cm of 3]{\tiny{$k-1$}};
\node[cnode] (6) [right=0.4cm of 5]{$k$};
\node[cnode] (10) [right=0.4cm of 6]{\tiny{$k-1$}};
\node[cnode] (13) [right=0.8cm of 10]{{$3$}};
\node[cnode] (14) [right=0.4cm of 13]{$2$};
\node[cnode] (17) [right=0.4cm of 14]{1};
\node[snode] (18) [below=0.5cm of 6]{2};
\node[snode] (20) [above=0.5cm of 6]{${\rm N}_f$};
\draw[-] (1) -- (2);
\draw[-] (2)-- (3);
\draw[dotted] (3) -- (5);
\draw[-] (5) --(6);
\draw[-] (6) -- (10);
\draw[dotted] (10) -- (13);
\draw[-] (13) -- (14);
\draw[-] (14) -- (17);
\draw[-] (6) -- (18);
\draw[dashed] (20) -- (6);
\end{tikzpicture}
\end{center}

\subsection{Theories of Class $\CS$ and the AGT Correspondence}
\label{sec:AGT}

In this paper, we are considering asymptotically free $\CN=2$ $SU(2)$ gauge theories with fundamental and adjoint hypermultiplets. These  are theories of class $\CS$.

Theories of class $\CS$ are a  set of 4D $\CN=2$ supersymmetric gauge theories 
with semi-simple, simply laced gauge groups. They are constructed by compactifying a  corresponding 6D $\CN=(0,2)$ theory on a Riemann surface $C$ with a topological twist that makes the theory independent of the scale of $C$. Because of this, the expectation value of SUSY operators in the 4D theory, which descend from SUSY operators in the 6D theory, are equal to the expectation value of a corresponding operator in the 2D theory on $C$ \cite{Alday:2009aq,Alday:2009fs}.


For theories of class $\CS$ with $SU(N)$ gauge group, the above construction is equivalent to wrapping a stack of $N$ M5-branes on $C$ with a topological twist. In this case, the corresponding 2D theory is $A_{N-1}$ Toda theory on the closure of $C$ denoted $\overline{C}$. Here, punctures of $C$ are associated with a flavor symmetry of 4D hypermultiplets and come with the data of a mass parameter specifying the 4D flavor symmetry. In the associated 2D Toda theory, each puncture corresponds to a vertex operator insertion in the path integral whose weight is determined by the  associated mass parameter \cite{Gaiotto:2010be,Alday:2009fs,Drukker:2009id,Drukker:2009tz}.

We are interested in computing the expectation value of magnetically charged line defects in the 4D theory. In theories of class $\CS$, line defects descend from strings in the 6D theory that wrap the 2-manifold $\gamma\times S^1_t\subset C\times (\IR^3\times S^1_t)$  where $\gamma$ is a closed 1-dimensional submanifold of $C$ that does not go into the punctures. 
The electromagnetic charge of the associated 4D line defect in an S-duality frame is determined by the homology class of $\gamma\subset C$ with respect to the weak coupling cut decomposition of $C$ corresponding to the S-duality frame. See \cite{Gaiotto:2010be,Drukker:2009tz,Drukker:2009id,Alday:2009fs} for more details.
 
In the 2D Toda theory, a line defect associated to a closed curve $\gamma$ corresponds to a  loop operator  $\CL_\gamma$. 
This can be computed by \cite{Alday:2009fs,Verlinde:1988sn,Moore:1989vd}
 \be
 \langle L_{\vec{p},0}\rangle_{T_4[SU(N),C]}=\left\langle
 \left( \prod_f V_{m_f} \right)\CL_{\gamma_{\vec{p}}}\right\rangle_{{\rm Toda}[A_{N-1},\overline{C}]}~,
 \ee
where $T_4[SU(N),C]$ is the type $SU(N)$ 4D theory of class $\CS$ corresponding to the Riemann curve $C$,  the $\{V_{m_f}\}$ are the vertex operators corresponding to the punctures of $C$ with mass parameters $\{m_f\}$, and $\gamma_{\vec{p}}$ is the curve corresponding to the operator $L_{\vec{p},0}$ \cite{Alday:2009fs,Drukker:2009tz,Drukker:2009id}. 

The expectation values of the line defects in theories of class $\CS$ are holomorphic functions on the Seiberg-Witten moduli space. Because of this, they can be expressed in terms of complexified Fenchel-Nielson coordinates which are defined as follows. Pick a weak coupling region of the Coulomb branch. Geometrically, this corresponds to picking a complex structure and maximal set of non-intersecting cycles $\{\gamma_{i}\}$ on $C$. 
Using the cuts $\{\gamma_i\}$, we can define a set of holomorphic functions on moduli space, denoted $\{\fa_i\}\in \ft_\IC$ by 
\be\label{adef}
\left\langle \left(\prod_f V_{m_f}\right) \CL_{\gamma_i}\right\rangle_{{\rm Toda}[A_{N-1},\bar{C}]}={\rm Tr}_N e^{\fa_i}~.
\ee
These functions can be seen to be Poisson commuting 
with respect to the standard, symplectic (2,0)-form $\Omega_J$ \cite{Kapustin:2006pk,Dimofte:2011jd}:
 \be
\Omega_J\left(\frac{\partial}{\partial\fa_i},\frac{\partial}{\partial\fa_j}\right)=0~,
\ee
and in fact, they form a maximal set of Poisson commuting holomorphic functions. They can be used as the first half of a set of holomorphic Darboux coordinates.  In terms of the parameters entering the path integral in a given weak coupling description, these coordinates have a semiclassical expansion
\be\label{semia}
\fa_i=i \theta_e^{(i)}- 2\pi  \beta Y^{(i)}_\infty+...,
\ee
where $A_{0,\infty}^{(i)}$ is the holonomy of the 4D gauge field $A^{(i)}_0$ of the $i^{th}$ factor of the 4D gauge group $G=\prod_i SU(N)_i$ along the thermal circle of radius $\beta$ at infinity and $Y_\infty^{(i)}$ is the real part of the Higgs vev $\zeta^{-1}\Phi_\infty^{(i)}=Y_\infty^{(i)}+i X_\infty^{(i)}$ of $i^{th}$ factor of the gauge group. In the above expression the ellipses (...) correspond to non-perturbative corrections \cite{BrDM2,Gaiotto:2008cd}.

Now we can define a set of dual coordinates $\{\fb_i\}$ with respect to the symplectic (2,0)-form $\Omega_J$ such that 
\be
\Omega_J=\frac{1}{\hbar}\sum_i {\rm Tr}_N\left(d\fa_i\wedge d\fb_i\right) ~.
\ee
 Note that this choice is only defined up to $\fb_i\mapsto \fb_i+f(\fa_i)$. However, in a weak coupling domain, we can fix $\fb_i$ by specifying its semiclassical limit
\be\label{semib}
\fb_i=i\theta_m^{(i)}+\frac{8\pi ^2\beta}{e^2}X^{(i)}_\infty-\vartheta \beta Y^{(i)}_\infty+...~,
\ee
where here $\theta_m^{(i)}$ is the magnetic theta angle of the $i^{th}$ gauge group factor, $X_\infty^{(i)}$ is the real part of the Higgs vev of the $i^{th}$ gauge group factor, and $\vartheta$ is the real part of the complex gauge coupling of the  4D gauge group. \footnote{Here we take the gauge coupling to be identical for all gauge groups by fixing the normalization of the Killing form.} Again, we have that the ellipses (...) indicate that $\fb_i$ has non-perturbative corrections. The non-perturbative corrections to $\fa_i,\fb_i$ can be understood as the consequence of holomorphy on all of the Seiberg-Witten moduli space.

For this paper, our main example will be the minimally charged 't Hooft defect in the 4D $\CN=2$ $SU(2)$ gauge theory with ${\rm N}_f=4$ fundamental hypermultiplets. In this case, the 2D theory is the Liouville theory on the 4-punctured sphere. Here the choice of S-duality frame corresponds to picking a pants decomposition \cite{Gaiotto:2009we}. We will label the punctures for the two different pairs of pants by $(m_1,m_2)$ and $(m_3,m_4)$ respectively. The minimally charged 't Hooft defect is then the simple loop that wraps around $m_1,m_3$. 

The expectation value of the minimal 't Hooft defect in this theory is of the form \cite{Ito:2011ea} 
\begin{align}\begin{split}\label{L10AGT}
&\big\langle L_{1,0}\big\rangle=(e^{ \fb}+e^{- \fb})F(\fa)+Z_{mono}~,\\
&F(\fa)=\left(\prod_\pm \frac{\prod_{f=1}^4 \sinh(\fa\pm m_f)}{\sinh^2(2 \fa)\sinh(2\fa\pm \epsilon_+)}\right)^{1/2}~,\\
&Z_{mono}=-4\sum_{\pm}\frac{\prod_{f=1}^4\sinh(\pm\fa-m_f-\epsilon_+)}{\sinh(2 \fa)\sinh(2\fa\mp2\epsilon_+)}+\cosh \left(\sum_{f=1}^4 m_f+2\epsilon_+\right)~. 
\end{split}\end{align}
where, by abuse of notation, $\fa,\fb$ in the above expression are one of the eigenvalues of the corresponding complexified Fenchel-Nielson coordinates. 

The key feature here is that there is an extra term $\cosh \left(\sum_{f} m_f+2\epsilon_+\right)$. As we will see, this term is special in the sense that it is the only term not reproduced in the standard Jeffrey-Kirwan residue prescription \cite{Ito:2011ea}. While the explicit form of $Z_{mono}(P,\vv)$ for 't Hooft defects with higher charge is generically unknown, \footnote{Note that there are an infinite class of known results for $\epsilon_+=0$. The reason is that $\langle L_{p,0}\rangle_{\epsilon_+=0}=\left(\langle L_{1,0}\rangle_{\epsilon_+=0}\right)^p$. In general, $\langle L_{p,0}\rangle_{\epsilon_+\neq 0}$ can be derived via repeated application of the Moyal product \cite{Ito:2011ea} 
\be
\langle L_{p,0}\rangle=\langle L_{1,0}\rangle \ast ...\ast \langle L_{1,0}\rangle ~,
\ee
where
\be
(f\ast g)(\fa,\fb)=e^{-\epsilon_+(\partial_b\partial_{a'}-\partial_a\partial_{b})}f(a,b)g(a',b')\big{|}_{\substack{a,a'=\fa\\b,b'=\fb}}~.
\ee
However, no general, closed form expression is known for $\langle L_{p,0}\rangle_{\epsilon_+\neq0}$ for arbitrary $p$. 
 See \cite{Gaiotto:2010be,Ito:2011ea,Drukker:2009id} for more details. } the presence of an ``extra term'' which is not captured by the Jeffrey-Kirwan residue prescription is a generic feature of minimally charged 't Hooft defects in $\CN=2$ $SU(N)$ gauge theories with ${\rm N}_f=2N$ hypermultiplets \cite{Ito:2011ea}.

\section{Localization for $Z_{mono}(P,\vv)$}

\label{sec:sec3}

Now we will attempt to compute the expectation value of 't Hooft defects by using localization.

Consider the 4D $SU(2)$ $\CN=2$ gauge theory on $\IR^3\times S^1$ with ${\rm N}_f\leq 4$ fundamental hypermultiplets. The field content of this theory consists of a $\CN=2$ $SU(2)$ vector multiplet ($\Phi,\psi_A,A_\mu)$ and N$_f$ fundamental hypermultiplets $(q_A^{(f)},\lambda^{(f)})$ with masses $m_f$ where $f=1,...,{\rm N}_f$. We will express these hypermultiplets as a single hypermultiplet $(q_A,\lambda)$ that transforms under the bifundamental representation of $G\times G_f=SU(2)\times SU({\rm N}_f)$ with a single mass parameter $m\in \ft_f\subset \fg_f$. Here $q_A,\psi_A$ are a scalar- and Weyl fermion-doublets transforming under the spin-$\half$ representation of $SU(2)_R$ and $\lambda$ is a Dirac fermion.

This theory is described by the Lagrangian \cite{Moore:2015szp,Brennan:2016znk}:
\begin{align}\begin{split}
&L=L_{vec}+L_{hyp}~,\\
&L_{vec}=\frac{1}{g^2}\int d^3x~\text{Tr}\Bigg( \half F^{\mu\nu}F_{\mu\nu}+|D_\mu \Phi|^2-\frac{1}{4}[\Phi,\Phi^\dagger]^2-2i\bar\psi^A \bar\sigma^\mu D_\mu\psi_A
-i \psi^A[\Phi^\dagger,\psi_A]+i \bar\psi^A[\Phi,\bar\psi_A]\Bigg)\\&\hspace{3cm}
+\frac{\vartheta}{8\pi^2}\int \text{Tr}\left(F\wedge F\right)~,\\
&L_{hyp}=\frac{1}{g^2}\int d^3x~\Bigg(|D_\mu q_A|^2+2i \bar\lambda  \slashed{D}\lambda+|m q_A|^2-i m q^{\dagger A} \Phi^\dagger q_A-i m^\ast q^{\dagger A}\Phi q_A-2m_R\bar\lambda \lambda+2i m_I\bar\lambda \gamma^5\lambda\\&\qquad\quad~-i \bar\lambda\Phi \lambda-i \lambda^T\Phi^\dagger \lambda^\ast
+2q^A \bar\lambda \Psi_A+2\bar\Psi^A q^\dagger_A \lambda+\half q^A\{\Phi,\Phi^\dagger\}q_A+\frac{1}{8}(q^{\dagger A}T^a(\tau_{s})_{A}^{~B} q_B)^2\Bigg)~,
\end{split}\end{align}
where $s=1,2,3$ is summed over, $(\tau_s)_A^{~B}$ are the $SU(2)_R$ generators, $\Psi^T_A=(\psi_A, \bar\psi_A)$ is a Dirac fermion, and $m=m_R+i m_I$.

The supersymmetry transformations of these fields are
\begin{align}\begin{split}\label{eq:SUSY}
&\delta_\xi  \Psi_A=-i \sigma^{\mu\nu}F_{\mu\nu}\xi_A+i \sigma^\mu D_\mu \bar\xi_A+\frac{i}{2}\xi_A[\Phi,\Phi^\dagger]~,\\
&\delta_\xi  \Phi=2\xi^A\psi_A\quad\,,\qquad 
\delta_\xi  A_\mu =\xi^A\sigma_\mu \bar\psi_A+\bar\psi^A\bar\sigma_\mu \psi_A~,\\
&\delta_\xi  q_A=2 \bar\Xi_A \lambda\quad~,\qquad\delta_\xi  \lambda=i \gamma^\mu \bar\Xi^A D_\mu q_A-(i \Phi^\dagger+m^\ast)q_A(\Xi^\ast)^A~,
\end{split}\end{align}
where $\Xi^T_A=(\xi_A, \bar\xi_A)$ is a Dirac-fermion doublet of SUSY transformation parameters that transforms in the spin $\half$-representation of $SU(2)_R$. 

Now let us include a (reducible) 't Hooft  operator specified by the data $(P,\vx=0,\zeta)$. The gauge field singularity at $\vx=0$  requires adding a local boundary term to the action specified  by $\zeta$: \footnote{Really we must take a sum of $p$ boundary terms (where the charge of the reducible 't Hooft defect is $P=p \, h^1$), each centered at $\vx^{(i)}$, and then take the limit as $\vx^{(i)}\to 0$. Each of these corresponds to the boundary condition for a constituent minimal 't Hooft defect inserted at $\vx^{(i)}$. To represent a single reducible 't Hooft defect, we require taking the limit $\vx^{(i)},\epsilon^{(i)}\to 0$ such that $|\vx^{(i)}|/\epsilon^{(i)} \to 0$ where the physical boundary term for each minimal 't Hooft defect is inserted on a 2-sphere of radius $\epsilon^{(i)}$ surrounding $\vx^{(i)}$. For simplicity, we will ignore this subtlety in the main discussion.}
\be
S_{bnd}=-\frac{1}{g^2}\lim_{\epsilon\to0}\int_{S^2_{\epsilon}(\vec0)} \text{Tr}\left(\text{Im}[\zeta^{-1}
\Phi]F+\text{Re}[\zeta^{-1}
\Phi]\ast F\right),
\ee
where $S^2_\epsilon(\vec0)$ is the 2-sphere of radius $\epsilon$ centered at $\vx=\vec0$. 

This insertion manifestly breaks $\half$-supersymmetry. The choice of $\zeta\in U(1)$ defines the conserved symmetries to be  generated by a parameter $\rho^A$ that is defined by 
\be
\xi^A=\zeta^\half(\rho^A+i \pi^A)~,
\ee
where $\rho^A,~\pi^A$ are symplectic-Majorana-Weyl fermions. \footnote{Symplectic-Majorana-Weyl fermions satisfy: $
\rho^A=\epsilon^{AB}\bar\sigma^0\bar\rho_B.$} 

In the upcoming discussion, we will choose to localize with respect to the conserved, real supercharge
\be\label{chargedef}
\CQ= \rho^A Q_A+\bar\rho_A \bar{Q}^A
~, 
\ee
where $Q_A$ is the complex supercharge of the full $\CN=2$ SUSY algebra. This 
$\CQ$ satisfies the relation 
\be
\CQ^2=H+\fa Q_\fa+\epsilon_+ J_++m \cdot F~,
\ee
where $H$ is the Hamiltonian, $Q_\fa$ is the charge associated with global gauge transformations with fugacity $\fa$, $J_+$ is the charge associated to supersymmetric rotations\footnote{These are spatial rotations with an $R$-charge rotation.}  in $\IR^3$ that we associate with $\epsilon_+$ in a $\half$-$\Omega$ background, and $F$ is the set of conserved flavor charges. 
We will write the conserved symmetry group $\tilde{T}=T_{gauge}\times U(1)_\epsilon\times T_f$ where $T_{gauge}$ is the maximal torus of the 4D gauge group, which describes the group of global gauge transformations, and $T_f$ is the maximal torus of the flavor symmetry group. 

Now we can attempt to compute the expectation value of the 't Hooft defect by using localization. The localization principle states that the expectation value of a $\CQ$-invariant operator is invariant under a $\CQ$-exact deformation of the Lagrangian
\be
\CL\to \CL+t \CQ\cdot V~.
\ee
Then, by studying the limit as $t\to \infty$, we see that the path integral localizes to the zeros of $V$ that are fixed under the action of $\tilde{T}$. As in \cite{Ito:2011ea,Gomis:2011pf}, if we make a choice
\be
V=({\CQ}\cdot \bar\lambda,\lambda)+({\CQ}\cdot \bar\psi_A,\psi^A)~,
\ee
then the path integral localizes to the zeroes of $\CQ\cdot \lambda$ and $\CQ\cdot \psi^A$. This reduces the path integral to an integral over (the $\tilde{T}$-invariant subspace of) the moduli space of BPS equations. Note that since shifting $t$ is a $\CQ$-exact deformation of the action, the localization behavior of the path integral is independent of the value of $t$. 

In our case, the associated  BPS equations (before $\half\Omega$-deformation) are given by \footnote{To regularize the path integral, we will need to turn on a $\half$-$\Omega$-deformation that modifies the BPS equations. However, the $\tilde{T}$-fixed locus of the moduli space of the deformed BPS equations will be identical to the $\tilde{T}$-fixed locus of the moduli space of the undeformed BPS equations. See 
 \cite{Gomis:2011pf,Brennan:2018yuj} for the explicit form and more details.}
\begin{align}\begin{split}\label{eq:BPS}
&D_i X=B_i\quad~,\qquad E_i=D_i Y~,\\
& D_t Y=0\qquad,\qquad D_t X-[Y,X]=0~,\\
&D_i q=0~\qquad,\qquad D_0 q+(Y+m_I)q+i(X-m_R)q=0~,
\end{split}\end{align}
where $B_i,E_i$ are the magnetic and electric field respectively and $m$ is rotated by the phase $\zeta$:  $\zeta^{-1}m=m_R+i m_I$. The solutions to these equations with respect to the 't Hooft defect \eqref{tHooftBC} and asymptotic boundary conditions \eqref{AsymptoticBC} are given exactly by singular monopole moduli space \cite{Ito:2011ea,Gomis:2011pf,Brennan:2018yuj}. 

Thus, the expectation value of the 't Hooft defect localizes to an integral over the $\tilde{T}$-fixed locus of singular monopole moduli space with measure determined by the 1-loop determinant times the exponential of the classical action. \footnote{To be precise, we are computing the expectation value of the 't Hooft operator with fixed electric and magnetic theta angle $\theta_e,\theta_m$. The electric theta angle is defined by fixing the holonomy of the gauge connection along the circle at infinity
\be
\oint_{S^1_\infty}A_t dt=\theta_e~.
\ee
The magnetic theta angle is defined as the Fourier dual of path integral with fixed magnetic charge $\langle L_{\vec{p},0}\rangle_{\gamma_m}$:
\be\label{fourierTheta}
\langle L_{\vec{p},0}\rangle_{\theta_m}=\sum_m \langle L_{\vec{p},0}\rangle_{\gamma_m} e^{- i (\gamma_m ,\theta_m)}~.
\ee 
Thus, by saying that the path integral ``localizes to singular monopole moduli space'', we mean that each term in the Fourier sum \eqref{fourierTheta} reduces to an integral over the reducible singular monopole moduli space $\fMM(P,\gamma_m;X_\infty)$. Due to the universality of the geometry of the transversal slices/bubblign SQMs, we will find that this subtlety is irrelevant for the calculation of $Z_{mono}(P,\vv)$. 
See \cite{Brennan:2018yuj} for more details. }
 In this integral, the classical action is determined by the effective bulk charge sourced by the 't Hooft defect. Since singular monopole moduli space decomposes as the disjoint union of bubbling sectors with different effective charges, the expectation value of the line defect reduces to a sum of integrals over the $\tilde{T}$-fixed locus of different strata of the bubbling locus. Thus, the expectation value is of the form
\be
\langle L_{\vec{p},0}\rangle=\sum_{\substack{|\vv|\leq |P|\\\vv\in \Lambda_{cr}+P}} Z(\fa,\fb,m_f,\epsilon_+;P,\vv)~,
\ee
where $Z(\fa,\fb,m_f,\epsilon_+;P,\vv)$ is the reduction of the localized integral over $\hfMM(P,\gamma_m;X_\infty)$ to the strata $\hfMM_{\tilde{T}}^{(s)}(\vv,\gamma_m;X_\infty)$ and the corresponding transverse slice $\CM(P,\vv)$. 

By integrating over the $\tilde{T}$-fixed subspace of each $\hfMM_{\tilde{T}}^{(s)}(\vv,\gamma_m;X_\infty)$, the computation for $Z(\fa,\fb,m_f,\epsilon_+;P,\vv)$ can be further reduced to a $\tilde{T}$-equivariant integral over the transversal slice of each strata, $\CM(P,\vv)$ \cite{Ito:2011ea,Gomis:2011pf,Brennan:2018yuj}. As shown in \cite{Brennan:2018yuj}, we can identify the universal coefficient of the integrand with $e^{(\vv,\fb)} Z_{1-loop}(\fa,m_f,\epsilon_+;\vv)$ and the remaining, integral dependent part as $Z_{mono}(\fa,m_f,\epsilon_+;P,\vv)$. This will lead to the form of the expectation value of the 't Hooft defect
\be
\langle L_{\vec{p},0}\rangle=\sum_{\substack{|\vv|\leq |P|\\\vv\in \Lambda_{cr}+P}} e^{ (\vv,\fb)} Z_{1-loop}(\fa,m_f,\epsilon_+;\vv) Z_{mono}(\fa,m_f,\epsilon_+;P,\vv)~,
\ee
where, $Z_{mono}(P,\vv)$ can be identified as:
\be\label{charint}
Z_{mono}(\fa,m_f,\epsilon_+;P,\vv)= \int_{\CM(P,\vv)}e^{\omega +\mu_{\tilde{T}}}\hat{A}_{\tilde{T}}(T\CM)\cdot C_{\tilde{T}}(\CV(\CR))~.
\ee
Here, $\hat{A}_{\tilde{T}}$ is the $\tilde{T}$-equivariant $\hat{A}$ genus, $C_{\tilde{T}}$ is a $\tilde{T}$-equivariant characteristic class that depends on the matter content of the theory,  $e^{\omega+\mu_{\tilde{T}}}$ is the equivariant volume form, and $\fa,m_f,\epsilon_+$ enter the expression as the equivariant weights under the $\tilde{T}$-action. It will be crucial to us that $Z_{mono}(P,\vv)$ is \textit{independent }of $\beta$. \footnote{$\fa$ is treated as independent of $\beta$. } See \cite{Moore:1997dj,Ito:2011ea,Brennan:2018yuj} for more details.

\subsection{Bubbling SQMs}
\label{sec:bubblingSQMs}

 As shown in \cite{Brennan:2018yuj},  $Z_{mono}(P,\vv)$ can be physically interpreted as the contribution of an SQM localized on the 't Hooft defect of charge $P$ that has an effective charge $\vv$. 
This leads to the interpretation of the  integral in \eqref{charint} as  the localized path integral of the bubbling SQM. 
Then, since the (twisted) path integral of a SQM is formally equal to its Witten index, the monopole bubbling contribution, $Z_{mono}(P,\vv)$ can be expressed as the Witten index of the corresponding bubbling SQM specified by the quiver $\Gamma(P,\vv)$ as in Section \ref{sec:singmono}:
\be\label{WittenInd}
Z_{mono}(P,\vv)=I_W(\Gamma(P,\vv)):=
{\rm Tr}_{\CH_{\Gamma(P,\vv)}}~(-1)^F e^{-\frac{\beta}{2} \{\CQ,\CQ\}+\fa Q_\fa+\epsilon_{_+} Q_\epsilon+m\cdot F}~,
\ee
where  $Q_\fa$ is the charge for the flavor symmetry associated with 4D global gauge transformations, $Q_\epsilon$ is an $R$-charge associated to the $\half\Omega$-deformed background, \footnote{$Q_\epsilon$ can also be understood as an $R$-symmetry charge in the bubbling SQM. The $\CN=(0,4)$ bubbling SQMs we are consider have an $SU(2)_R$ $R$-symmetry and an $SU(2)_{r}$ outer-automorphism ``$R$-symmetry''. Here the $Q_\epsilon$ is diagonal combination of the Cartans: $Q_{\epsilon}=Q_R-Q_{r}$. See \cite{Tong:2014yna} for more details. 
} and $F$ is the set of conserved flavor charges.

 
 \begin{figure}
 \begin{center}
 \includegraphics[scale=1,trim=6cm 21.5cm 2cm 1cm,clip]{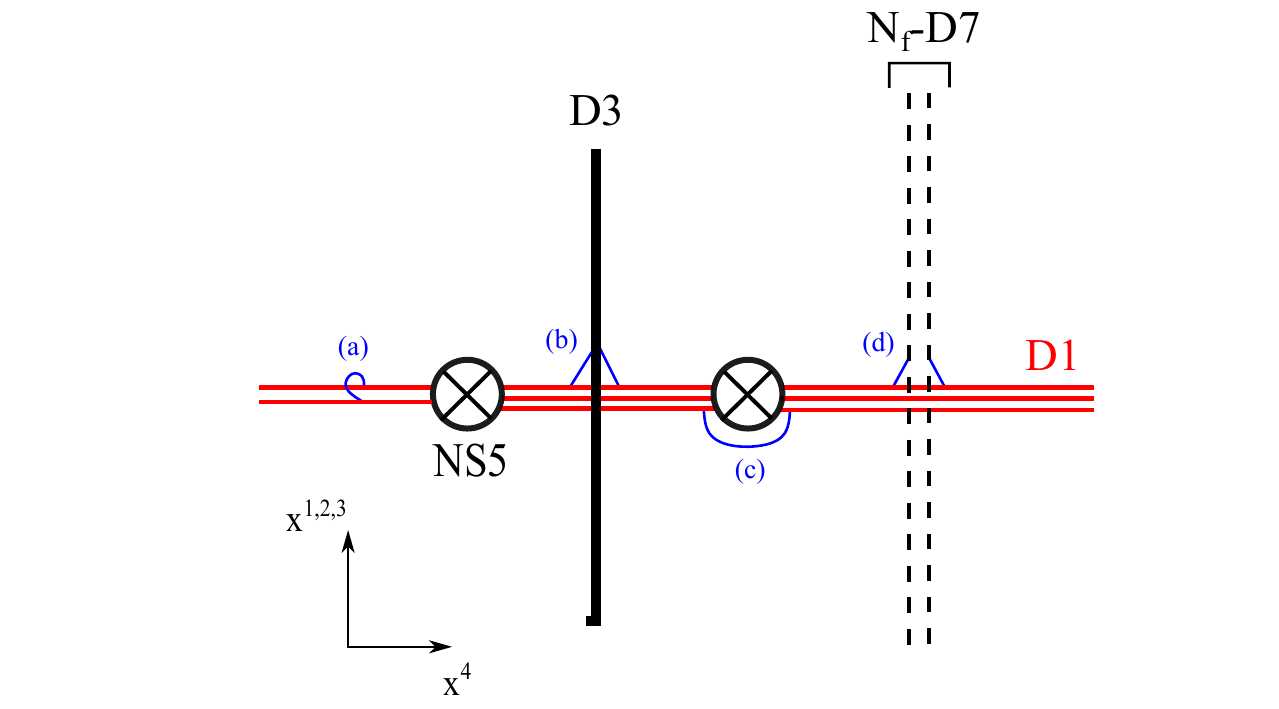}
 \end{center}
 \caption{This figure illustrates which strings give rise to the various fields in the bubbling SQM. (a) describes the D1-D1 strings that give rise to a $\CN=(0,4)$ vector multiplet with fields $(v_t,\sigma,\lambda^A)$, (b) describes the D1-D3 strings that give rise to $\CN=(0,4)$ fundamental hypermultiplets with fields $(\phi^A,\psi_I)$, (c) illustrates D1-D1' strings that give rise to $\CN=(0,4)$ bifundamental hypermultiplets with fields $(\uphi^A,\upsi_I)$, and (d) describes D1-D7 strings that  give rise to the short $\CN=(0,4)$ Fermi multiplets with fields $(\eta,G)$.  }
 \label{fig:SQMStrings}
 \end{figure}

The bubbling SQM specified by the quiver $\Gamma(P,\vv)$  is given by compactifying the 2D $\CN=(0,4)$ quiver gauge theory. Let us use the notation $G=\prod_{i=1}^{p-1} U(k^{(i)})$   for the gauge group of the SQM such that the corresponding Lie algebra $\fg$ decomposes as $\fg=\bigoplus_{i=1}^{n-1}\fg^{(i)}=\bigoplus_{i=1}^{n-1} \fu(k^{(i)})$ and its Cartan subalgebra $\ft=\bigoplus_{i=1}^{n-1}\ft^{(i)}=\bigoplus_{i=1}^{n-1}\fu(k^{(i)})$ where $P=n \,\hat{h}^1$, $\hat{h}^I\in \Lambda_{cochar}$. \footnote{Recall that here we use the notation $L_{p,0}$ for $P=p h^1$, $h^1\in \Lambda_{cochar}$. }

 Each gauge node corresponds to a $\CN=(0,4)$ vector multiplet with constituent fields $(\sigma^{(i)},\lambda^{A(i)},v_t^{(i)})$, where $i=1,..., n-1$ indexes the gauge nodes. In the string theory interpretation of Section \ref{sec:branes}, these vector multiplets arise from the D1$_i$-D1$_i$ strings on the stack of D1$_i$-branes stretched between the NS5$_i$- and NS5$_{i+1}$-branes. Additionally there are $\CN=(0,4)$ fundamental hypermultiplets  with constituent fields $(\phi^{(i)},\psi^{(i)})\oplus (\tildephi^{(i)},\tildepsi^{(i)})$ that come from D1$_i$-D3 strings and $\CN=(0,4)$ bifundamental hypermultiplets with constituent fields $(\uphi^{(i)},\upsi^{(i)})\oplus(\tilde\uphi^{(i)},\tilde\upsi^{(i)})$ that come from the D1$_{i}$-D1$_{i+1}$ strings at NS5-branes. 
Also, in the case of theories with 4D fundamental hypermultiplets, there are additional $\CN=(0,4)$ short Fermi-multiplets with constituent fields $(\eta^{(i)},G^{(i)})$ coming from D3-D7
 strings. See Figure \ref{fig:SQMStrings}. 
Additionally,  see \cite{Tong:2014yna,Hori:2014tda} for more details on $\CN=(0,4)$ SQMs. 


The bubbling SQM  has a Lagrangian that decomposes as a sum of terms
\be
L=L_{vec}+L_{Fermi}+L_{f}+L_{bf}~,
\ee
which describe the contributions from vector multiplets, Fermi-multiplets, fundamental hypermultiplets, and bifundamental hypermultiplets respectively. These contributions can be found in Appendices \ref{app:A} and \ref{app:B}. 
%
Here we will pick the convention where the gauge couplings for each factor in the gauge group are equal to $e^2$ by fixing a universal normalization of the Killing form for the SQM Lie algebra. 

Now we wish to compute the Witten index of the bubbling SQM. This requires an understanding of the spectrum of the bubbling SQM.  We can infer an approximate version of the spectrum from the classical moduli space and its surrounding potential. In this SQM, the potential energy is of the form 
\begin{align}\begin{split}\label{potential}
U&=\frac{1}{e^2}\sum_i \left(|\sigma^{(i)}\cdot\phi^{(i)}|^2+|\sigma^{(i)}\cdot\tildephi^{(i)}|^2\right)\\
&+\frac{1}{e^2}\sum_i\left(|(\sigma^{(i+1)}-\sigma^{(i)})\cdot\uphi^{(i)}|^2+|(\sigma^{(i+1)}-\sigma^{(i)})\cdot\tilde\uphi^{(i)}|^2\right)\\&+
\frac{1}{2e^2}\sum_i\left(|\phi^{(i)}|^2-|\tilde\phi^{(i)}|^2-|\uphi^{(i)}|^2+|\tilde\uphi^{(i)}|^2+|\uphi^{(i-1)}|^2-|\tilde\uphi^{(i-1)}|^2\right)^2\\&
+\frac{1}{e^2}\sum_i\left|\phi^{(i)}\tilde\phi^{(i)}-\uphi^{(i)}\tilde\uphi^{(i)}+\uphi^{(i-1)}\tilde\uphi^{(i-1)}\right|^2~,
\end{split}\end{align}
where here we are using scalar contraction. 
Thus, the 
moduli space is 
defined by the solutions to the equations:
\begin{align}\begin{split}\label{groundstatebare}
&0=|\sigma^{(i)}\phi^{(i)}|^2\qquad\qquad\quad~, \quad 0=|\sigma^{(i)}\tildephi^{(i)}|^2~,\\
&0=|(\sigma^{(i+1)}-\sigma^{(i)})\uphi^{(i)}|^2\quad, \quad 0=|(\sigma^{(i+1)}-\sigma^{(i)})\tilde\uphi^{(i)}|^2~,\\
&0=|\phi^{(i)}|^2-|\tilde\phi^{(i)}|^2-|\uphi^{(i)}|^2+|\tilde\uphi^{(i)}|^2+|\uphi^{(i-1)}|^2-|\tilde\uphi^{(i-1)}|^2~,\\
&0=\phi^{(i)}\tilde\phi^{(i)}-\uphi^{(i)}\tilde\uphi^{(i)}+\uphi^{(i-1)}\tilde\uphi^{(i-1)}~,
\end{split}\end{align}
for each $i$. 

The solutions of these equations can be divided into Coulomb, Higgs, and mixed branches
\be
\CM_{vac}=\CM_{C}\cup\CM_{H}\cup \CM_{mix}~,
\ee
where 
\begin{align}\begin{split}\label{vacbranchesC}
&\CM_{C}=\{\sigma^{(i)}\in\ft^{(i)}~,~ \sigma^{(i)}\neq \sigma^{(i+1)}~,~\phi^{(i)},\tildephi^{(i)},\uphi^{(i)},\tilde\uphi^{(i)}=0\}~,\\
&\CM_{H}=\Bigg\{\begin{array}{l}|\phi^{(i)}|^2-|\tilde\phi^{(i)}|^2-|\uphi^{(i)}|^2+|\tilde\uphi^{(i)}|^2+|\uphi^{(i-1)}|^2-|\tilde\uphi^{(i-1)}|^2=0\\
\phi^{(i)}\tilde\phi^{(i)}-\uphi^{(i)}\tilde\uphi^{(i)}+\uphi^{(i-1)}\tilde\uphi^{(i-1)}=0\end{array}~,~\sigma^{(i)}=0\,\Bigg\}\Bigg{/}\CG~,
\end{split}\end{align}
 and $\CG$ is the group of gauge transformations. The mixed branch is significantly more complicated to write down in full generality, but it should be thought of as having 
asymptotic directions as in the Coulomb branch for some subset of directions of $\sigma\in \ft$ and some hypermultiplet scalars with non-zero expectation value. 
Because of this hybrid quality, the mixed branch, like the Coulomb branch, is non-compact and, like the Higgs branch,  is a singular manifold. 

We can additionally add an FI-deformation to the theory
\be
L_{FI}=-(\xi ,D)=-\sum_i \xi^{(i)} D^{(i)}~.
\ee
This contribution changes the potential to 
\begin{align}\begin{split}\label{defpotential}
U&=\frac{1}{e^2}\sum_i \left(|\sigma^{(i)}\cdot\phi^{(i)}|^2+|\sigma^{(i)}\cdot\tildephi^{(i)}|^2\right)\\
&+\frac{1}{e^2}\sum_i\left(|(\sigma^{(i+1)}-\sigma^{(i)})\cdot\uphi^{(i)}|^2+|(\sigma^{(i+1)}-\sigma^{(i)})\cdot\tilde\uphi^{(i)}|^2\right)\\&+
\frac{1}{2e^2}\sum_i\left(|\phi^{(i)}|^2-|\tilde\phi^{(i)}|^2-|\uphi^{(i)}|^2+|\tilde\uphi^{(i)}|^2+|\uphi^{(i-1)}|^2-|\tilde\uphi^{(i-1)}|^2-e^2\xi^{(i)}\right)^2\\&
+\frac{1}{e^2}\sum_i\left|\phi^{(i)}\tilde\phi^{(i)}-\uphi^{(i)}\tilde\uphi^{(i)}+\uphi^{(i-1)}\tilde\uphi^{(i-1)}\right|^2~.
\end{split}\end{align}
This  lifts the classical vacua associated to the Coulomb branch  along with certain non-compact directions in the mixed branch by modifying the D-term vacuum equation to
\begin{align}\begin{split}
e^2\xi^{(i)}=|\phi^{(i)}|^2-|\tilde\phi^{(i)}|^2-|\uphi^{(i)}|^2+|\tilde\uphi^{(i)}|^2+|\uphi^{(i-1)}|^2-|\tilde\uphi^{(i-1)}|^2~.
\end{split}\end{align}
Consequently, when $\xi^{(i)}\neq 0$, the hypermultiplet scalar fields cannot all simultaneously satisfy $\phi^{(i)},\tilde\phi^{(i)},\uphi^{(i)},\tilde\uphi^{(i)}=0$. 

Additionally, the FI-deformation resolves the singularities of the mixed and Higgs branches and lifts certain directions in the mixed branch. Now the Higgs branch can be written as a (resolved) \hk quotient
\be
\CM_{H}=\vec\mu^{\,-1}(\vec\xi\,)\Big{/}\CG\quad,\quad\vec\xi=(\xi_\IR,\xi_\IC)=(e^2\xi,0)~,
\ee
where 
\begin{align}\begin{split}
&\mu_\IR=|\phi^{(i)}|^2-|\tilde\phi^{(i)}|^2-|\uphi^{(i)}|^2+|\tilde\uphi^{(i)}|^2+|\uphi^{(i-1)}|^2-|\tilde\uphi^{(i-1)}|^2~,\\
&\mu_\IC=\phi^{(i)}\tilde\phi^{(i)}-\uphi^{(i)}\tilde\uphi^{(i)}+\uphi^{(i-1)}\tilde\uphi^{(i-1)}~.
\end{split}\end{align}

Now in order to couple the Witten index to flavor fugacities, let us add masses for the hypermultiplet fields. These can be defined as flat connections coming from an associated flavor symmetry. We will choose to turn on mass parameters corresponding to the $\half$-$\Omega$ deformation with a mass parameter $\epsilon_+={\rm Im}[\epsilon_+/\beta]$ and to a fugacity for 4D global gauge symmetry with mass parameter $a={\rm Im}[\fa/\beta]$.  \footnote{This identification allows us to work with a unitary theory. We can then derive the full Witten index by analytic continuation. See \cite{Festuccia:2011ws} for more details.} These mass deformations modify the mass terms in the potential \eqref{defpotential}:
\begin{align}\begin{split}\label{massdefvac}
U&=\sum_i \frac{1}{e^2}\left|(\sigma^{(i)}+a Q_\fa+\epsilon_+ Q_\epsilon)\cdot\phi^{(i)}\right|^2+\frac{1}{e^2}\left|(\sigma^{(i)}+a Q_\fa+\epsilon_+ Q_\epsilon)\cdot\tildephi^{(i)}\right|^2\\
&+\frac{1}{e^2}\sum_i\left|\big(\sigma^{(i+1)}-\sigma^{(i)}+a Q_\fa+\epsilon_+ Q_\epsilon\big)\cdot\uphi^{(i)}\right|^2+\left|\big(\sigma^{(i+1)}-\sigma^{(i)}+a Q_\fa+\epsilon_+ Q_\epsilon\big)\cdot\tilde\uphi^{(i)}\right|^2\\&+
\frac{1}{2e^2}\sum_i\left(|\phi^{(i)}|^2-|\tilde\phi^{(i)}|^2-|\uphi^{(i)}|^2+|\tilde\uphi^{(i)}|^2+|\uphi^{(i-1)}|^2-|\tilde\uphi^{(i-1)}|^2-e^2\xi^{(i)}\right)^2\\&
+\frac{1}{e^2}\sum_i\left|\phi^{(i)}\tilde\phi^{(i)}-\uphi^{(i)}\tilde\uphi^{(i)}+\uphi^{(i-1)}\tilde\uphi^{(i-1)}\right|^2~.
\end{split}\end{align}
where $Q_\fa\cdot \Phi$ and $Q_\epsilon\cdot \Phi$ encode the $Q_\fa,Q_\epsilon$ charges of the field $\Phi$.  See Appendices \ref{app:A}, \ref{app:B} for details about the charges of the fields.

The mass deformation lifts most of the  Higgs and mixed branch vacua except at a collection of intersecting hyperplanes where hypermultiplet scalars become massless. This reduces the Higgs branch to a collection of points while reducing the mixed branch so that it only has non-compact directions coming from vector multiplet scalars. 
 The mass deformations additionally give a mass of $4\epsilon_+$ to the fermionic component $\lambda^2$ which breaks SUSY $\CN=(0,4)\to \CN=(0,2)$ under which the $\CN=(0,4)$ vector multiplet decomposes as a $\CN=(0,2)$ vector multiplet $(v_t,\sigma,\lambda^1,D)$ and a $\CN=(0,2)$ Fermi-multiplet $(\lambda^2,F)$.  With this choice, $Q=\rho^A Q_A$ is the preserved complex supercharge.

Due to the form of \eqref{massdefvac}, the potential around each of the vacuum branches is quadratically confining. In the limit $e^2\to 0$, this potential becomes infinitely steep and states become exactly localized on the moduli spaces.  Since the Higgs branch is given by a collection of points, in the limit $e^2\to 0$, this supports an infinite, discrete spectrum of harmonic oscillator-like states. However, the mixed branch, which has non-compact directions, supports both a discrete spectrum of bound states and a continuum of scattering states. 

In addition, there are also states localized on the classically lifted Coulomb and mixed branches. Even though the potential energy on these branch is no longer zero, it is bounded. Again the potential in the normal direction is quadratically confining such that in the limit $e^2\to 0$, the states become exactly localized on the lifted branches. Thus, the the Coulomb and lifted mixed branches constitute non-compact directions in field space with finite potential energy which can also support both a discrete spectrum of bound states and a continuum of scattering states. See Section \ref{sec:HiggsCoulomb} for further discussion of Higgs, mixed, and Coulomb Branch states. 


\subsubsection{Localization}

Now we will attempt to compute the partition function of this theory by using localization. While parts of the following analysis have also been done using similar methods in \cite{Benini:2013nda,Hori:2014tda}, we will find it instructive and physically insightful to present the full derivation of the localization computation.  

The key to using localization in this setting is that the action of these theories is $Q$-exact. That is to say, we can rewrite the Lagrangian
\be
L=\frac{1}{e^2} Q\cdot V_{vec}+\frac{1}{e^2} Q\cdot V_{matter}~, 
\ee
with
\begin{align}\begin{split}
&V_{vec}=\sum_i\left(\bar{Q}\cdot \bar\lambda^{\left(i\right)}_A,\lambda^{\left(i\right)A}\right)~,\\
&V_{matter}=\sum_i\Bigg[\left(\bar{Q}\cdot \bar\psi^{\left(i\right)},\psi^{\left(i\right)}\right)+\left(\bar{Q}\cdot\bar\tildepsi^{\left(i\right)},\tildepsi^{\left(i\right)}\right)+\left(\bar{Q}\cdot\bar\eta^{\left(i\right)},\eta^{\left(i\right)}\right)\Bigg]
\end{split}\end{align}
Thus, shifting the value of $e$ is a supersymmetric deformation of the theory. This means that the result of localization should be independent of $e$ and therefore we will take $e$ to be generic and strictly positive. \footnote{Note that these are actually dimensionful quantities in the 1D SQM. These have dimension $[e^2]=\ell^{-3}$. Thus to take the ``$e\to 0$'' limit, we must take $\ell^3 e^2\to 0$ where $\ell$ is some fixed length scale. In our discussion we will use the FI-parameter $\xi$ as our fixed length scale since in the upcoming discussion we want $\beta$ to be variable.}

Now by the localization principle,  the partition function reduces to an integral over the 
 the moduli space of the time independent BPS equations: \footnote{See \eqref{SQMSUSY1}-\eqref{SQMSUSY2} for the full SUSY transformations. }
\be
[\sigma,v_t]=0\quad ,\quad M_r=0~.
\ee
These BPS equations have a moduli space of solutions given  by 
\footnote{Note that this rescaling enforces the periodicity condition $\varphi\sim \varphi+2\pi i\lambda$  for $\lambda\in \Lambda_{cr}$.} 
\be
\CM_{BPS}=\Big\{\varphi=  \beta (\sigma+i v_t)\big{|}_{t=0}\in ( \ft\times T)/W\Big\}\cong (\ft_\IC/\Lambda_{cr})/W=\widehat\CM/W~, \ee
where $\ft$ is the Lie algebra corresponding to the torus $T$ of the SQM gauge group as defined by the quiver $\Gamma(P,\vv)$. Note that this $\varphi$ is not to be confused with the hypermultiplet fields $\phi_i,\tilde\phi_i,\uphi_i,\tilde\uphi_i$. 
Now as in \cite{Hori:2014tda,Benini:2013nda}, the Wick rotated path integral is reduced to 
\be
Z^{(Loc)}=\int_{\CM_{BPS}} \frac{d^{2r}\varphi}{(2\pi i )^r}  Z_{det}(\varphi) = \frac{1}{|W|} \int_{\widehat\CM} \frac{d^{2r}\varphi}{(2\pi i )^r} \,Z_{det}(\varphi) ~, 
\ee
where $r={\rm rnk}\,\fg$ and $\fg$ is the Lie algebra of the gauge group of the quiver SQM. 
The 1-loop determinant $Z_{det}$ can now be computed 
in the background given by the zero-mode $\varphi$. For quiver SQMs this is of the form \cite{Hori:2014tda}
\be
Z_{det}(\varphi)=\int_{\ft} d^rD \,Z_{int}(\varphi,D)~{\rm exp}\left\{-\frac{\pi \beta(D,D)}{e^2}+2\pi i \beta(\xi ,D)\right)~,
\ee
where 
\be\label{Zint}
Z_{int}(\varphi,D)=Z_{vec}(\varphi)\cdot Z_{Fermi}(\varphi)\cdot Z_{hyp}(\varphi,D)~. 
\ee
Here $Z_{vec}$, $Z_{Fermi},$ and $Z_{hyp}$ are the 1-loop determinants from the vector-, Fermi-, and hyper- multiplet fields respectively. 

First consider the vector multiplet contribution. This term originates solely from  vector multiplet fermions. 
The reason is that there are no propagating modes of $\varphi$ due to Gauss's law and there are no propagating modes of $D$ due to the lack of a kinetic term \cite{Hori:2014tda}. Thus, the contributions to $Z_{vec}$ come from integrating over the non-zero modes of $\lambda^A$. Note that $\lambda^2$ does not have any zero modes because it has a generic, non-zero mass due to the $\Omega$-deformation. 

Explicitly, the vector multiplet fermions give the contribution
\begin{align}\begin{split}\label{veccont}
Z_{vec}
&= \prod_{i=1}^{n-1} \prod_{\alpha \in \Delta_{adj}^{(i)}}\sinh(\alpha(\varphi^{(i)})+ q_i)\prod_{\substack{\alpha \in \Delta_{adj}^{(i)}\\ \alpha\neq 0
}}\sinh(\alpha(\varphi^{(i)}))~,
\end{split}\end{align}
where $\Delta_{adj}^{(i)}$ are the weights of the adjoint representation of the $i^{th}$ simple summand of the gauge group and $q_i$ represent the coupling to all global charges associated to the $\CN=(0,2)$ Fermi-multiplet of the $\CN=(0,4)$ vector multiplet (since only $\CN=(0,2)$ SUSY is preserved).

Similarly, the contribution from the Fermi-multiplet is given by only by the 1-loop determinant of the fermions which can be written as 
\begin{align}\begin{split}\label{fermicont}
Z_{Fermi}&=\prod_{f=1}^{{\rm N}_f} \prod_{\mu\in \Delta_{fund}^{(f)}}\sinh(\mu(\varphi^{(f)})+ q_f)~,
\end{split}\end{align}
where $q_f$ encodes the coupling to all global charges and $\varphi^{(f)}$  is the  complex vector multiplet scalar that couples to the $f^{th}$ Fermi multiplet. 

Now consider the contribution from hypermultiplets. This term can be divided into two parts 
\be\label{hypermultIntegrand}
Z_{hyp}=Z_{hyp}^{(kin)}\cdot Z_{hyp}^{(Yuk)}~,\ee 
where $Z_{hyp}^{(kin)}$ comes from kinetic terms of the hypermultiplet fields and $Z_{hyp}^{(Yuk)}$ comes from integrating out Yukawa interactions. Explicitly, these are of the form 
\begin{align}\begin{split}
Z_{hyp}^{(kin)}&=\prod_j \prod_{\mu\in \Delta_{hyp}^{(j)}}\prod_{ m\in \IZ}\frac{(\pi m-i\mu(\bar\varphi^{(j)})-i\bar{q}_j)}{(|\pi m+i\mu(\varphi^{(j)})+iq_j|^2+i  \mu(D)}\\
&=\prod_j \prod_{\mu\in \Delta_{hyp}^{(j)}}\frac{\sinh(\mu(\bar\varphi^{(j)})+\bar{q}_j)}{\sinh(\alpha_{j,\mu}^+)\sinh(\alpha_{j,\mu}^-)}~,\\
Z_{hyp}^{(Yuk)}&= {\rm det}\,h^{ab}(\varphi,D)~,
\end{split}\end{align}
where $j$ indexes the  set of fundamental and bifundamental $\CN=(0,2)$ chiral multiplets making up the $\CN=(0,4)$ hypermultiplets  and $\varphi^{(j)}$ is the complex vector multiplet scalar that couples to the $j^{th}$ hypermultiplet where 
\begin{align}\begin{split}
&\alpha_{j,\mu}^\pm= i\,{\rm Im}[\mu(\varphi^{(j)})+q_j]\pm\sqrt{{\rm Re}[\mu(\varphi^{(j)})+q_j]^2+i\mu(D)}~,\\
&h^{ab}(\varphi,D)=\sum_j\sum_{\mu\in \Delta_{hyp}^{(j)}}\sum_{m\in \IZ} \frac{\langle \mu, H_{I(a)}\rangle \langle \mu, H_{I(b)}\rangle}{\left(|\pi m+i\mu(\varphi^{(j)})+iq_j|^2+i \mu(D)\right)(\pi m-i\mu(\bar\varphi^{(j)})-i\bar{q_j})}~,
\end{split}\end{align}
and $H_{I(a)}$ runs over the simple coroots of $\fg$. 

Although the Yukawa coupling of is order $O(e)$, it is required to soak up the $\lambda^1$ zero modes. Thus, all other contributions from expanding the exponential of the Yukawa term will be suppressed by additional positive powers of $e$. Since these higher order terms do not contribute in the limit $e\to 0$, they must evaluate to zero by the localization principle.

Therefore, putting all of these elements together, the total 1-loop determinant is given by 
\begin{align}\begin{split}
Z_{det}(\varphi)=&\int_{\ft} d^rD \prod_{f=1}^{{\rm N}_f} \prod_{\mu\in \Delta_{fund}^{(f)}}\sinh(\mu(\varphi^{(f)})+ q_f) ~{\rm exp}\left\{-\frac{\pi \beta(D,D)}{e^2}+2\pi i \beta(\xi ,D)\right\}\\&\times
\prod_{i=1}^{n-1} \prod_{\alpha \in \Delta_{adj}^{(i)}}\sinh(\alpha(\varphi^{(i)})+ q_i)\prod_{\substack{\alpha \in \Delta_{adj}^{(i)}\\ \alpha\neq 0
}}\sinh(\alpha(\varphi^{(i)}))\\&
\times \prod_j \prod_{\mu\in \Delta_{hyp}^{(j)}}\frac{\sinh(\mu(\bar\varphi^{(j)})+\bar{q}_j)}{\sinh(\alpha_{j,\mu}^+)\sinh(\alpha_{j,\mu}^-)}
\cdot {\rm det}\,h^{ab}(\varphi,D)~.
\end{split}\end{align}

\subsubsection{Regularization} 

As it turns out, this integral is singular and requires regularization. Physically, this arises because the bosonic part of the Euclidean action is of the form 
\be
S_{bos}=\frac{1}{e^2}\int dt \left(\bar\phi|\partial_t+\varphi+m|^2\phi+iD(|\phi|^2-e^2\xi)+\half D^2\right)~,
\ee
for a generic bosonic hypermultiplet field $phi$  where $m$ is its mass which is generically dependent on $\fa,\epsilon_+$. 
Thus, there is a bosonic zero mode when $\varphi=-m,D=0$. This makes the path integral infinite due to the co-dimension $3r$ singularity. 

Therefore, consider the local behavior near finite singularities. These singularities come from the hypermultiplet contribution to the 1-loop determinant where $m=0$ and are given by the a collection of intersecting singular hyperplanes in $\ft_\IC/\Lambda_{cr}$ located at 
\be
H_{\mu,j}=\Big\{\varphi\in \ft_\IC/\Lambda_{cr} ~|~ D=0~,~\mu(\varphi^{(j)})+q_j=0~,~\mu\in \Delta_{hyp}^{(j)}\Big\} ~.
\ee
In order to see that this singularity leads to a divergent integral, it is sufficient to study the singularity from a single hyperplane in a transverse plane. In a local coordinate $z$ centered at the hyperplane, this singularity is of the form
\be
\int_{B_R^2}d^2 z\int_{-L}^L dx\, \frac{1}{(|z|^2+i x)^2}~,
\ee
where $x=\mu(D^{(j)})$ and $B_R^2$ is an 2D ball of radius $R$ around the origin and $L$ is some finite cutoff. This integral is singular. Therefore, we need to regularize  this integral. 

One way we can regulate this expression is by shifting the contour of integration for $D$ by $\ft\to \ft+i\eta$ for $\eta\in \ft$. In this case the singular integral becomes 
\be
\int_{B_R^2}d^2 z\int_{-L+i y}^{L+i y} dx\, \frac{1}{(|z|^2+i x)^2}=\int_{B_R^2}d^2 z\int_{-L}^L dx\, \frac{1}{(|z|^2-y+i x)^2}~,
\ee
where $y=\mu(\eta^{(j)})$. This resolves the singularities where $\mu(\eta^{(j)})<0$. However, the integrand is still singular along a circle in the complex plane for the case for those $\mu$ such that $\mu(\eta^{(j)})>0$. 
This can further be regulated by cutting out the disks $B^{(sing)}_\delta$ of radius $\sqrt{|\mu(\eta^{(i)})|}+\delta$ around the ring singularity and then send $\delta\to0$ . There are subtleties associated with taking the limit $\delta\to0$ which will also require taking $\eta\to 0$, however we will postpone a discussion until later. 
 Now the regularized path integral is given by 
 \be\label{regularized}
 I_W^{(Loc)}=\int_{(\ft_\IC/\Lambda_{cr})\backslash B^{(sing)}_\delta}\frac{d^{2r} \varphi }{(2\pi i )^r}\int _{\ft+i \eta} d^rD\, Z_{int}~{\rm exp}\left\{-\frac{\pi \beta(D,D)}{e^2}+2\pi i \beta(\xi ,D)\right\}~,
 \ee
 where $B_\delta^{(sing)}$ is a union of $\delta$-neighborhoods of the singularities of the integrand.

There can also be singularities arising from the infinite volume over $\ft$ and $\ft_\IC/\Lambda_{cr}$. Thus, let us examine the behavior of the integrand at $D\to \partial \ft$. Here, the Gaussian factor will exponentially suppress the integrand and hence there will be no singularity from the $D$-field. 

Now let us examine the behavior of the integrand near $\partial (\ft_\IC/\Lambda_{cr}))$. 
Consider the integrand in the limit 
\be
\tau\to \infty \quad {\rm  where }\quad \varphi=\tau \uu\quad, \quad \uu\in \ft
\ee
where $\ft$ is the Lie algebra of $\fg$ which itself decomposes as $\ft=\bigoplus_{i=1}^{p-1}\ft^{(i)}$. As shown in \ref{app:C}, 
the integrand $Z_{int}$ has the limiting form 
\be
|Z_{int}|\stackrel[\substack{\tau\to \infty\\\varphi=\tau \uu}]{}{\large \lesssim}\prod_{i=1}^{n-1}{\rm exp}\Bigg\{2\tau\left(s(i)-2-2\delta_{s(i),1}-4\delta_{s(i),2}+\frac{{\rm N}_f}{2}\delta_{i,i_m}\right)\sum_{a=1}^{k^{(i)}}|\uu_a^{(i)}|\Bigg\}~,
\ee
where 
\be
s(i)=2k^{(i)}-k^{(i+1)}-k^{(i-1)}\quad, \quad i_m=\frac{n}{2} -1~. 
\ee
Using the fact that $s(i_m)=0$ or 2 and the fact that ${\rm N}_f\leq4$, we see that 
the exponential factors can at most completely cancel as $\tau\to \infty$. 
In this case, the behavior of the 1-loop determinant at infinity will be polynomially suppressed by the Yukawa terms for the hypermultiplet fields to order 
 $O\left(\prod_i\tau^{-3k^{(i)} }\right)$. Therefore,  since the measure goes as $\prod_i \tau^{2k^{(i)}-1}$, we have that the product of the integrand and measure will vanish as $O\left(\prod_i \tau^{-k^{(i)}-1}\right)$ and does not contribute infinitely to the localized path integral.

Therefore, excising the $\delta$-neighborhoods $B_\delta^{(sing)}$ clearly resolves all singularities in the integrand and renders its integral finite. However, since we are making a choice of regularization, it is unclear how the resulting integral is related to the true path integral. Therefore, we will refer to this as the localized Witten index, $I_{W}^{(Loc)}$, to emphasize how it is distinct from the true Witten index. 

\rmk.
The $D$-contour deformation is physically well motivated because introducing a FI-parameter is equivalent to shifting the saddle point of the $D$ integral to $i e^2\xi$.  In our regularization prescription, the localization result will generically be dependent on $\eta,\xi$. This dependence even persists in the limit $\eta\to 0$, $\xi/\beta\to 0$ as the dependence on the chamber of $\eta,\xi\in \ft^\ast$ defined by the charges of the hypermultiplet scalars $\mu_i\in \ft^\ast$ as in the Jeffrey-Kirwan residue prescription. \footnote{See \ref{footnote:JK} for more details.} This dependence encodes wall crossing in the SQM as studied in \cite{Hori:2014tda}. Thus, since the saddle point occurs at $\eta=e^2\xi$, we will restrict $\eta,\xi\in \ft^\ast$ to be in the same chamber. This is most easily achieved by assuming $\eta=c \xi$ for some positive constant $c\in \IR^+$. 


\subsection{Reduction to Contour Integral}

Now that we have a well defined volume integral over $\ft_\IC\slash \Lambda_{cr}\times (\ft+i \eta)$, we can utilize the identity 
\be\label{identity}
\frac{\partial}{\partial \bar\varphi_a}Z_{hyp}^{(kin)}=-i D_b h^{ab} Z_{hyp}~,
\ee
where $a,b$ are indices for a basis of simple coroots, to reduce the volume  integral to a contour integral. This allows us to write the 1-loop determinant as a total derivative
\be
	Z_{int}=\left(\prod_a\frac{1}{iD_a} \bar\partial_{\bar\varphi_a} \right)Z_{int}^{(kin)}\quad, \quad Z_{int}^{(kin)}=Z_{vec}\cdot Z_{Fermi}\cdot Z_{hyp}^{(kin)}~,
\ee
such that the volume integral over $\ft_\IC\slash \Lambda_{cr}$ can be reduced to a contour integral   over the boundaries of the excised $\delta$-neighborhoods and boundary $\partial\ft_\IC\slash\Lambda_{cr}$
\begin{align}\begin{split}\label{ILoc}
&I_{W}^{(Loc)}=\int_{(\ft_\IC/\Lambda_{cr})\backslash B^{(sing)}_\delta}\frac{d^{2r} \varphi }{(2\pi i )^r}\int _{\ft+i \eta} d^rD\, Z_{int}~{\rm exp}\left\{-\frac{\pi \beta(D,D)}{e^2}+2\pi i \beta(\xi ,D)\right\}~,\\
&=\int_{(\ft_\IC/\Lambda_{cr})\backslash B^{(sing)}_\delta}\frac{d^{2r} \varphi }{(2\pi i )^r}\int _{\ft+i \eta}  \prod_a\frac{dD_a}{iD_a}\,\partial_{\bar\varphi}^{(r)}\left( Z_{int}^{(kin)}\right)~{\rm exp}\left\{-\frac{\pi \beta(D,D)}{e^2}+2\pi i \beta(\xi ,D)\right\}~,\\
&=\oint_{\substack{\partial B^{(sing)}_\delta\\\cup(-\partial (\ft_\IC/\Lambda_{cr}))}}\frac{d\varphi_1\wedge...\wedge d\varphi_r}{(2\pi i )^r}\int _{\ft+i \eta} 
\prod_a\frac{dD_a}{iD_a}\, Z_{int}^{(kin)}~{\rm exp}\left\{-\frac{\pi \beta(D,D)}{e^2}+2\pi i \beta(\xi ,D)\right\}~,
\end{split}\end{align}
where $r={\rm rnk}\fg$ and $a$ indexes the simple coroots of $\ft$. 
Here $B_{\delta}^{(sing)}$ is the neighborhood of radius $\sqrt{|\mu(\eta)|}+\delta$ surrounding  each ring singularity in the integrand (where $\mu(\eta)>0$) and $\partial \ft_\IC/\Lambda_{cr}$ is the (asymptotic) boundary of $\ft_\IC/\Lambda_{cr}$.  The identity \eqref{identity} is a consequence of supersymmetry 
\cite{Moore:1997dj,Moore:1998et,Hori:2014tda,Benini:2013nda,Benini:2015noa,Benini:2013xpa}.

Consider the contributions from the contour integral around the excised  $B_\delta^{(sing)}$. These terms are non-zero due to the poles in the 1-loop determinant from the bosonic fields of the hypermultiplets which are of the form:
\be
Z_{int}\sim\prod_j \prod_{\mu\in \Delta_{hyp^{(j)}}}\frac{1}{(|\mu(\varphi^{(j)})+q_j|^2+\mu(\eta)+i \mu(D'))^2}\quad, \quad D'=D+i \eta~,
\ee
for $\mu(\eta)>0$. In this case, the contour integral over the excised disk of radius $\sqrt{|\mu(\eta)|}+\delta=r+\delta$ where $D'=0$ is of the form
\be
\oint_{\partial B_\delta}\frac{\bar\varphi d\varphi}{|\varphi|^2-r^2}=2\pi i\frac{(r+\delta)^2}{2r\delta+\delta^2}~. 
\ee
Now we need to take $\delta\to 0$ as a regulator of the singularity at $|\varphi|^2=|\mu(\eta)|$. Note that the integral above is infinite unless we take $\sqrt{|\mu(\eta)|}\to0$ faster than $\delta$. Therefore, we will define the regularization of the localized path integral with $\sqrt{|\mu(\eta)|}\to0$, $\delta\to 0$, such that $\sqrt{|\mu(\eta)|}/\delta\to0$.  In this limit, we find that the boundary integrals are equivalent to computing the residue at the singularity with $\sqrt{|\mu(\eta)|},D=0$. 

Now we can evaluate the terms in the integral \eqref{ILoc} attributed to the poles $\partial B_\delta^{(sing)}$. By using the fact 
\be
\lim_{|\eta_a|\to 0}\frac{1}{D_a+i \eta_a}=P\left(\frac{1}{D_a}\right)-i\pi\,{\rm sign}(\eta_a)\delta(D_a)~, 
\ee
we get 
\begin{align}\begin{split}
I_{W}^{(Loc),sing}=&\oint_{\partial B^{(sing)}_\delta}\frac{d\varphi_1\wedge...\wedge d\varphi_r}{(2\pi i )^r}\int _{\ft+i \eta} 
\prod_ad D_a\delta(D_a) , Z_{int}^{(kin)}~{\rm exp}\left\{-\frac{\pi \beta(D,D)}{e^2}+2\pi i \beta(\xi ,D)\right\}\\
&+\Big\{{\rm Principal~Terms ~in~} 1/D\Big\}~,
\end{split}\end{align}
where principal terms are those that have a principal value of some $D_a$. Here, the principal value term vanishes because integrand  does not have a singularity of sufficiently high codimension in $\varphi$ and hence the contour integral over $\varphi$ is identically zero.  Therefore, we find that the terms coming from the excised disks is exactly 
\be
I_{W}^{(Loc),sing}=\oint_{\partial B^{(sing)}_\delta}\frac{d\varphi_1\wedge...\wedge d\varphi_r}{(2\pi i )^r}Z_{int}^{(kin)}(\varphi,D=0)~,
\ee
which reduces to a sum over residues of $Z_{int}^{(kin)}(\varphi,D=0)$. 

This sum over residues is equivalent to the Jeffrey-Kirwan residue prescription  \cite{JK}. The reason is that the contour integral simply picks out tuples of poles for which $\mu(\eta)>0$ -- or equivalently it picks poles corresponding to given tuples of $\{\mu_p\}_{p=1}^{{\rm rnk} \fg}$ such that $\mu_p(\eta)>0$, $\forall p$. By mapping $\eta\in \ft$ to $\eta^\vee\in \ft^\ast$ by the Killing form, this is equivalent to the statement that the contour integral includes tuples of poles such that $\eta^\vee$ is in the positive cone defined by the $\{\mu_p\}$. This is the definition of the JK residue prescription \cite{JK}. \footnote{\label{footnote:JK}
The Jeffrey-Kirwan residue prescription selects a contour that such that the integral evaluates to a sum of residues corresponding a particular set of poles specified by a parameter $\vec\xi\in \ft^\ast$. These are selected as follows. Consider a contour integral over an $r$-complex dimensional space. The poles of the integrand  are solutions of the equations
\be\label{pole}
Q_i( \varphi)+f_i(q)=0~,
\ee
for some set of $Q_i\in \ft^\ast$ and $f_i(q)$ functions of some parameters $q_j$. 
Each of these poles defines a  hyperplane in $\ft$ along which the integrand is singular. To each hyperplane specified by the solution of \eqref{pole}, we associate the charge $Q_i\in \ft^\ast$. 
Any set of $r$ linearly independent $\{Q_i\}\in \ft^\ast$ defines a positive cone in $\CC_{\{Q_i\}}\subset\ft^\ast$. Each such cone corresponds to the intersection of $r$ hyperplanes, which has a non-trivial residue.

 The Jeffrey-Kirwan  prescription specified by the $\xi\in \ft^\ast$ picks a contour such that the contour integral evaluates to the sum of residues associated to all cones $\CC_{\{Q_i\}}$ such that $\xi\in \CC_{\{Q_i\}}$ weighted by the sign of the determinant $sgn(Q_{i_1}\wedge...\wedge Q_{i_r})$. 
}

%

\subsubsection{Boundary Terms at Infinity}

Now consider the contributions to the contour integral from the boundary $\partial \ft_\IC/\Lambda_{cr}$. For simplicity we will consider only the case of a $U(1)$ gauge theory as it is our main example. However, the following analysis in the next two sections generalizes to generic gauge groups. We will comment more on this later and continue to use notation that accommodates this generalization. 

 Here we are considering the integral
\begin{align}\begin{split}
Z_{bnd}=\oint_{-\partial (\ft_\IC/\Lambda_{cr})}\frac{d\varphi}{2\pi i}\int_{\IR+i \eta}\frac{dD}{iD} Z_{int}^{(kin)}(\varphi,D)e^{-\frac{\pi \beta D^2}{e^2}+2\pi i \beta \xi D}
\end{split}\end{align}
where \footnote{Note that the contribution from the vector multiplet fermions is only given by $\sinh(2\epsilon_+)$ since the adjoint action is trivial for a $U(1)$ gauge group. }
\begin{align}\begin{split}
Z_{int}^{(kin)}(\varphi,D)=&2\sinh(2\epsilon_+)\prod_{f=1}^{{\rm N}_f}\sinh(\mu_f(\varphi)- m_f)\times\\&
\prod_{j=1}^4\frac{\sinh(\mu_j(\bar\varphi)+ \bar{q}_j)}{\cosh(2i\, {\rm Im}[\mu_j(\varphi)+ q_j])-\cosh(2 \sqrt{{\rm Re}[\mu_j(\varphi)+q_j]^2+i \mu_j(D)})}~.
\end{split}\end{align}
Here $j$ indexes over the representations of the 4 different $\CN=(0,4)$ fundamental chiral multiplets making up the two $\CN=(0,4)$ fundamental hypermultiplets. 

In the limit ${\rm Re}[\varphi]\to \pm\infty$, the function $Z_{int}^{(kin)}(\varphi,D)$ is independent of $D$ and is the $0^{th}$ order coefficient of the Laurent expansion in $e^{\varphi }$. 
Thus,  the boundary integral, which is evaluated in the limit $Re[\varphi]\to \pm\infty$, is given by 
\begin{align}\begin{split}\label{boundary1}
&Z_{bnd}=\sum_\pm \lim_{Re[\varphi]\to \pm  \infty}\left(\pm Z_{int}^{(kin)}\right)\left(c(\eta)-{\rm erf}\left(\sqrt{\pi \beta}e\xi\right)\right)~,\\
&~=\sum_\pm  \pm \lim_{Re[\varphi]\to \pm  \infty}\left(2\sinh(2\epsilon_+)e^{\left(\sum_f|\mu_f|-\sum_j |\mu_j|\right)|Re[\varphi]|}\right)e^{\pm \sum_f{\rm sign}(\mu_f)\bar{q}_f\mp \sum_j{\rm sign}(\mu_j)q_j}\\
&\quad\times\left(c(\eta)-{\rm erf}\left(\sqrt{\pi \beta}e\xi\right)\right)~,
\end{split}\end{align}
where we have applied the formula in for the integral in Appendix \ref{app:D} and erfc$^+(x)$ is the error function
\be
{\rm erf}(x)=\frac{2}{\sqrt{\pi}}\int_{0}^x du \,e^{-u^2}~.
\ee
Here
\be
c(\eta)=\begin{cases}
1& \eta>0\\
-1& \eta<0
\end{cases}~.
 \ee 
 See Appendix \ref{app:D} for more details. 
 
By using the fact that in our models
\be
\sum_f |\mu_f|-\sum_i |\mu_i|={\rm N}_f-4~,
\ee
we see that this boundary term is only non-zero when N$_f=4$. \footnote{We can additionally consider the effect of including Chern-Simons terms as in Section \ref{sec:CS}. A Chern-Simons term with level $k$ shifts the argument of \eqref{boundary1} by a factor $e^{2k{\rm Re}[\varphi]}$. Thus we have that this boundary term is only non-zero when 
\be
\sum_f |\mu_f|-\sum_i |\mu_i|+2k={\rm N}_f-4+2k=0~.
\ee
}

In summary, by carefully performing the localization computation of $I_{W}\to I_{W}^{(Loc)}$ by regularization, we find that 
\be\label{bulk}
I_{W}^{(Loc)}=Z^{JK}+Z_{bnd}~,
\ee
where $Z^{JK}$ is the result from the Jeffrey-Kirwan residue prescription and $Z_{bnd}$ is the boundary computation computed in \eqref{boundary1}.

Note that $Z_{bnd}$ has explicit $\beta,e,\xi$ dependence. Generically one would expect that the answer is independent of these parameters since the Lagrangian is $Q$-exact and hence variations of $\beta,e,\xi$ are supersymmetric deformations of the action. However, this dependence can arise from a continuous spectrum of states which allows for a spectral asymmetry between bosonic and fermionic states \cite{Akhoury:1984pt}. As we previously discussed, our models have such a continouous spectrum of states arising from the non-compact directions in the mixed and Coulomb branches.

\rmk.~ Note that if we had instead identified $\eta=e^2\xi$, regularity would have required us to take  the limit $e^2\to0$. Then we would find that $I_W^{(Loc)}=Z^{JK}+Z_{bnd}(\beta=0)$. This matches with the analysis of \cite{Lee:2016dbm} in which the authors found that the localization computation of the Witten index, under a specific choice of regulator, can be identified with the computation of the Witten index in the limit $\beta\to 0$. 


\subsubsection{Comparison with Literature}

Let us take a moment to compare our results with that of the literature \cite{Benini:2013nda,Benini:2013xpa}. In these papers, the authors give a physical derivation of the Jeffrey-Kirwan residue prescription for the elliptic genus of 2D $\CN=(0,2)$ gauge theories. Here the authors consider the localized path integral in the limit $e^2\to 0$ over $(\ft_\IC/\Lambda_{r})\times \ft$ which they decompose as 
\begin{align}\begin{split}
Z&=\int_{\ft_\IC/\Lambda_{cr}}\frac{d^{2r}\varphi}{(2\pi i)^r}\int_\ft d^r DZ_{int}(\varphi,D)e^{-\pi \frac{\beta D^2}{e^2}+2\pi  i\beta \xi D}\\
&=\int_{(\ft_\IC/\Lambda_{cr})\backslash  B_\delta^{(sing)}}\frac{d^{2r}\varphi}{(2\pi i)^r}\int_\ft d^r DZ_{int}(\varphi,D)e^{-\frac{\pi \beta D^2}{e^2}+2\pi  i\beta \xi D}\\&
\quad+\int_{  B_\delta^{(sing)}}\frac{d^{2r}\varphi}{(2\pi i)^r}\int_\ft d^r DZ_{int}(\varphi,D)e^{-\frac{\pi \beta D^2}{e^2}+2\pi  i\beta \xi D}~,
\end{split}\end{align}
where $B_\delta^{(sing)}$ is the collection of $\delta$-neighborhoods of the singularities as before.

After dropping the singular term, which the authors argue can be regularized to zero, the path integral to
\be\label{reducedint}
Z=\int_{(\ft_\IC/\Lambda_{cr})\backslash B_\delta^{(sing)}}\frac{d^{2r}\varphi}{(2\pi i)^r}\int_\ft d^r DZ_{int}(\varphi,D)e^{-\frac{\pi \beta D^2}{e^2}+2\pi  i\beta \xi D}~,
\ee
which is the same as \eqref{regularized}. The authors then also deform the $D$-contour and reduce the path integral to a contour integral around $\partial B_\delta^{(sing)}$. They then show by contour integral methods that \eqref{reducedint} reduces to a sum of residues according to the Jeffrey Kirwan residue prescription as above. 

In our analysis we also take into account the possibility of an additional contribution coming from the asymptotic boundary of $\partial(\ft_\IC/\Lambda_{cr})$, while the models studied in  \cite{Benini:2013nda,Benini:2013xpa} have a compact target space so such terms do not arise. Similar boundary terms are also discussed for some models in \cite{Hori:2014tda,Benini:2015noa}. However, the analysis of these papers is not directly applicable to our model. 

\subsection{Coulomb and Higgs Branch States}
\label{sec:HiggsCoulomb}

In our discussion we often use the terminology such as ``Higgs branch states'' and ``Coulomb branch states.'' Here we will define this terminology precisely. 

\begin{figure}
\begin{center}
\includegraphics[scale=1.2,trim=2cm 22.5cm 2.5cm 1.25cm,clip]{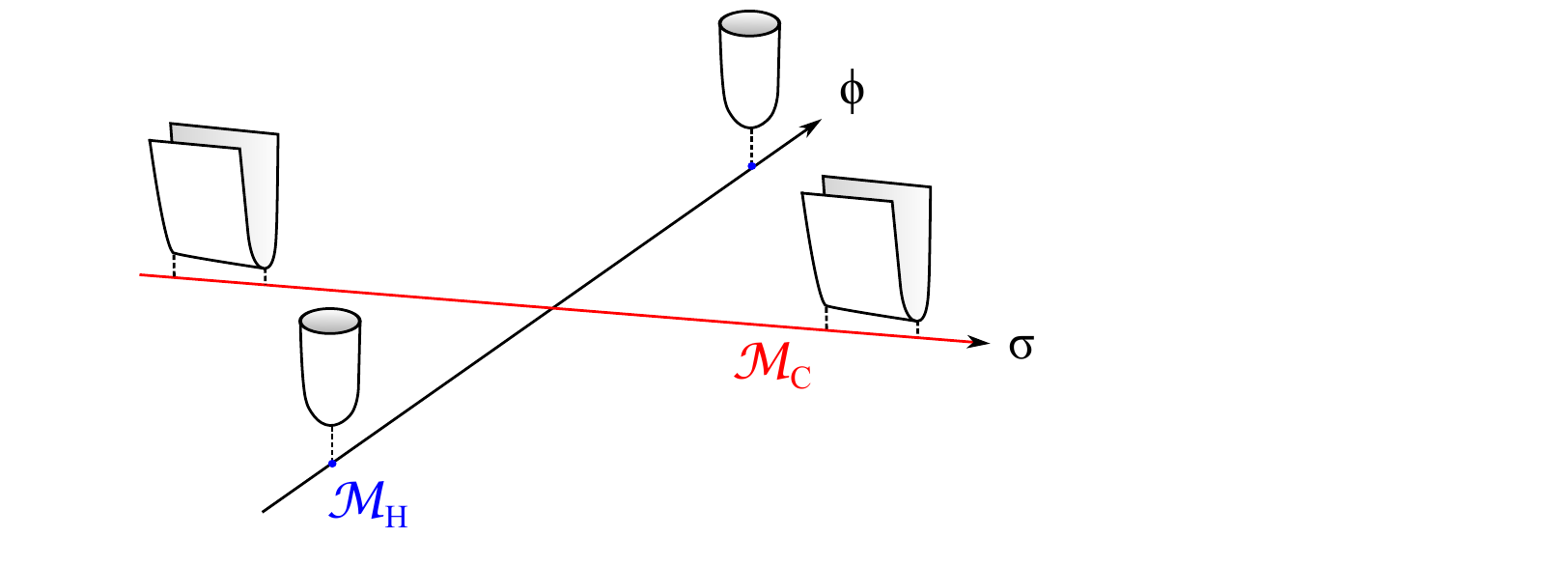}
\end{center}
\caption{This figure illustrates the behavior of the potentials on the Higgs (blue) and Coulomb  (red) branch.}
\label{fig:Limit}
\end{figure}

Pick a bubbling SQM and consider the family of quantum systems defined by the varying with respect to $e,\beta,\xi$. 
Due to the localization principle, the states that survive in the limit $e^2|\xi|^3\to 0$ should be the only ones that give non-canceling contributions to the Witten due to the localization principle. \footnote{Here we are taking our fixed length scale to be set by $\xi$ in anticipation of the next section where we allow $\beta$ to vary. } 
 As we monotonically approach $e^2|\xi|^3\to 0$ with $\xi$ fixed, the potential energy function of these families approaches an infinite value on all of field space except along the Higgs, Coulomb, and mixed  branches as discussed in Section \ref{sec:bubblingSQMs}. See Figure \ref{fig:Limit}. 

As we decrease $e^2|\xi|^3\to 0$, the potential around each component of the Higgs branch (which is topologically a collection of points) approaches an infinitely steep harmonic oscillator potential for all fields. In this limit, the potential additionally becomes flat along the Coulomb and mixed branches (which are non-compact) while simultaneously approaching an infinitely steep harmonic potential in the transverse directions. 

Now consider the spectral decomposition of the Hilbert space defined by the 
family of Hamiltonians parametrized by $e,\xi$. Due to the behavior of the potential function near the different  classical vacua, there will be orthogonal projection 
operators $P^{H}, P^C, P^{mix}$ 
onto a space of eigenstates of the Hamiltonian 
such that the wave functions of states in this subspace have support 
localized near the classical Higgs, Coulomb, and mixed vacua respectively. We will refer to states in the 
image of $P^{H}$  as ``Higgs branch states," those in the image of $P^C$ as ``Coulomb branch states," and those in the image of $P^{mix}$ as ``mixed branch states."


In fact,  we believe there are projectors $P^{C,+}\geq P^{C}$ and $P^{mix,+}\geq P^{mix}$ 
to a subspace on which the Hamiltonian has a continuous spectrum but such that 
all the states in the image have wave functions with support   localizing to a neighborhood 
of the Coulomb branch or mixed branch respectively.

Of course,  in the spectral decomposition of the Hilbert space defined by the 
Hamiltonian, there will additionally be a projection operator $P^{M}$ to a subspace on which 
the spectrum of the Hamiltonian is continuous and bounded below by a large constant $M$ 
such that states in the image of $P^{M}$ will  have support throughout field space and are not 
in any sense localized near either branch. However, the mass gap $M$ to the unlocalized, continuum of states goes to infinity as $e^2|\xi|^3\to 0$ and hence gives vanishing contribution to the Witten index.

 \subsubsection{Physical Interpretation of Jeffrey-Kirwan Residues}
 
The Jeffrey-Kirwan prescription for computing the path integral counts the BPS states  that are localized on the Higgs branch. 
The reason is that the residues that are summed over in the JK residue prescription are in one-to-one correspondence with the unlifted Higgs branch vacua. 

Consider the integrand $Z_{int}$ in \eqref{Zint}. This has poles along the hyperplanes 
\be
H_{\mu,j}=\left\{\varphi\in \ft_\IC/\Lambda_{cr}~|~\mu_j(\varphi)+q_j=0\right\}~,~ \mu_j\in \Delta_{hyper}^{(j)}~,
\ee
where $q_j$ is the global charge for the $j^{th}$ hypermultiplet (or equivalently its mass). The JK residue formula specified by the FI parameter $\xi\in \ft^\ast$ then selects the residue given by $r$-tuples of poles corresponding to a codimension $r$ intersection of $r$ hyperplanes $H_{\mu,j}$ such that 
\be
(\xi,\mu_j)>0~,~{\rm for~each~}\mu_j~. 
\ee

Physically, the hyperplanes $H_{\mu,j}$ define the locus in field space where the corresponding hypermultiplet field becomes massless:
\begin{align}\begin{split}
&0=\left|(\mu_j(\sigma)+a Q_\fa+\epsilon_+ Q_\epsilon)\cdot\Phi_j\right|^2~,
\end{split}\end{align}
where $j$ indexes over the fundamental and bifundamental scalar fields $\Phi_j\in \{\phi^{(i)},\tilde\phi^{(i)},\uphi^{(i)},\tilde\uphi^{(i)}\}$ that have charge $\mu_j$ and $(\Phi_j,\bar{\tilde{\Phi}}_j)$ forms an $SU(2)_R$ doublet. 
Since ${\rm rnk}\fg=r$, and each $\Phi_i$ has a different mass, there can only be $r$ simultaneously massless hypermultiplet fields. This corresponds to the statement that there are at most codimension $r$ intersections of the hyperplanes $H_{\mu,j}$. 

Now consider the D- and F-term equations for the Higgs branch. These can be written as
\begin{align}\begin{split}
0=&-
\sum_{j\,:\,(\xi^{(i)},\mu_j)>0}|\Phi_j|^2+\sum_{j\,:\,(\xi^{(i)},\mu_j)<0}|\Phi_j|^2+e^2|\xi^{(i)}|\\
0=&\sum_{j\,:\,(\xi^{(i)},\mu_j)>0}\Phi_j \tilde\Phi_j-\sum_{j\,:\,(\xi^{(i)},\mu_j)<0}\Phi_j\tilde\Phi_j~.
\end{split}\end{align}
As in the JK-prescription, the solutions of these equations where at most $r$ $\Phi_j$ are massless are enumerated by an $r$-tuple $\Phi_j$ which obey 
\be
(\xi,\mu_j)>0~{\rm for ~each~}j~. 
\ee
This enumerates the entire resolved Higgs branch with respect to an FI parameter $\xi\in \ft^\ast$. 
Therefore, the JK-residues are in one-to-one correspondence with the points on the Higgs branch. 

Now note that the Jeffrey-Kirwan residue computation is independent of the value of $e^2$. Thus, in the limit $e^2\to 0$, we can identify states as being localized to a single vacuum branch in field space. Thus, we can identify each residue of the JK-prescription as counting the states that are localized on the corresponding point of the Higgs branch in the limit $e^2\to 0$:
\be
Z^{JK}=I_{Higgs}~.
\ee

\rmk.~
Note that the interpretation of the $Z^{JK}$ as an object counting the contribution of Higgs branch states matches with the previous analysis by taking a limit of $e^2\beta^3\to0$, $\xi/\beta\to\infty$ with $\beta$ fixed such that $e^2\xi =\xi'$ is constant. \footnote{Note that this is different from the rest of our analysis where we take $\xi$ to be the fixed length scale.} In the effective SQM on the non-compact branches, the mass of the ground states is given by 
\be
m\sim\sqrt{e|\xi|}\to \infty\quad {\rm ~as}\quad e^2\beta^3,\beta/\xi\to0~,~ {\rm with~}e^2\xi=\xi'~,~ \beta~{\rm fixed}~. 
\ee
Thus in taking this limit, all states on the non-compact branches are killed and $I_{asymp}\to 0$. Similarly, if we were to compute the standard Witten index, taking this limit kills the boundary terms. Thus 
\be
\lim_{\substack{e^2\beta^3\to 0\\\xi/\beta\to \infty\\e^2\xi=\xi'}}I_W=I_{Higgs}=Z^{JK}~. 
\ee
We stress that this is not the appropriate limit for computing $Z_{mono}(P,\vv)$. See \cite{Hori:2014tda} for more details regarding the computation of the Witten index in this limit. 

\subsection{Ground State Index}

As shown in \cite{Sethi:1997pa,Akhoury:1984pt,Yi:1997eg},  the Witten index has to be handled with care in the case of a SQM with a continuous spectrum. As in  our case, we have found that when there is a continuous spectrum, there can be a spectral asymmetry that gives rise to non-trivial $\beta$ ,$e,$ and $\xi$ dependence. Note that in order to compute the Witten index, we introduced the FI-parameter $\xi$. In the 4D picture this corresponds to separating the insertions of the minimal 't Hooft defects that make up the reducible 't Hooft defect. Thus, to compute $Z_{mono}(P,\vv)$, we want to take the limit ``$\xi\to 0$'' which is formally given by the limit $\xi/\beta\to 0$. However, this is computationally indistinguishable from taking $\beta/\xi\to \infty$ with $\xi$ fixed. Thus, $Z_{mono}(P,\vv)$ can be identified with  the Witten index in the limit ``$\beta\to \infty$:"
\be
Z_{mono}(P,\vv)=I_{\CH_0}:=\lim_{\beta\to \infty}
{\rm Tr}_{\CH}~(-1)^F e^{-\frac{\beta}{2} \{\CQ,\CQ\}+\fa Q_\fa+\epsilon_+ J_++m \cdot F}~.
\ee


In the limit $\beta \to \infty$, contributions from all non-BPS states are completely suppressed. This effectively  restricts the Witten index to a trace over the Hilbert space of BPS ground states. We will refer to the Witten index in this limit, $I_{\CH_0}$, as the \textit{ground state index}. This matches with the fact that the AGT computation is independent of $\beta$, suggesting that we should only have contributions from BPS states. 

As in the case of the Witten index, the ground state index can be computed as (the limit of) a path integral. Thanks to  supersymmetry, one can attempt to compute that path integral by using localization. 
Again, using localization requires a choice of regularization\footnote{We will be taking the same choice of regularization as in the case of the Witten Index.} and hence we will refer to  the result of the localization computation as $I_{\CH_0}^{(Loc)}$ to distinguish it from the true ground state index.

This limit  of the Witten index can be easily computed using our analysis from the previous section. Recall that $I_{W}^{(Loc)}=Z^{JK}+Z_{bnd}$.  Since the $Z^{JK}$ term is independent of $\beta$, taking the limit $\beta\to \infty$ only affects $Z_{bnd}$. 
The limit of the boundary term can be computed
\begin{align}\begin{split}
\lim_{\beta\to \infty}Z_{bnd}=\lim_{\beta\to \infty}\sum_\pm \lim_{Re[\varphi]\to \pm  \infty}\left(\pm Z_{int}^{(kin)}\right)\left(c(\eta)+{\rm erf}\left(\sqrt{\pi \beta}e\xi\right)\right)=0~,
\end{split}\end{align}
where
\be\label{cfunction}
c(\eta)=\begin{cases}
1& \eta>0\\
-1& \eta<0
\end{cases}~.
\ee
Using the fact that 
\be
\lim_{\beta\to \infty}\left(c(\eta)-{\rm erf}\left(\sqrt{\pi \beta}e\xi\right)\right)=\begin{cases}
0&\xi\times \eta>0\\
-2 &\xi>0>\eta\\
2  & \eta>0>\xi
\end{cases}~,
\ee
we see that 
\be\label{GSIBoundary}
\lim_{\beta\to \infty}Z_{bnd}=2\sinh(2\epsilon_+)\sinh\left(\sum_f m_f\right)\times\begin{cases}
0&\xi\times \eta>0\\
-2 &\xi>0>\eta\\
2  & \eta>0>\xi
\end{cases}~.
\ee
By identifying $\eta\sim e^2 \xi$, we find that 
\be
\lim_{\beta\to \infty} Z_{bnd}=0~, 
\ee
and hence that the localization computation of the ground state index is given by \be\label{finalLocEQ}
I_{\CH_0}^{(Loc)}=Z^{JK}~.\ee

\rmk.~ Although we have only shown that the boundary contribution vanishes for a SQM with a $U(1)$ gauge group, this result holds in general. One can see that the boundary contributions vanish more generally in the limit $\beta\to \infty$ as follows.  Decompose $\partial \widehat \CM$ into a disjoint union of open sets of increasing codimension $\partial\widehat\CM=\partial \ft_\IC/\Lambda_{cr}=\coprod_i(\partial \ft_\IC/\Lambda_{cr})^{(i)}$. For each boundary component,  the contour integral is of the form 
 the contour integral is of the form 
\be
Z_{bnd}^{(i)}
=\oint_{(\partial \ft_\IC/\Lambda_{cr})^{(i)}}\frac{d\varphi_1\wedge...\wedge d\varphi_r}{(2\pi i )^r} \int_{\ft+i \eta}\prod_a \frac{dD_a}{D_a}  Z_{int}^{(kin)}(\varphi,D)e^{-\frac{\pi\beta(D,D)}{ e^2}+2\pi i \beta (\xi,D)}~,
\ee
where $a$ indexes over the simple coroots of $\ft$.
On each component, there exists a simple root $\alpha\in \Phi^+$ such that $|\langle \alpha,\varphi\rangle|\to \infty$ on $(\partial \ft_\IC/\Lambda_{cr})^{(i)}$.
Thus, in each such integral, $Z_{det}^{(kin)}$ will be   independent of $\langle \alpha,D\rangle$ for some positive root $\alpha\in \Phi^+$. This means that  each boundary integral will be proportional to a factor of 
 \be
 Z_{bnd}^{(i)}\sim \left(c(\langle \alpha,\eta\rangle)-{\rm erf}\left(\sqrt{\pi \beta}e\langle \alpha,\xi\rangle
\right)\right)~,
\ee
for some postive root $\alpha\in \Phi^+$  where $c(\langle \alpha,\eta\rangle)$ is defined in \eqref{cfunction}. By identifying $\eta \sim e^2\xi$, this factor completely suppresses all boundary terms in  the limit $\beta\to \infty$. Thus, the boundary terms vanish in the localization computation of the ground state index
\be
\lim_{\beta\to \infty}Z_{bnd}^{(i)}=0 \quad \forall i~, 
\ee and therefore the ground state index  is generally given by the Jeffrey-Kirwan residue formula 
\be
I_{\CH_0}^{(Loc)}=Z^{JK}~. 
\ee

\subsection{Summary}

In this section we reviewed the localization computation for the Witten index of bubbling SQMs. In summary, we found:
\begin{enumerate}
\item The localized integral over the BPS moduli space is not well defined: it requires regularization. 
In general, the  regularized Witten index, $I_W^{(Loc)}$, will differ from the true Witten index, $I_W$.  
\item Under the choice of regularization we have presented, one arrives at the JK residue prescription plus a $\beta$-dependent boundary term that indicates the existence of a continuous spectrum of excited states: $I_{W}^{(Loc)}=Z^{JK}+Z_{bnd}$. 
\item Since the AGT computation shows that $Z_{mono}(P,\vv)$  is independent of $\beta$, we conjecture that $Z_{mono}(P,\vv)$ should only count contributions from BPS states.  Therefore,  we  identify $Z_{mono}(P,\vv)$ as the ground state index $I_{\CH_0}$ which eliminates contributions from non-ground states by taking the limit as $\beta\to \infty$ of the Witten index. By direct computation, we find that in this limit, the localization computation of the ground state index is given by  the Jeffrey-Kirwan residue prescription: 
\be
I_{\CH_0}^{(Loc)}=\lim_{\beta\to\infty}\left(Z^{JK}+Z_{bnd}\right)=Z^{JK}~,
\ee
which we identify as counting the states localized on the Higgs branch
\be
I_{\CH_0}=Z^{JK}=I_{Higgs}~.
\ee

After regularization with $\eta\sim e^2\xi$, we have 
\be
I_{\CH_0}^{(Loc)}= \oint_{JK(\xi)}\frac{d\varphi_1\wedge...d\varphi_r}{(2\pi i)^r} Z_{vec}(\varphi) Z_{Fermi}(\varphi)\tilde{Z}_{hyper}^{(kin)}(\varphi)~,
\ee
where
\begin{align}\begin{split}
Z_{vec}
&= \prod_{i=1}^{n-1} \prod_{\alpha \in \Delta_{adj}^{(i)}}\sinh(\alpha(\varphi^{(i)})+ q_i)\prod_{\substack{\alpha \in \Delta_{adj}^{(i)}\\\alpha\neq 0
}}\sinh(\alpha(\varphi^{(i)}))~,\\
Z_{Fermi}&=\prod_{f=1}^{{\rm N}_f} \prod_{\mu\in \Delta_{fund}^{(f)}}\sinh(\mu(\varphi^{(f)})+ q_f)~,\\
\tilde{Z}_{hyp}^{(kin)}&=\prod_j \prod_{\mu\in \Delta_{hyp}^{(j)}}\frac{1}{\sinh(\mu(\varphi^{(j)})+q_j)}~,
\end{split}\end{align}
and $\xi\in \ft^\ast$. 
 
\end{enumerate}

\subsection{Examples: $SU(2)$ ${\rm N}_f=4$ Theory}
\label{sec:NullEx} 

Now we have eliminated the $\beta$ dependence of the localization computation of $Z_{mono}(P,\vv)$ by identifying $Z_{mono}=I_{\CH_0}$. Nevertheless, in general,  the localization computation $I_{\CH_0}^{(Loc)}$ still does not generically agree with $Z_{mono}(P,\vv)$ as computed from AGT. 
We will now illustrate this claim with 
several non-trivial examples in the $SU(2)$ N$_f=4$ theory to show that the localization calculation for the ground state index $I_{\CH_0}^{(Loc)}=Z^{JK}$ does not match with the results from the AGT computations \cite{Ito:2011ea}. 
These examples are an explicit realization of a generic feature of 't Hooft defects in $\CN=2$ $SU(N)$ gauge theories with N$_f=2N$ fundamental hypermultiplets.

\subsubsection{$Z_{mono}(1,0)$}

Consider the $L_{1,0}$ (minimal) 't Hooft defect in the $SU(2)$ ${\rm N}_f=4$ theory. This has 't Hooft charge 
\be
P=h^1=2\hat{h}^1={\rm diag}(1,-1)\quad, \quad h^1\in \Lambda_{mw}~~,~~\hat{h}^1\in \Lambda_{cochar}~.
\ee
From general considerations, the expectation value of $L_{1,0}$ is of the form 
\be
\langle L_{1,0}\rangle=\left(e^{ \fb}+e^{- \fb}\right)F(\fa,m_f,\epsilon_+)+Z_{mono}(1,0)~,
\ee 
where $Z_{mono}(1,0)$ corresponds to the bubbling with $\vv={\rm diag}(0,0)$. Here 
\begin{align}\begin{split}
&F(\fa,m_f,\epsilon_+)=\left(\frac{\prod_\pm \prod_{f=1}^4 \sinh (\fa\pm m_f)}{\sinh^2(2 \fa)\prod_\pm \sinh(2\fa\pm 2\epsilon_+)}\right)^\half~.
\end{split}\end{align}

In this example, the  monopole bubbling contribution can be computed as the Witten index of the $\CN=(0,4)$ SQM described by the 
quiver:
\begin{center}
\begin{tikzpicture}[node distance=2cm,cnode/.style={circle,draw,thick,minimum size=10mm},snode/.style={rectangle,draw,thick,minimum size=10mm}]
\node[cnode] (1) {1};
\node[snode] (4) [below of=1]{2};
\node[snode] (5) [above of=1]{4};
\draw[dashed,thick] (1) -- (5);
\draw[-] (4) -- (1);
\end{tikzpicture}
\end{center}

Now the path integral from the previous section reduces to the contour integral 
\be
Z_{mono}^{(Loc)}(1,0)=\oint_{JK(\xi)}\frac{d\varphi}{2\pi i}Z_{vec}(\varphi,\fa,\epsilon_+)Z_{hyper}(\varphi,\fa)Z_{Fermi}(\varphi,\epsilon_+,m_i)~. 
\ee
where $JK(\xi)$ is the Jeffrey-Kirwan residue prescription specified by a choice of $\xi\in \ft^\ast\cong\IR$ \cite{JK,Lee:2016dbm,Hori:2014tda}. 

The general contributions of the different $\CN=(0,2)$ multiplets for a SQM labeled by gauge nodes $(k^{(1)},...,k^{(n-1)})$ and fundamental hypermultiplet nodes $(w_1,...,w_p)$ are given by \cite{Hwang:2014uwa,Brennan:2018yuj}
\begin{align}\begin{split}\label{eq:localizationcomp}
&Z_{vec}(\varphi,\epsilon_+)=\prod_{i=1}^n\left[\prod_{a\neq b=1}^{k^{(i)}}\right]^\prime2\sinh(\varphi_{ab}^{(i)})\times\prod_{i=1}^n\prod_{a\neq b=1}^{k^{(i)}}\sinh(\varphi_{ab}^{(i)}+2\epsilon_+)~,\\
&Z_{Fermi}(\varphi,m_f,\epsilon_+)=\prod_{f=1}^4 \prod_{a=1}^{k^{(f)}} 2\sinh(\varphi_a^{(f)} - m_f)~, \\
&Z_{fund}(\varphi,\fa,\epsilon_+)=\prod_{j} \prod_{a=1}^{k^{(j)}}\prod_{\ell=1}^{w_j}\prod_\pm \frac{1}{2\sinh(\pm(\varphi_a^{(j)}- \fa_\ell)+ \epsilon_+)}~,\\&
Z_{bifund}(\varphi,\fa,\epsilon_+)
= \prod_{i=1}^{n-1}\prod_{a=1}^{k^{(i+1)}}\prod_{b=1}^{k^{(i)}}\prod_\pm\frac{1}{2\sinh(\pm(\varphi_a^{(i+1)}-\varphi_b^{(i)})+ \epsilon_+)}~,
\end{split}\end{align}
where the product $\left[\prod_{I,J}\right]^\prime$ omits factors of 0, $\fa={\rm diag}(\fa_1,\fa_2)\in {\rm Lie}[SU(2)]$, the fundamental Fermi-multiplets couple to the $f^{th}$ gauge group, and $j$ indexes over the fundamental hypermultiplets (which couple to $\varphi^{(j)}$).
For our SQM, this reduces to 
\be
Z_{mono}^{(Loc)}(1,0)=\oint_{JK(\xi)}\frac{d\varphi}{2\pi i} 2\sinh(2\epsilon_+)\frac{\prod_{f=1}^4\sinh(\varphi- m_f)}{\prod_{\pm}\sinh(\varphi\pm  \fa+\epsilon_+)\sinh(-\varphi\pm \fa+\epsilon_+)}~. 
\ee
Here the JK residue prescription is determined by a choice of $\xi\in \IR$ which corresponds to introducing an FI parameter in the SQM. 
As shown in \cite{Hori:2014tda}, the Witten index of an SQM can generically have wall crossing as $\xi$ jumps between $\xi\in \IR^+$  and $\xi \in \IR^-$. 

Using this, the localization computation becomes
\be\label{ZmonoLoc10}
Z^{(Loc)}_{mono}(1,0)=-4\frac{\prod_f \sinh\,(\fa-{\rm m}_f\mp\epsilon_+)}{\sinh(2\fa)\sinh\,(2\fa\mp2\epsilon_+)}-4\frac{\prod_f \sinh\,(\fa+{\rm m}_f\pm\epsilon_+)}{\sinh(2\fa)\sinh\,(2\fa\pm2\epsilon_+)}\quad,\quad {\rm for }~\pm\xi>0~.
\ee
This function is not symmetric under the action of the Weyl group of the flavor symmetry group \footnote{The Weyl group under the $SO(8)$ flavor symmetry is generated by $m_i\leftrightarrow m_{i+1}$ and $m_3\leftrightarrow -m_4$ \cite{Ito:2011ea,Seiberg:1994aj}.} and is not invariant under the choice of $\xi\in \IR$. Therefore, this cannot be the correct form of $Z_{mono}(1,0)$.

From the AGT result presented above, we know that the correct $Z_{mono}(1,0)$ is given by 
\begin{align}\begin{split}\label{Zmono10}
Z_{mono}(1,0)=-4\frac{\prod_f \sinh(\fa-{\rm m}_f\mp\epsilon_+)}{\sinh(2\fa)\sinh(2\fa\mp2\epsilon_+)}-4\frac{\prod_f \sinh(\fa+{\rm m}_f\pm\epsilon_+)}{\sinh(2\fa)\sinh(2\fa\pm2\epsilon_+)}\qquad\\
+2\cosh\left(\sum_f m_f\pm2\epsilon_+\right)\quad,\quad {\rm for }~\pm\xi>0~.
\end{split}\end{align}
This answer for $Z_{mono}(1,0)$ is surprisingly independent of the choice of $\xi$:
\be
Z_{mono}(1,0;\xi>0)-Z_{mono}(1,0;\xi<0)=0~.
\ee
Clearly the AGT result for $Z_{mono}(1,0)$ in \eqref{Zmono10} does not match the localization result \eqref{ZmonoLoc10} due to the ``extra term'' $2 \cosh\left(\sum_f m_f\pm2\epsilon_+\right)$. 
Therefore, as noted in \cite{Ito:2011ea}, there is a discrepancy between the localization and AGT computation for $Z_{mono}(1,0)$. 

\subsubsection{$Z_{mono}(2,1)$}
\label{sec:Zmono21}

Now consider the $L_{2,0}$ line defect. This defect has 't Hooft charge 
\be
P=2h^1=4\hat{h}^1={\rm diag}(2,-2)\quad, \quad h^1\in \Lambda_{mw}~~,~~\hat{h}^1\in \Lambda_{cochar}~.
\ee
The expectation value of this line defect has two different monopole bubbling contributions:
\be
\langle L_{2,0}\rangle=\left(e^{2\fb}+e^{-2\fb}\right)F(\fa,m_f)^2+\left(e^{\fb}+e^{-\fb}\right)F(\fa,m_f)Z_{mono}(2,1)+Z_{mono}(2,0)~,
\ee
where $Z_{mono}(2,v)$ is $Z_{mono}(\fa,m_f,\epsilon_+;P,\vv)$ for $\vv={\rm diag}(v,-v)$. Here we will only be interested in the term $Z_{mono}(\fa,m_f,\epsilon_+;2,1)$. In this case,  the relevant SQM is given by\\

\begin{center}\begin{tikzpicture}[node distance=2cm,cnode/.style={circle,draw,thick,minimum size=10mm},snode/.style={rectangle,draw,thick,minimum size=10mm}]
\node[cnode] (1) {1};
\node[cnode] (2) [right of=1]{1};
\node[cnode] (3) [right of=2]{1};
\node[snode] (4) [below of=1]{1};
\node[snode] (6) [below of=3]{1};
\node[snode] (5) [below of=2]{4};
\draw[double,double distance=6pt,thick] (1) -- (2);
\draw[double,double distance=6pt,thick] (3) -- (2);
\draw[dashed,thick] (2) -- (5);
\draw[double,double distance=6pt,thick] (4) -- (1);
\draw[double,double distance=6pt,thick] (6) -- (3);
\path[dashed,every loop/.style={looseness=5}] (1)
         edge  [in=120,out=60,loop] node {} (); 
\path[dashed,every loop/.style={looseness=5}] (2)
         edge  [in=120,out=60,loop] node {} (); 
\path[dashed,every loop/.style={looseness=5}] (3)
         edge  [in=120,out=60,loop] node {} (); 
\end{tikzpicture}\end{center}

The contour integral from localization of this SQM is of the form
\begin{align}\begin{split}\label{eq:L20int}
Z^{JK}&(2,1)=\half\oint_{JK(\vec{\xi})} \left(\prod_{i=1}^3 \frac{d\varphi_i}{2\pi i}\right)\sinh^3(2\epsilon_+)\frac{\prod_{f=1}^4\sinh(\varphi_2- m_f)}{\sinh(\pm\varphi_{21}+ \epsilon_+)\sinh(\pm\varphi_{32}+\epsilon_+)}\times\\&
\frac{1}{\sinh(\pm(\varphi_1-\fa_2)+\epsilon_+)\sinh(\pm(\varphi_3-\fa_1)+\epsilon_+)}~.
\end{split}\end{align}

Evaluating the above contour integral requires a choice of parameter $\vec\xi\in \IR^3$ that specifies the JK residue prescription. Due to the intricate dependence on the choice of $\vec\xi$, we will examine this in the simple sectors of $\xi_i>0$ and $\xi_i<0$. 

For the choice of $\xi_i>0$, the Jeffrey-Kirwan prescription sums over the residues associated to four poles specified by the triples:
\begin{align}\begin{split}\begin{array}{rlll}
{\rm I.}~&~\varphi_1=-\fa-\epsilon_+~&~ \varphi_2=-\fa-2\epsilon_+~&~\varphi_3=\fa+\epsilon_+\\
{\rm II.}~&~\varphi_1=\fa-3\epsilon_+~&~ \varphi_2=\fa-2\epsilon_+~&~ \varphi_3=\fa-\epsilon_+\\
{\rm III.} ~&~ \varphi_1=-\fa-\epsilon_+ ~&~\varphi_2=\fa - 2\epsilon_+~&~ \varphi_3=\fa-\epsilon_+\\
{\rm IV.} ~&~ \varphi_1=-\fa-\epsilon_+ ~&~ \varphi_2=-\fa-2\epsilon_+~&~ \varphi_3=-\fa-3\epsilon_+
\end{array}
\end{split}\end{align}
Summing over the associated residues, the above contour integral evaluates to:
\begin{align}\begin{split}\label{ZmonoLoc21-1}
Z_{mono}^{(Loc)}(2,1)=
-\frac{4\prod_f\sinh(\fa+{\rm m}_f+2\epsilon_+)}{\sinh(2\fa+2\epsilon_+)\sinh(2\fa)}
-\frac{4\prod_f \sinh(\fa-{\rm m}_f-2\epsilon_+)}{\sinh(2\fa-2\epsilon_+)\sinh(2\fa-4\epsilon_+)}\\
-\frac{4\prod_f \sinh(\fa-{\rm m}_f-2\epsilon_+)}{\sinh(2\fa)\sinh(2\fa-2\epsilon_+)}-\frac{4\prod_f \sinh(\fa+{\rm m}_f+2\epsilon_+)}{\sinh(2\fa+2\epsilon_+)\sinh(2\fa+4\epsilon_+)}~,
\end{split}\end{align}
for $\xi_i>0$. 
Similarly, for $\xi_i<0$, the Jeffrey-Kirwan prescription sums over the residues associated to the poles
\begin{align}\begin{split}\begin{array}{rlll}
{\rm I.}~&~\varphi_1=-\fa+\epsilon_+~&~ \varphi_2=-\fa+2\epsilon_+~&~\varphi_3=\fa-\epsilon_+\\
{\rm II.}~&~\varphi_1=\fa+3\epsilon_+~&~ \varphi_2=\fa+2\epsilon_+~&~ \varphi_3=\fa+\epsilon_+\\
{\rm III.} ~&~ \varphi_1=-\fa+\epsilon_+ ~&~\varphi_2=\fa + 2\epsilon_+~&~ \varphi_3=\fa+\epsilon_+\\
{\rm IV.} ~&~ \varphi_1=-\fa+\epsilon_+ ~&~ \varphi_2=-\fa+2\epsilon_+~&~ \varphi_3=-\fa+3\epsilon_+
\end{array}
\end{split}\end{align}
In the case $\xi_i<0$, summing over the residues associated to these poles computes the contour integral to be 
\begin{align}\begin{split}\label{ZmonoLoc21-2}
Z_{mono}^{(Loc)}(2,1)=
-\frac{4\prod_f\sinh(\fa+{\rm m}_f-2\epsilon_+)}{\sinh(2\fa-2\epsilon_+)\sinh(2\fa)}
-\frac{4\prod_f \sinh(\fa-{\rm m}_f+2\epsilon_+)}{\sinh(2\fa+2\epsilon_+)\sinh(2\fa+4\epsilon_+)}\\
-\frac{4\prod_f \sinh(\fa-{\rm m}_f+2\epsilon_+)}{\sinh(2\fa)\sinh(2\fa+2\epsilon_+)}
-\frac{4\prod_f \sinh(\fa+{\rm m}_f-2\epsilon_+)}{\sinh(2\fa-2\epsilon_+)\sinh(2\fa-4\epsilon_+)}~.
\end{split}\end{align}
Now we can make use of the fact that the expectation value of line defects form a ring under the Moyal product \cite{Ito:2011ea}
\be
\langle L_{2,0}\rangle=\langle L_{1,0}\rangle \ast \langle L_{1,0}\rangle ~,
\ee
with respect to the (2,0) symplectic form $\Omega_J$:
\be
(f\ast g)(\fa,\fb)=e^{-\epsilon_+(\partial_b\partial_{a'}-\partial_a\partial_{b})}f(a,b)g(a',b')\big{|}_{\substack{a,a'=\fa\\b,b'=\fb}}~,
\ee
to compute $Z_{mono}(2,1)$ from the AGT computation for $\langle L_{1,0}\rangle$. This gives the result 
\begin{align}\begin{split}\label{Zmono21}
Z_{mono}&(2,1)=
-\frac{4\prod_f\sinh(\fa\pm{\rm m}_f+2\epsilon_+)}{\sinh(2\fa+2\epsilon_+)\sinh(2\fa)}
-\frac{4\prod_f \sinh(\fa\pm{\rm m}_f+2\epsilon_+)}{\sinh(2\fa+2\epsilon_+)\sinh(2\fa+4\epsilon_+)}\\&
-\frac{4\prod_f \sinh(\fa\mp{\rm m}_f-2\epsilon_+)}{\sinh(2\fa)\sinh(2\fa-2\epsilon_+)}
-\frac{4\prod_f \sinh(\fa\mp{\rm m}_f-2\epsilon_+)}{\sinh(2\fa-2\epsilon_+)\sinh(2\fa-4\epsilon_+)}\\
&+2\cosh\left(\sum_f m_f\pm6\epsilon_+\right)+2\cosh\left(\sum_f m_f\pm 2\epsilon_+\right)\quad, \quad \pm \xi_i>0~,~\forall i~.
\end{split}\end{align}
As before, this answer is independent of the choice of sign of $\xi_i$. \footnote{In fact, this is independent of the choice of $\vec\xi\in \IR^3\backslash\{0\}$. }

Again, we see that this does not match the localization computations for $Z_{mono}(2,1)$ for either choice of $\xi$ due to the ``extra term'' $+2\cosh\left(\sum_f m_f\pm6\epsilon_+\right)+2\cosh\left(\sum_f m_f\pm 2\epsilon_+\right)$. Further, the localization result is not independent of the choice of   $\vec\xi$ and for both choices of $\vec\xi$ (and indeed  for all other choices of $\vec\xi$), $Z_{mono}(2,1)$ is not invariant under Weyl symmetry of the $SO(8)$ flavor symmetry group. Therefore, as in the previous example, we find that the localization computation cannot be correct.

\section{Proposed Resolution: Coulomb Branch States}
\label{sec:Coulomb}

As we have shown, there is a discrepency between the localization and AGT result for the expectation value of 't Hooft defects in $SU(2)$ ${\rm N}_f=4$ supersymmetric gauge theory. Let us write the AGT result for the expectation value as
\begin{align}\begin{split}
&\langle L_{[P,0]}\rangle_{AGT}=\sum_{|\vv|\leq |P|} e^{(\vv,\fb)} \big(F(\fa)\big)^{|\vv|} Z_{mono}(\fa,m_i,\epsilon_+;P,\vv)~, \\
&Z_{mono}(\fa,m_i,\epsilon_+;P,\vv)=Z_{mono}^{(Loc)}(\fa,m_i,\epsilon_+;P,\vv)+Z_{mono}^{(extra)}(\fa,m_i,\epsilon_+;P,\vv)~,
\end{split}\end{align}
where $Z_{mono}^{(Loc)}=I_{\CH_0}^{(Loc)}$ is the localization computation for $Z_{mono}(P,\vv)$ and $Z_{mono}^{(extra)}(P,\vv)$ is some extra term that is the difference between $Z_{mono}(P,\vv)$ and $Z_{mono}^{(Loc)}=Z^{JK}$. 

We  now would like to understand what is the origin of the extra term $Z_{mono}^{(extra)}(P,\vv)$ that we must add to the localization computation to give the full result for $Z_{mono}(P,\vv)$. 
As we will now show, these extra contributions come from states that are not counted by localization.

\subsection{Witten Indices with Continuous Spectra}

\label{sec:toys}

As shown in \cite{Sethi:1997pa,Akhoury:1984pt,Yi:1997eg,Lee:2016dbm}, computing  the Witten index is much more subtle for theories with a continuous spectrum of states. In that case, the supercharges are non-Fredholm operators and thus the Witten index, which is still well defined, cannot be understood as the index of a supercharge operator. In order to illustrate some  features of the computation of the Witten index in these cases, we will take a brief aside to study a toy model that is closely related to the bubbling SQMs we are studying.

\subsubsection{Toy Model on Semi-Infinite Line}

Here we will examine a simplified model of the effective SQM on the Coulomb branch. Consider a supersymmetric particle on a semi-infinite line with a smooth potential 
$h(x)$. 
This theory is described by the  Hamiltonian \cite{Akhoury:1984pt,Hori:2014tda}
\be
H=\frac{e ^2p^2}{2}+\frac{e^2}{2} h^2(x)+\frac{e^2}{2}[\psi^\dagger,\psi]h'(x)~,
\ee
where $x,\psi$ are superpartners that satisfy the commutation relations 
\be
[x,p]=i\quad, \quad \{\psi^\dagger,\psi\}=1~.
\ee
These fields satisfy the supersymmetry transformations
\begin{align}\begin{split}
&\delta_\eta x=\frac{1}{\sqrt{2}}\eta \,\psi+\frac{1}{\sqrt{2}}\eta^\dagger \psi^\dagger~,\\
&\delta_\eta \psi=\frac{1}{\sqrt{2}}  \eta^\dagger( ip+h(x))~,
\end{split}\end{align}
which are generated 
by the supercharge
\be
Q=\frac{1}{\sqrt{2}}\psi^\dagger(ip+ h)~.
\ee
Let us consider a toy model of the effective SQM on an asymptotic Coulomb branch where 
$h(x)=h_0+\frac{q}{x}$ where $q\neq 0$.  In our applications, $2q$ will be an integer. 

Generic states in this theory are described by wave functions of the form  
\be
\psi(x)=f_+(x)|0\rangle+f_-(x)\psi^
\dagger|0\rangle~
\ee
where we define the Clifford vacuum by 
$\psi|0\rangle=0$.

We are interested in computing the ground state index of this theory. This can be derived from the spectrum of the Hamiltonian. The Hamiltonian can be written  as a diagonal operator 
\be
H=-\frac{e^2}{2}\partial_x^2+\frac{e^2(q^2\pm q)}{2x^2}+\frac{e^2q h_0}{x}+\frac{e^2}{2} h_0^2~,
\ee
on a basis of states $\{f_+(x)|0\rangle,f_-(x)\psi^\dagger|0\rangle\}$. 
Thus, eigenstates of the Hamiltonian solve the differential equation
\be
\left(-\partial_x^2+\frac{q^2\pm q}{x^2}+\frac{2q h_0}{x}+h_0^2-\frac{2E}{e^2}\right)f_\pm(x)=0~.
\ee
 The $L^2$-normalizable solutions of this equation are given by \footnote{
Note that we could also solve for the space of BPS states by writing
\be
\psi(x)=f_+(x)|0\rangle+f_-(x)\psi^
\dagger|0\rangle~\longrightarrow~\psi(x)=\left(\begin{array}{c}
f_+(x) \\
f_-(x) \end{array}\right)~,
\ee
so that the real supercharge  $\CQ =\big(Q+\bar{Q}\big)$ acts as the differential operator
\be
\CQ=\frac{1}{\sqrt{2}}\left(\begin{array}{cc}
0&\partial_x+h(x)\\
-\partial_x+h(x)&0
\end{array}\right)=\frac{1}{\sqrt{2}}\big(i\sigma^2\partial_x+ \sigma^1h(x)\big)~. 
\ee
It is now straightforward to solve for the kernel of $\CQ$ and impose normalizability. When one further takes into account boundary conditions, equation \eqref{GSIToy} is reproduced.  } 
\be
f_s(x)=c\, e^{-\kappa x}x^{j_{s_1}}~_1F_1\left(j_{s_1} +\frac{s_2|q|\, |h_0|}{\kappa},2s_1j_{s_1} ,2\kappa x\right)\quad, \quad s,s_i=\pm 1~,
\ee
where $c$ is a constant,  $\kappa=\sqrt{h_0^2-\frac{2E}{e^2}}$, $s_1=s\times {\rm sign}(q)$, $s_2={\rm sign}(q)\times {\rm sign}(h_0)$, and 
\be
j_{s_1}=|q|+\frac{1+s_1}{2}\quad {\rm or}\quad j_{s_1}=-|q|+\frac{1+s_1}{2}~. 
\ee
 
Further, due to the large $x$-behavior of the confluent hypergeometric function of the first kind
\be
_1F_1(m,n,x) \underset{ x \to \infty}{\sim} \Gamma(n)e^{x}\frac{x^{m-n}}{\Gamma(m)}\quad, \quad m\notin \IZ_+~,
\ee 
$L^2$-normalizability implies that 
\be
j_{s_1}> -\half\quad, \quad j_{s_1} +\frac{s_2|q|\, |h_0|}{\kappa}=-n \quad, \quad n\in \IZ_+~.
\ee
The first condition comes from imposing regularity at $x=0$ whereas the second condition comes from imposing regularity at $x\to \infty$. Together, these conditions imply that there is a global minimum of the potential energy at some $x>0$ that supports a bound state. 


From regularity at $x\to \infty$, we can solve for the discrete spectrum of the Hamiltonian for generic $h_0,q$
\be
E_{n}=\left(1-\frac{4 q^2}{(1+2n+2|q|+s_1)^2}\right)\frac{e^2h_0^2}{2}~.
\ee
In the case of BPS ground states ($E=0$), the $L^2$-normalizability constraints imply that the ground states of this theory solve
\be\label{SecondCon}
(1+s_2)|q|+\half(1+s_1)=-n~,
\ee
or 
\be
(s_2-1)|q|+\half(1+s_1)=-n\quad{\rm and}\quad |q|<\half~. 
\ee
This implies that the allowed solutions are those with $s_1,s_2<0$ or $s_1<0<s_2$ with $|q|<\half$. 

As in the case of the Coulomb branch, this theory  has a continuum of scattering states.  These occur due to the non-compact direction in field space where the potential energy approaches a finite value $\lim_{x\to \infty}U(x)=\frac{e^2h_0^2}{2}$. In this case, the gap to the continuum is given by $E_{\rm gap}=\frac{e^2h_0^2}{2}$. 
Thus, the full spectrum of states is similar to that of the Hydrogen atom. There is a discrete spectrum of states for energies $E<E_{\rm gap}$ that accumulate at $E_{\rm gap}$ and a continuous spectrum of states for energies $E\geq E_{\rm gap}$.

However, there is an additional subtlety to this model. Due to the presence of the boundary at $x=0$, we additionally have to worry about the real supercharge $\CQ=Q+\bar{Q}$ being Hermitian:
\be\label{toyHerm}
\langle \Psi_1|\CQ\Psi_2\rangle=\langle \CQ \Psi_1|\Psi_2\rangle~. 
\ee
If we consider two generic states 
\be
|\Psi_i\rangle=f_i(x)|0\rangle+g_i(x)\psi^\dagger|0\rangle~\mapsto~|\Psi_i\rangle=\left(\begin{array}{c}f_i(x)\\g_i(x)\end{array}\right)~,
\ee
then the real supercharge operator $\CQ$ acts as 
\be
\CQ=\frac{1}{\sqrt{2}}\left(\begin{array}{cc}
0&\partial_x+h(x)\\
-\partial_x+h(x)&0
\end{array}\right)
\ee
The constraint that $\CQ$ be Hermitian \eqref{toyHerm}, then reduces to 
\be
\left[\bar{f}_1g_2-\bar{g}_1f_2\right]_{x=0}=0~. 
\ee
There are 3 different types of restrictions we can impose on the Hilbert space so that $\CQ$ is Hermitian:
\begin{enumerate}
\item Impose $f(x)=0$, $\forall x\in \IR^+$. In this case the Hilbert space is reduced so that wave functions are only of the form $\CH={\rm span}_{L^2(\IR_+)}\{\psi^\dagger|0\rangle\}$. 
\item Impose $g(x)=0$, $\forall x\in \IR^+$. In this case the Hilbert space is reduced so that wave functions are only of the form $\CH={\rm span}_{L^2(\IR_+)}\{|0\rangle\}$. 
\item Impose $f(0)=0$, $g(0)=0$ or $f(0)$ and $g(0)=0$. In this case we restrict the form of the wave functions allowed in the Hilbert space $\CH={\rm span}_{\substack{L_2(\IR^+)\\\langle x|\psi\rangle|_{x=0}=0}}\{|0\rangle\,,\,\psi^\dagger |0\rangle\}$. 
\end{enumerate}

These three different choices give the different answers. They are each given by 
\begin{align}\begin{split}\label{GSIToy}
&{\rm 1.)}\quad I_{\CH_0}=\begin{cases}
-1&h_0<0~,~q>-\half\\
0&{\rm else}
\end{cases}\quad, \quad {\rm 2.)}\quad
I_{\CH_0}=\begin{cases}
1&h_0>0~,~q<\half\\
0&{\rm else}
\end{cases}~, \\&
{\rm 3.)}\quad
I_{\CH_0}=
\begin{cases}
0& h_0\times q >0\\
-1&q>0> h_0\\
1&h_0>0>q
\end{cases}
\end{split}\end{align}

Now let us try to use localization to compute the ground state index.  This SQM can be described by the Lagrangian 
\be
L=\frac{1}{2e^2}\left(\dot{x}^2+\psi^\dagger \dot\psi+D^2\right)+D h(x)-\frac{e^2}{2} h'(x)[\psi^\dagger ,\psi]~,
\ee
on the half space $x>0$. As before, this Lagrangian is $Q$-exact. Thus, by studying the limit $e\to 0$, we see that the path integral  localizes to the $Q$-fixed points $\dot{x}=0$. This reduces the path integral to an integral over the line \footnote{Here we fix the normalization of the path integral so that the ground state index is an integer.}
\be
Z=\int_\IR dD \int_{\IR^+} dx \int [d\psi d\psi^\dagger] \, e^{-S}~.
\ee
In this case, the 1-loop determinant  comes from integrating over the fermion zero modes. The partition function then reduces to 
\be\label{toypart1}
Z=-\int_\IR dD\int_{\IR^+} dx \,\beta \,h'(x) e^{-\frac{\pi \beta D^2}{ e^2}+2\pi i\beta D h}~. 
\ee
This can be evaluated by first integrating over $D$:
\be
Z=-\int_{\IR^+}dx \,\beta \,h'(x)\sqrt{\frac{e^2}{\beta}}e^{-\pi \beta e^2 h^2(x)}~.
\ee
We can then evaluate the partition by making a change of variables. This produces the result
\be\label{Ztoy}
Z=\half\begin{cases}
-1+{\rm erf}\left(\sqrt{\pi\beta e^2} h_0\right)& q>0\\
1+{\rm erf}\left(\sqrt{\pi\beta e^2} h_0\right)& q<0
\end{cases}
~.
 \ee


Now we can compute the ground state index by taking the limit as $\beta\to \infty$. In this case we find that the localization computation for the ground state index is exactly given by 
\be \label{Zqplus}
I_{\CH_0}^{(Loc)}=
\begin{cases}
0& h_0\times q >0\\
-1&q>0> h_0\\
1&h_0>0>q
\end{cases}
\ee
Which matches with the explicit computation in the SQM for the third choice of boundary condition. 
Note that in the case $q=0$ the Witten index is identically zero because the fermionic fields are non-interacting and massless. This is also reflected in the identically vanishing of the path integral since there are no fermion insertions. However, any correction can lift this exact degeneracy and give rise to a possibly non-trivial Witten index. 

\subsubsection{Effective Coulomb Branch SQM	}
Now let us apply this computation to the effective SQM on the Coulomb branch for the bubbling SQM of the minimal 't Hooft defect in the $SU(2)$ $\CN=2$ gauge theory with ${\rm N}_f=4$ fundamental hypermultiplets. See Section \ref{sec:examples} and Appendix \ref{app:A}.  We will refer to this theory as $T_{\CM_C}$. This theory is again  a supersymmetric particle moving in a potential 
\be\label{explicith}
h(\sigma)=\langle D\rangle(\sigma)=\xi+\frac{1}{2}\sum_{i=1}^2 \left(\frac{1}{\omega_i}-\frac{1}{\tilde\omega_i}\right)~.
\ee
However, we are now taking the theory 
on two semi-infinite intervals $I_+=\{\sigma>a+\epsilon\}$ and $I_-=\{\sigma<-a-\epsilon\}$ 
where $\omega_i,\tilde\omega_i$ are the effective masses of the integrated out hypermultiplet fields
\be
\omega_i=|\sigma+(-1)^i a+\epsilon|\quad, \quad \tilde\omega_i=|\sigma+(-1)^ia-\epsilon|~,
\ee 
for parameters $a,\epsilon\in \IR_+$.  On each of these intervals, the vacuum state has Fermion number $-1$ and has flavor charges $+1$ on $I_+$ and $-1$ on $I_-$. Note that this differs slightly from the effective SQM in Appendix \ref{app:A}. They are related by a different choice of normalization for the wave function describing the matter fields. \footnote{See Appendix \ref{app:A} for more details.} 

The localization computation of the ground state index proceeds as before except that there are now two boundary contributions at $\sigma=\pm(a+\epsilon).$ 
Due to the limiting behavior of $h(\sigma)$, the localization result  for the ground state index, analogous to \eqref{Ztoy}, is 
 given by 
\begin{align}\begin{split}\label{BObulk}
I_{\CH_0}^{(Loc)}&\left(T_{\CM_C}\right)=-\lim_{\beta\to \infty}\sinh(2\epsilon_+)e^{\sum_f m_f}\left(1-{\rm erf}\left(\sqrt{\pi\beta}e \xi\right)\right)\\&-\lim_{\beta\to \infty}\sinh(2\epsilon_+)e^{-\sum_f m_f}\left(1+{\rm erf}\left(\sqrt{\pi\beta}e \xi\right)\right)~.
\end{split}\end{align}
Here, the factor of $2\sinh(2\epsilon_+)$ comes from the decoupled Fermi-multiplet in the $\CN=(0,4)$ vector multiplet described by $\lambda^2,\bar\lambda_2$ and the $e^{\pm\sum_f m_f}$ comes flavor charge of the ground state on $I_+$ and $I_-$ respectively. By using the explicit form of $h(\sigma)$ \eqref{explicith}, we obtain the result
\be\label{ZCoulomb}
I_{\CH_0}^{(Loc)}\left(T_{\CM_C}\right)=-
2\sinh(2\epsilon_+)e^{\mp \sum_f m_f}\quad, \quad \pm \xi>0~.
\ee


Solving for the entire spectrum of this theory is much more difficult than in the previous example. However, only the BPS states contribute to the ground state index. These are computed in the Born-Oppenheimer approximation in Appendix \ref{app:A}. 
 In summary, we find that there are over 10 different types of restrictions on the Hilbert space in each interval that make $\CQ$ Hermitian: we will make a symmetric choice. 
Two distinguished choices lead to
\begin{align}\begin{split}\label{GSI}
&{\rm 1.) }\quad 
I_{\CH_0}\left(T_{\CM_C}\right)=2\cosh\left(\sum_f m_f\pm 2\epsilon_+\right)\quad, \quad \pm \xi>0~,\\
&{\rm 2.)}\quad 
I_{\CH_0}\left(T_{\CM_C}\right)=-
2\sinh(2\epsilon_+)e^{\mp \sum_f m_f}\quad, \quad \pm \xi>0~.
\end{split}\end{align}
Now we see that the localization computation matches the explicit computation for the second choice of boundary condition.

However, now recall the localization expression for $Z_{mono}(1,0)$ in the $SU(2)$ ${\rm N}_f=4$ theory in the expectation value of $L_{1,0}$ from Section \ref{sec:NullEx}. There we showed that $Z_{mono}^{(Loc)}(1,0)\neq Z_{mono}(1,0)$ as computed via AGT. In fact, they differed by a term 
\be
Z_{mono}^{(extra)}(1,0):=Z_{mono}(1,0)-Z_{mono}^{(Loc)}(1,0)=2\cosh\left(\sum_f m_f\pm 2\epsilon_+\right)\quad, \quad \pm \xi>0~.
\ee
However, this is exactly the computation of the ground state index of the BPS states localized on the Coulomb branch with the first choice \eqref{GSI}. As it turns out, there is a unique choice of boundary conditions if we restict to the case of pure Neumann or Dirichlet.

This result is in fact very natural. Recall that in Section \ref{sec:HiggsCoulomb} we explained that the Hilbert space of BPS states of a generic bubbling SQM theory can be decomposed into states localized on the Higgs, Coulomb, and mixed branches:
\be
\CH_{BPS}=\CH_{BPS}^{(Higgs)}\oplus \CH_{BPS}^{(Coulomb)}\oplus\CH_{BPS}^{(mixed)}~. 
\ee
Thus, the ground state index should similarly decompose as
\be
I_{\CH_0}=I_{Higgs}+I_{Coulomb}+I_{mixed}~. 
\ee
In our case there is no mixed branch so that the summand $\CH_{BPS}^{(mixed)}$ is trivial and $I_{mixed}=0$. However, since we have identified the Jeffrey-Kirwan residue, $Z^{JK}=I_{Higgs}$, as counting the Higgs branch states, we have that $I_{\CH_0}^{(Loc)}=I_{Higgs}$ has no contribution from Coulomb branch states. Therefore, it is clear that we need to add a term
\be
I_{asymp}=I_{Coulomb}+I_{mixed}~,
\ee
which counts the BPS ground states on the non-compact Coulomb and mixed branches. 


\subsection{Proposal}
Thus far we have been able to show that the localization computation of $I_{\CH_0}$, $I_{\CH_0}^{(Loc)}$, reproduces the $JK$-prescription for the path integral, but that this does not correctly reproduce $Z_{mono}(P,\vv)$ at least with the regularization procedure for localization that we have adopted.  
Further, by identifying $Z^{JK}$ as counting Higgs branch states, we were able to conclude that  $I_{\CH_0}^{(Loc)}$ does not count any contributions from the ground states along the Coulomb and mixed branches.

Therefore, we propose that $Z_{mono}^{(extra)}(P,\vv)$ is the contribution of 
BPS states along the non-compact vacuum branches in the bubbling SQM. Mathematically, this can be phrased as 
\be
Z_{mono}=I_{\CH_0}=I_{\CH_0}^{(Loc)}+I_{asymp}=Z^{JK}+I_{asymp}~,
\ee
where $I_{asymp}:=I_{Coulomb}+I_{mixed}$ is the ground state index evaluated on the states localized along the Coulomb and mixed branches and $Z^{JK}=I_{Higgs}$ is the Jeffrey-Kirwan sum over residues \cite{JK}.

 Note that $I_{asymp}$ is fundamentally distinct from the defect term $\delta I_{\CH_0}$ which similarly can be appended to $\lim_{\beta\to \infty}I_{W}^{(Loc)}$ to correct the localization result \cite{Sethi:1997pa,Yi:1997eg,Lee:2016dbm}. As shown in section \ref{sec:sec3} and \ref{sec:toys}, our computation of $I_{\CH_0}^{(Loc)}$ has already taken 
the defect term into account. Rather, we propose that one must add an additional term $I_{asymp}$ that corrects for the omitted ground states localized on the non-compact vacuum branches.

These states can be computed in the effective theory on the relevant vacuum branches in the Born-Oppenheimer approximation. As we will see, this definition is independent of the choice of $\xi$ in all known examples. Since the Born-Oppenheimer approximation is only valid for $|\sigma/a|,|\sigma/\epsilon_+|>>0$, we must make a choice of effective boundary conditions at $|\sigma|=|a|+|\epsilon_+|$. Unitarity 
then restricts the types of allowed  boundary conditions. In each of the following examples, there exists a (sometimes unique) boundary condition such that $I_{asymp}=Z_{mono}^{(extra)}$. We have chosen to use this boundary condition in all cases. The cases in which $I_{asymp}=0$ do not require such a choice.

\subsubsection{Relation to Defect Contribution}

\label{sec:defect}

The correction of the ground state index by $I_{asymp}$ at first glance appears to be similar to the work of \cite{Yi:1997eg,Sethi:1997pa} in which the authors compute the ground state index by adding a ``defect term'' or ``secondary term'' to the Witten index. However, the two stories are quite different. The definition of the  ``defect term" 
relies on rewriting 
\be
I_{\CH_0}=I_{\CH_0}^{Bulk}(\beta_0)+\delta I_{\CH_0}(\beta_0)~,
\ee
where
\begin{align}\begin{split}
I_{\CH_0}^{Bulk}&=I_W(\beta_0)={\rm Tr}_\CH(-1)^F e^{-\beta_0 H+...}~,\\
\delta I_{\CH_0}&=\int_{\beta_0}^\infty d\beta\,\partial_\beta\left({\rm Tr}_\CH(-1)^F e^{-\frac{\beta}{2}\{\CQ,\CQ\}+...}\right)~,
\end{split}\end{align}
which we will call the \textit{bulk} and \textit{defect} terms respectively. This is a trivial rewriting by making use of the fundamental theorem of calculus. When we write the path integral as an integral over field space, we can use supersymmetry to rewrite the $\partial_\beta$ in the defect term as  
\begin{align}\begin{split}
\partial_\beta\left({\rm Tr}_\CH (-1)^F e^{-\frac{\beta}{2}\{\CQ,\CQ  \}+...}\right)=-{\rm Tr}_\CH (-1)^F \CQ ^2 e^{-\frac{\beta}{2}\{\CQ ,\CQ \}+...}={\rm Tr}_\CH \CQ (-1)^F \CQ  e^{-\frac{\beta}{2}\{\CQ ,\CQ \}+...}~.
\end{split}\end{align}
Then by integrating by parts inside the path integral \cite{Akhoury:1984pt,Sethi:1997pa}, this is equal to a derivative on field space 
\begin{align}\begin{split}
{\rm Tr}_\CH \CQ (-1)^F \CQ  e^{-\frac{\beta}{2}\{\CQ ,\CQ \}+...}&={\rm Tr}_\CH(-1)^F \CQ ^2 e^{-\beta  H+...}+{\rm Tr}_\CH \partial_{\phi^i}\left(\psi^i  (-1)^F \CQ  e^{-\frac{\beta}{2}\{\CQ ,\CQ \}+...}\right)\\&
=\half{\rm Tr}_\CH \partial_{\phi^i}\left(\psi^i  (-1)^F \CQ  e^{-\frac{\beta}{2}\{\CQ ,\CQ \}+...}\right)~,
\end{split}\end{align}
where $\psi^i\partial_{\phi_i} $ are the derivative terms in the supercharge which in turn can be written as a boundary integral in field space. One might therefore hope that the defect term is a feasible computation. 

The utility of this rewriting is in the fact that the bulk term $I_{\CH_0}^{Bulk}$ can be computed exactly in the limit as $\beta_0\to 0$ by heat-kernel techniques. In this way, one might try to compute the ground state index $I_{\CH_0}$. 

In this paper we are using a different decomposition of $I_{\CH_0}$. 
Here, we want to compute the ground state index directly by using localization. Since we can compute the Witten index via localization for generic $\beta$, we find that we can simply take the limit as $\beta\to \infty$ to obtain the ground state index
\be
I_{\CH_0}^{(Loc)}=\lim_{\beta\to \infty}I_W^{(Loc)}(\beta)~. 
\ee
Unfortunately, once we have used what appears to be the most natural way of regularizing the localized integral, we find that the localization expression for the ground state index does not agree with AGT. Further, we find that this can be corrected by adding the contribution from BPS states that are localized along non-compact directions in field space with finite asymptotic potential. 

In summary, the difference between our proposal and the defect term of \cite{Sethi:1997pa,Yi:1997eg} is that the defect term $\delta I_{\CH_0}$ is the difference between the Witten index at $\beta=0$ and the ground state index whereas the asymptotic contribution $I_{asymp}$ counts the BPS states that are omitted in the implementation of localization to compute the ground state index.

\subsection{Examples}
\label{sec:examples}

In this section, we will provide several non-trivial examples to show that $Z_{mono}^{(extra)}(P,\vv)$ is indeed reproduced by the ground state index of the Coulomb and mixed branch BPS states. 
In these examples we will study components of the expectation value of two line defects: $L_{1,0}$ and $L_{2,0}$. Specifically, we will be again interested in $Z_{mono}(1,0)$ and $Z_{mono}(2,1)$. 

Although we are only performing the computation for examples of abelian gauge groups, there is no fundamental obstruction for performing the analogous computations for non-abelian gauge groups. The computation would be analogous to that of Appendices \ref{app:A} and \ref{app:B} with increased computational complexity. We believe that in the case of a non-abelian bubbling SQM, $I_{asymp}$  may have a more interesting form and could potentially depend on the gauge fugacity $\fa$.

\subsubsection{$SU(2)$ ${\rm N}_f=4$ Theory}

In this theory we will study components of the expectation value of two line defects: $L_{1,0}$ and $L_{2,0}$. Specifically, we will be interested in $Z_{mono}(1,0)$ and $Z_{mono}(2,1)$. \\

\noindent$\mathbf{Z_{mono}(1,0)}$\\

In the full expression for the expectation value of the $L_{1,0}$ 't Hooft line defect, there are two terms that contribute to $Z_{mono}(1,0)$:
\be
Z_{mono}(1,0)=Z_{mono}^{(Loc)}(1,0)+Z_{mono}^{(extra)}	(1,0)~.
\ee
As shown in the previous section, the localization result for this $Z_{mono}(1,0)$ is given by  \eqref{ZmonoLoc10} whereas the full expression for $Z_{mono}(1,0)$, as we know from AGT, is given by \eqref{Zmono10}. 
%
This means that $Z_{mono}^{(extra)}(1,0)$ is given  by 
\be
Z_{mono}^{(extra)}(1,0)=\begin{cases} 2\cosh\left(\sum_f m_f+2\epsilon_+\right)& \xi>0\\
2\cosh\left(\sum_f m_f-2\epsilon_+\right)& \xi<0
\end{cases}
\ee 
We conjecture that this should be exactly reproduced by the Witten index of the ground states on the Coulomb branch. 

As we have shown in Appendix \ref{app:A}, this is indeed exactly reproduced by the Witten index of the asymptotic states on the Coulomb branch: 
\be
Z_{mono}^{(extra)}(1,0)=I_{asymp}(1,0)=\begin{cases} 2\cosh\left(\sum_f m_f+2\epsilon_+\right)& \xi>0\\
2\cosh\left(\sum_f m_f-2\epsilon_+\right)& \xi<0
\end{cases}
\ee

\noindent$\mathbf{Z_{mono}(2,1)}$\\

Again by comparing the localization expressions \eqref{ZmonoLoc21-1}-\eqref{ZmonoLoc21-2} with the full expression from AGT \eqref{Zmono21} for $Z_{mono}(2,1)$, we find that 
\be
Z_{mono}^{(extra)}(2,1)=
2\cosh\left(\sum_f m_f\pm 6\epsilon_+\right)+2\cosh\left(\sum_f m_f\pm2\epsilon_+\right)\quad, \quad \pm \xi_i>0~,~ \forall i~. 
\ee
%
As shown in Appendix \ref{app:B}, $Z_{mono}^{(extra)}(2,1)$ is exactly reproduced by the ground state index of the effective super quantum mechanics on the Coulomb branch ($I_{asymp})$:
\be
I_{asymp}=
2\cosh\left(\sum_f m_f\pm 6\epsilon_+\right)+2\cosh\left(\sum_f m_f\pm2\epsilon_+\right)\quad, \quad \pm \xi_i>0~,~ \forall i~.
\ee

%
By explicit computation, one can see that the Coulomb branch terms restore Weyl-invariance\footnote{The Weyl group under the $SO(8)$ flavor symmetry is generated by $m_i\leftrightarrow m_{i+1}$ and $m_3\leftrightarrow -m_4$
.} and invariance of $Z_{mono}(2,1)$ under the choice of $\vec\xi$. 

Note that the bubbling SQM for this example has a non-trivial mixed branch. However, we conjecture that there are no states localized there. See \ref{app:mixed} for more details.

\subsubsection{$SU(2)$ ${\rm N}_f=2$ Theory}

Now consider the $L_{1,0}$ (minimal) 't Hooft defect in the $SU(2)$ ${\rm N}_f=2$ theory. As in the case of the $SU(2)$ ${\rm N}_f=4$ theory, this has 't Hooft charge 
\be
P=h^1=2\hat{h}^1={\rm diag}(1,-1)\quad, \quad h^1\in \Lambda_{mw}~~,~~\hat{h}^1\in \Lambda_{cochar}~.
\ee
Similarly, the expression for its expectation value is of the form 
\be
\langle L_{1,0}\rangle=\left(e^{\fb}+e^{- \fb}\right)F(\fa,m_f)+Z_{mono}(1,0)~,
\ee
where 
\be
F(\fa,m_f)=\left(\frac{\prod_\pm\prod_{f=1}^2 \sinh\,(\fa\pm m_f)}{\sinh(2\fa)\prod_{\pm}\sinh\,(2\fa\pm 2\epsilon_+)}\right)^\half~. 
\ee
The monopole bubbling contribution  can be computed as the Witten index of the $\CN=(0,4)$ SQM described by the $\CN=(0,2)$ quiver:
\begin{center}
\begin{tikzpicture}[node distance=2cm,cnode/.style={circle,draw,thick,minimum size=10mm},snode/.style={rectangle,draw,thick,minimum size=10mm}]
\node[cnode] (1) {1};
\node[snode] (4) [below of=1]{2};
\node[snode] (5) [above of=1]{2};
\draw[dashed,thick] (1) -- (5);
\draw[double,double distance=6pt,thick] (4) -- (1);
\path[dashed,every loop/.style={looseness=5}] (1)
         edge  [in=120,out=60,loop] node {} (); 
\end{tikzpicture}
\end{center}
The Witten index of this quiver SQM reduces to the contour integral 
\be
Z^{(Loc)}_{mono}(1,0)=\oint_{JK(\xi)}\frac{d\varphi}{2\pi i}Z_{vec}(\varphi,\fa,\epsilon_+)Z_{hyper}(\varphi,\fa,\epsilon_+) Z_{Fermi}(\varphi,\fa,\epsilon_+)~,
\ee
which is explicitly given by 
\be
Z_{mono}^{(Loc)}(1,0)=\half\oint_{JK(\xi)}\frac{d\varphi}{2\pi i} \sinh(2\epsilon_+)\frac{\prod_{f=1}^2\sinh(\varphi- m_f)}{\prod_\pm\sinh(\varphi\pm \fa+\epsilon_+)\sinh(-\varphi\pm \fa+\epsilon_+)}~.
\ee
This integral evaluates to \footnote{Note that this required fixing the overall sign of the Jeffrey-Kirwan residue computation. The reason is that the JK prescription does not give a derivation of the overall sign.
}
\be
Z^{(Loc)}_{mono}(1,0)=\frac{\prod_f \sinh(\fa-{\rm m}_f\mp\epsilon_+)}{\sinh(2\fa)\sinh(2\fa\mp2\epsilon_+)}+\frac{\prod_f \sinh(\fa+{\rm m}_f\pm\epsilon_+)}{\sinh(2\fa)\sinh(2\fa\pm2\epsilon_+)}\quad,\quad {\rm for }~\pm\xi>0~.
\ee

From carefully taking the limit of $\langle L_{1,0}\rangle$ in the ${\rm N}_f=4$ theory to the ${\rm N}_f=2$ theory\footnote{This requires taking $Re[m_4],Re[m_3]\to \infty$ such that $sgn(Re[m_4]/Re[m_3])=-1$. This is a very subtle point that we will discuss in section \ref{sec:CS}. See \cite{Intriligator:1997pq} for more details about the analogous issues in 5D SYM.}, we can see that the correct $Z_{mono}(P,\vv)$ contribution is given exactly by 
\be
Z_{mono}(1,0)=\frac{\prod_f \sinh(\fa-{\rm m}_f\mp\epsilon_+)}{\sinh(2\fa)\sinh(2\fa\mp2\epsilon_+)}+\frac{\prod_f \sinh(\fa+{\rm m}_f\pm\epsilon_+)}{\sinh(2\fa)\sinh(2\fa\pm2\epsilon_+)}\quad,\quad {\rm for }~\pm\xi>0~.
\ee
Thus, we see that 
\be
Z_{mono}^{(extra)}(1,0)=0~. 
\ee
As seen in Appendix \ref{app:A}, there are no Coulomb branch states and thus:
\be
Z_{mono}^{(extra)}(1,0)=I_{asymp}(1,0)=0~. 
\ee
Note that the fact that $Z_{mono}^{(extra)}(1,0)=0$ is consistent with the fact that $Z_{mono}$(1,0) is 
actually invariant under the choice of $\xi$ and under the Weyl symmetry of the flavor symmetry group. 

\subsubsection{$SU(2)$ SYM Theory}

Here we will again be interested in the expectation value of the minimal 't Hooft defect that experiences monopole bubbling. This line defect has 't Hooft charge 
\be
P=2h^1=2\hat{h}^1=\half{\rm diag}(2,-2)\quad, \quad h^1\in \Lambda_{mw}~~,~~\hat{h}^1\in \Lambda_{cochar}~,
\ee
and hence is the next to minimal 't Hooft defect: $L_{2,0}$. 

 Its expectation value takes a similar form to $\langle L_{1,0}\rangle$ of the $SU(2)$ ${\rm N}_f=4$ theory:
\be
\langle L_{2,0}\rangle=\left(e^{2 \fb}+e^{-2 \fb}\right)F(\fa,m_f)+Z_{mono}(2,0)~,
\ee
where 
\be
F(\fa,m_f)=\left(\frac{1}{\sinh(2\fa)\prod_{\pm}\sinh(2\fa\pm 2\epsilon_+)}\right)^\half~. 
\ee
The monopole bubbling contribution ($Z_{mono})$ can be computed as the Witten index of the $\CN=(0,4)$ SQM described by the $\CN=(0,2)$ quiver:
\begin{center}
\begin{tikzpicture}[node distance=2cm,cnode/.style={circle,draw,thick,minimum size=10mm},snode/.style={rectangle,draw,thick,minimum size=10mm}]
\node[cnode] (1) {1};
\node[snode] (4) [below of=1]{2};
\draw[double,double distance=6pt,thick] (4) -- (1);
\path[dashed,every loop/.style={looseness=5}] (1)
         edge  [in=120,out=60,loop] node {} (); 
\end{tikzpicture}
\end{center}
The Witten index of this quiver SQM reduces to the contour integral 
\be
Z^{(Loc)}_{mono}(2,0)=\oint_{JK(\xi)}\frac{d\varphi}{2\pi i}Z_{vec}(\varphi,\fa,\epsilon_+)Z_{hyper}(\varphi,\fa,\epsilon_+) 
~,
\ee
which is explicitly given by 
\be
Z_{mono}^{(Loc)}(2,0)=\frac{1}{8}\oint_{JK(\xi)}\frac{d\varphi}{2\pi i} \sinh(2\epsilon_+)\prod_\pm\frac{1
}{\sinh(\varphi\pm \fa+\epsilon_+)\sinh(-\varphi\pm \fa+\epsilon_+)}~.
\ee
This integral evaluates to 
\be
Z^{(Loc)}_{mono}(2,0)=-\frac{1
}{4\sinh(2\fa)\sinh(2\fa\mp2\epsilon_+)}-\frac{1
}{4\sinh(2\fa)\sinh(2\fa\pm2\epsilon_+)}\quad,\quad {\rm for }~\pm\xi>0~.
\ee

From carefully taking the limit of $\langle L_{2,0}\rangle$ in the ${\rm N}_f=4$ theory to the ${\rm N}_f=0$ theory\footnote{As before, this requires a bit of care by taking $Re[m_i]\to \infty$ such that two masses go to $+\infty$ and two go to $-\infty$.}, we can see that the correct $Z_{mono}(2,0)$ contribution is given exactly by 
\be
Z_{mono}(2,0)=-\frac{1
}{4\sinh(2\fa)\sinh(2\fa\mp2\epsilon_+)}-\frac{1
}{4\sinh(2\fa)\sinh(2\fa\pm2\epsilon_+)}\quad,\quad {\rm for }~\pm\xi>0~.
\ee
Thus, we see that 
\be
Z_{mono}^{(extra)}(2,0)=0~. 
\ee
As seen in Appendix \ref{app:A}, there are no Coulomb branch states and thus
\be
Z_{mono}^{(extra)}(2,0)=I_{asymp}(2,0)=0~. 
\ee
This is again consistent with the fact that $Z_{mono}^{(Loc)}(2,0)$ is independent of the choice of $\xi$ and invariant under the action of the Weyl symmetry of the flavor symmetry group. 

\subsubsection{$SU(2)$ $\CN=2^\ast$ Theory}

Here we will again be interested in the expectation value of the minimal 't Hooft defect that exhibits monopole bubbling. As in the case of $SU(2)$ SYM theory, 
this line defect has 't Hooft charge 
\be
P=2h^1=2\hat{h}^1=\half{\rm diag}(2,-2)\quad, \quad h^1\in \Lambda_{mw}~~,~~\hat{h}^1\in \Lambda_{cochar}~,
\ee
and hence is the next to minimal 't Hooft defect: $L_{2,0}$.   As in the case of the $SU(2)$ SYM theory, this is of the form 
\be
\langle L_{2,0}\rangle=\left(e^{2\fb}+e^{-2\fb}\right)F(\fa,m_f)+Z_{mono}(2,0)~,
\ee
where 
\be
F(\fa)=\left(\frac{\prod_{s_1,s_2=\pm}\sinh(2\fa+s_1m+s_2\epsilon_+)}{\prod_\pm \sinh^2(2\fa)\sinh(2\fa\pm 2\epsilon_+)}\right)^\half~.
\ee
The monopole bubbling contribution ($Z_{mono})$ can be computed as the Witten index of the (mass deformed) $\CN=(4,4)$ SQM described by the quiver:
\begin{center}
\begin{tikzpicture}[node distance=2cm,cnode/.style={circle,draw,thick,minimum size=10mm},snode/.style={rectangle,draw,thick,minimum size=10mm}]
\node[cnode] (1) {1};
\node[snode] (4) [below of=1]{2};
\draw[thick] (4) -- (1);
\end{tikzpicture}
\end{center}
The Witten index of this quiver SQM reduces to the contour integral 
\be
Z^{(Loc)}_{mono}(2,0)=\oint_{JK(\xi)}\frac{d\varphi}{2\pi i}Z_{vec}(\varphi,\fa,\epsilon_+)Z_{hyper}(\varphi,\fa,\epsilon_+) 
~,
\ee
where  the contributions of the different $\CN=(0,2)$  multiplets for this SQM are given by 
\begin{align}\begin{split}
&Z_{mono}^{(Loc)}(2,0)=\oint_{JK(\xi)}\frac{d\varphi}{2\pi i}\frac{\sinh(2\epsilon_+)}{2\prod_\pm\sinh(m\pm \epsilon_+)}\prod_\pm\frac{ \sinh(\pm (\varphi+\fa)+ m)\sinh(\pm (\varphi-\fa)+ m)}{ \sinh(\pm (\varphi+\fa)+\epsilon_+)\sinh(\pm (\varphi-\fa)+\epsilon_+)}~.
\end{split}\end{align}
 Using this we can compute 
\be
Z_{mono}^{(Loc)}(2,0)=\frac{\prod_{s=\pm}\sinh(2\fa+ sm+\epsilon_+)}{\sinh(2\fa)\sinh(2\fa+2\epsilon_+)}+\frac{\prod_{s=\pm}\sinh(2\fa+sm-\epsilon_+)}{\sinh(2\fa)\sinh(2\fa-2\epsilon_+)}\quad,\quad \pm\xi>0~.
\ee
As shown in \cite{Ito:2011ea}, the AGT computation produces
\be
Z_{mono}(2,0)=\frac{\prod_{s=\pm}\sinh(2\fa+ sm+\epsilon_+)}{\sinh(2\fa)\sinh(2\fa+2\epsilon_+)}+\frac{\prod_{s=\pm}\sinh(2\fa+sm-\epsilon_+)}{\sinh(2\fa)\sinh(2\fa-2\epsilon_+)}\quad,\quad \pm\xi>0~,
\ee
and therefore that 
\be
Z_{mono}^{(extra)}(2,0)=0~. 
\ee
As shown in Appendix \ref{app:A}, there is a complete cancellation between Coulomb branch states such that $Z_{mono}^{(extra)}(2,0)$ is reproduced by the Witten index of the asymptotic Coulomb branch states:
\be
Z_{mono}^{(extra)}(2,0)=I_{asymp}(2,0)=0~. 
\ee
This is again consistent with the fact that $Z_{mono}^{(Loc)}(2,0)$ is independent of the choice of $\xi$ and invariant under the action of the Weyl symmetry of the flavor symmetry group.

\rmk.~ In general, the ``extra'' terms can be dependent on $\fa$ as well as $m_f$ and $\epsilon_+$. The reason is that because for non-abelian SQM gauge groups, there are generically non-trivial contributions from mixed branches which we expect can give rise to $\fa$ dependence.

\subsection{Comment on 4D $\mathcal{N}$=2 Quiver Gauge Theories}
\label{sec:quiverGT}

It is interesting to ask how this analysis applies to 4D $\CN=2$ quiver gauge theories with gauge group $SU(2)$ at each node. \footnote{Here we consider only $SU(2)$ gauge groups due to additional subtleties with higher rank simple gauge groups with ${\rm N}_f\geq 4$ fundamental hypermultiplets. See upcoming work for additional details.} We believe there is no fundamental obstruction 
to applying this analysis to such theories beyond increasing computational complexity. There is also the additional complication that the brane configuration presented here does not simply generalize to the case of quiver gauge theories and hence can not be used to derive the quivers for the bubbling SQM. 

However, the bubbling SQM can be deduced from the following arguments. Consider the bubbled defect in a quiver gauge theory with gauge group $G=\prod_i SU(2)_i$ specified by the data $(P,\vv)\in \Lambda_{mw}\times \Lambda_{mw}$ which decomposes as a sum over gauge group factors 
\be
P=\bigoplus_i P_i\quad, \quad \vv=\bigoplus_i \vv_i~. 
\ee
Further, let us define the quiver $\Gamma_i$ which specifies the bubbling SQM associated to the pair $(P_i,\vv_i)$ with appropriate matter interactions. The full bubbling SQM is then derived by taking into account the 4D bifundamental hypermultiplets which lead to extra fermi and/or chiral multiplets connecting nodes between different $\Gamma_i$. The precise couplings can be obtained from demanding $U(1)_K$-invariance of the full quiver. \footnote{See upcoming work for additional details.}

Unfortunately, testing our hypothesis in this setting would be quite difficult as the necessary AGT computations also become increasingly difficult with increasing gauge group rank. The computation of the bubbling contribution to the expectation value of 't Hooft defects in quiver gauge theories is of interest for many reasons. One reason is the potential utilitity in exploring the 
deconstruction of the 6D $\CN=(0,2)$ theory \cite{ArkaniHamed:2001ie}. 

Consider a $\CN=2$ superconformal ring quiver gauge theory with $G=\prod_{i=1}^N SU(2)_i$. 
The deconstruction hypothesis conjectures that in the limit $N\to \infty$ and $g_{YM}\to \infty$, the UV completion of this 4D theory is that of the 6D $A_1$ $\CN=(0,2)$ theory. In this limit the 't Hooft defects become surface defects that interact with tensionless strings \cite{Seiberg:1996vs}. Thus, the correct computation of expectation 't Hooft defects in quiver gauge theories can be used as a probe for understanding the 6D $\CN=(0,2)$ theory. 

Therefore, let us demonstrate that our analysis applies to the computation of the monopole bubbling contribution of an 't Hooft defect in  the simplest example of quiver gauge theory of higher rank. 
Consider the case of a superconformal $\CN=2$ quiver gauge theory with $G=SU(2)_1\times SU(2)_2$ with fundamental matter:
\begin{center}
\begin{tikzpicture}[node distance=2cm,
cnode/.style={circle,draw,thick,minimum size=9mm},snode/.style={rectangle,draw,thick,minimum size=9mm}]
\node[snode] (1) {2};
\node[cnode] (2) [right of=1] {SU(2)};
\node[cnode] (3) [right of=2] {SU(2)};
\node[snode] (4) [right of=3] {2};
\draw[-] (1) -- (2);
\draw[-] (2) -- (3);
\draw[-] (3) -- (4);
\end{tikzpicture}
\end{center}
Now consider the bubbling sector where 
\be
P=\bigoplus_{i=1}^2 P_i\quad, \quad \vv=\bigoplus_{i=1}^2 \vv_i\quad,\quad (P_i,\vv_i)=\big({\rm diag}(1,-1),{\rm diag}(0,0)\big)~.
\ee
In this case, the $\CN=(0,2)$ bubbling SQM is of the form
\begin{center}\begin{tikzpicture}[node distance=2cm,cnode/.style={circle,draw,thick,minimum size=10mm},snode/.style={rectangle,draw,thick,minimum size=10mm}]
\node[cnode] (1) {1};
\node[cnode] (2) [right of=1,xshift=1cm]{1};
\node[snode] (3) [below of=1]{2};
\node[snode] (4) [below of=2]{2};
\node[snode] (5) [above of=1]{2};
\node[snode] (6) [above of=2]{2};
\draw[double,double distance=6pt,thick] (1) -- (2);
\draw[double,double distance=6pt,thick] (1) -- (3);
\draw[double,double distance=6pt,thick] (2) -- (4);
\draw[dashed,thick] (1) -- (5);
\draw[dashed,thick] (2) -- (6);
\draw[dashed,thick] (3) -- (2);
\draw[dashed,thick] (4) -- (1);
\path[dashed,every loop/.style={looseness=5}] (1)
         edge  [in=210,out=150,loop] node {} (); 
\path[dashed,every loop/.style={looseness=5}] (2)
         edge  [in=30,out=330,loop] node {} (); 
\end{tikzpicture}\end{center}
The localization contribution to $Z_{mono}\big((1,0)\oplus(1,0)\big)$  is then given by the contour integral
\begin{align}
\begin{split}
Z_{mono}^{(Loc)}((1,0)\oplus(1,0))=&\oint_{JK(\xi_1,\xi_2)}\frac{d\varphi_1d\varphi_2}{(2\pi i)^2}\frac{\sinh^2(2\epsilon_+)\prod_{f=1}^2\sinh(\varphi_1-m_f)\sinh(\varphi_2-m_{f+2})}{\prod_{i=1}^2\prod_{\pm}\sinh(\pm (\varphi_i-\fa_i)+\epsilon_+)\sinh(\pm (\varphi_i+\fa_i)+\epsilon_+)}\\&
\times \frac{
\prod_{\pm}\sinh(-\varphi_1\pm \fa_2+m+\epsilon_+)\sinh(\varphi_2\pm \fa_1+m+\epsilon_+)
}{4\sinh(\varphi_2-\varphi_1)\sinh(\varphi_1-\varphi_2+2\epsilon_+)}~.
\end{split}\end{align}
Let us choose $\xi_1,\xi_2>0$. In this case there are 8 poles contributing  to this path integral:
\begin{align}\begin{split}
{\rm I:}&~\varphi_1=\fa_1-\epsilon_+\quad~~, \quad \varphi_2=\fa_2-\epsilon_+~,\\
{\rm II:}&~\varphi_1=\fa_1-\epsilon_+\quad~~ ,\quad \varphi_2=-\fa_2-\epsilon_+~,\\
{\rm III:}&~\varphi_1=-\fa_1-\epsilon_+\quad, \quad \varphi_2=\fa_2-\epsilon_+~,\\
{\rm IV:}&~\varphi_1=-\fa_1-\epsilon_+\quad, \quad \varphi_2=-\fa_2-\epsilon_+~,\\
{\rm V:}&~\varphi_1=\fa_1-\epsilon_+\quad~~\,, \quad \varphi_2=\fa_1-\epsilon_+~,\\
{\rm VI:}&~\varphi_1=-\fa_1-\epsilon_+\quad, \quad \varphi_2=-\fa_1-\epsilon_+~,\\
{\rm VII:}&~\varphi_1=\fa_2-3\epsilon_+\quad~, \quad \varphi_2=\fa_2-\epsilon~,\\
{\rm VIII:}&~\varphi_1=-\fa_2-3\epsilon_+~\, ,\quad \varphi_2=-\fa_2-\epsilon~.
\end{split}\end{align}
See Appendix \ref{app:E} for the full expression of $Z_{mono}^{(Loc)}((1,0)\oplus(1,0))$ computed with $\xi_1,\xi_2>0$. 

One can check that the localization result for $Z_{mono}((1,0)\oplus(1,0))$ from residues associated to these poles is not invariant under the Weyl symmetry of the $SU(2)\times SU(2)$ flavor symmetry group. 
Therefore, this cannot be the full, correct monopole bubbling contribution $Z_{mono}\big((1,0)\oplus(1,0)\big)$ for  this 't Hooft defect. Thus, we expect that the true $Z_{mono}\big((1,0)\oplus(1,0)\big)$ has an extra contribution coming from Coulomb and mixed branch states that are missed in the standard localization computation.

\section{Decoupling Flavors}
\label{sec:CS}

Another way we can check our hypothesis is to see whether it is compatible with decoupling matter from $\CN=2$ theories. Consider decoupling fundamental hypermultiplets from the $SU(2)$ N$_f$=4 theory. 
The expressions for $\langle L_{p,0}\rangle_{{\rm N}_f<4}$ can be obtained by taking the limit as $m_i\to \infty$ while holding $\fa,\fb$ fixed, provided we allow for a multiplicative renormalization of $\langle L_{p,0}\rangle_{{\rm N}_f=4}$. 
This kind of limit was described in \cite{Gaiotto:2010be} Section 9. 
 Note that this limit is not the decoupling limit in \cite{Seiberg:1994aj}. \footnote{The limit from  \cite{Seiberg:1994aj} takes 
\be
\Lambda_{{\rm N}_f=3}=64 q^{1/2} m_4~, 
\ee
fixed with $u,m\to \infty$ with
\be
u^{{\rm N}_f=3}=u+\frac{1}{3} m_4^2~,
\ee
held constant where $\Lambda_{{\rm N}_f=3}$ is the UV cutoff of the ${\rm N}_f=3$ theory. Further decoupling to the ${\rm N}_f\leq 2$ proceeds analogously. 
The limit from \cite{Seiberg:1994aj} is different from the limit we take here because in their limit, the expressions for $\fa,\fb$ from \eqref{semia}, \eqref{semib} diverge (and non-perturbative corrections are small). }

This allows us to compute the expectation value of the line defects in 4D $SU(2)$ gauge theories with ${\rm N}_f\leq 3$ fundamental hypermultiplets  by taking the decoupling limit of the ${\rm N}_f=4$ theory. 
However, in order to use our prescription to compute $I_{asymp}$, we must take into account the effect of the decoupling limit on the bubbling SQM. 
When we decouple the fundamental hypermultiplets we are integrating out fundamental Fermi-multiplets coupled to the gauge field in bubbling SQM. This will generically introduce a Chern-Simons term (or in this case a Wilson line) determined by the way we decouple the masses \cite{Redlich:1983kn,Redlich:1983dv,Niemi:1983rq,AlvarezGaume:1984nf}
\be
q=\half \sum_{f={\rm N}_f+1}^4 {\rm sgn}(Re[m_f])~.
\ee
This means that the ${\rm N}_f=1,3$ theories must necessarily have a Chern-Simons term of level $\frac{2n+1}{2}$, $n\in \IZ$ to be well defined. \footnote{An analogous system was studied in 5D in \cite{Intriligator:1997pq}.} The necessity of these half integer Chern-Simons terms is reflected in bubbling SQMs as a gauge anomaly for $q\in \IZ$. In general, the allowed values of the Chern-Simons levels is consistent with the condition that the bubbling SQM be anomaly free $q\in \IZ+\frac{{\rm N}_f}{2}$ as noted in \cite{Hori:2014tda}.

\subsection{Examples}

Now we can determine the value of $Z_{mono}(P,\vv)$ in the general ${\rm N}_f$ theory by taking certain limits of $Z_{mono}(P,\vv)$ in the ${\rm N}_f=4$ theory. This also gives an additional check of our hypothesis  as we will now demonstrate in the case of the ${\rm N}_f=2,3$ theories. Similar results also hold for the ${\rm N}_f=0,1$ theories. 

\subsubsection{$\langle L_{1,0}\rangle$ in the ${\rm N}_f=3$ Theory}

Recall that for the expectation value of the $\langle L_{1,0}\rangle$ 't Hooft defect in the ${\rm N}_f=4$ theory, $Z_{mono}(1,0)$ is of the form
\begin{align}\begin{split}
Z_{mono}^{{\rm N}_f=4}(1,0)=-4\frac{\prod_f \sinh(\fa-{\rm m}_f\mp\epsilon_+)}{\sinh(2\fa)\sinh(2\fa\mp2\epsilon_+)}-4\frac{\prod_f \sinh(\fa+{\rm m}_f\pm\epsilon_+)}{\sinh(2\fa)\sinh(2\fa\pm2\epsilon_+)}\\
+2\cosh\left(\sum_f m_f\pm2\epsilon_+\right)~,
\end{split}\end{align}
for $\pm\xi>0$. 
Note that this actually invariant under the choice of $\xi\in \IR_+$. 

By taking the limit $Re[m_4]\to \pm \infty$ we can decouple the 4$^{th}$ fundamental hypermultiplet and find a result for the ${\rm N}_f=3$ theory. This produces the result 
\begin{align}\begin{split}
&Z_{mono}^{{\rm N}_f=3}(1,0;q=\frac{s}{2})=\lim_{{\rm Re}[m_4]\to s \infty }e^{s  m_4}Z_{mono}({\rm N}_f=4)\\
&=2se^{-s\fa\pm s\epsilon_+}\frac{\prod_f \sinh(\fa-{\rm m}_f\mp\epsilon_+)}{\sinh(2\fa)\sinh(2\fa\mp2\epsilon_+)}-2se^{s\fa\pm s\epsilon_+} \frac{\prod_f \sinh(\fa+{\rm m}_f\pm\epsilon_+)}{\sinh(2\fa)\sinh(2\fa\pm2\epsilon_+)}~,\\&\quad
+e^{s\sum_f m_f\pm2s \epsilon_+}
\quad,\quad {\rm for }~\pm\xi>0~.\\
\end{split}\end{align}
Note that $Z_{mono}^{{\rm N}_f=3}(1,0;q=\frac{1}{2})\neq Z_{mono}^{{\rm N}_f=3}(1,0;q=-\frac{1}{2})$. 

In this case we see that ${\rm Re}[m_4]\to \pm \infty$ corresponds to $q=\frac{s}{2}$. This modifies the localization integrand by including a factor of $e^{2q\varphi}$:
\be
Z_{mono}^{(Loc)}(1,0;q=\frac{s}{2})=\oint_{JK(\xi)}\frac{d\varphi}{2\pi i} \frac{2e^{s \varphi}\sinh(2\epsilon_+)\prod_{f=1}^3\sinh(\varphi- m_f)}{\sinh(\varphi\pm \fa+\epsilon_+)\sinh(-\varphi\pm \fa+\epsilon_+)}~.
\ee
As before, this can be evaluated as
\be
Z^{(JK)}_{mono}(1,0;q=\frac{s}{2})=2se^{-s\fa\pm s\epsilon_+}\frac{\prod_f \sinh(\fa-{\rm m}_f\mp\epsilon_+)}{\sinh(2\fa)\sinh(2\fa\mp2\epsilon_+)}-2se^{s\fa\pm s\epsilon_+} \frac{\prod_f \sinh(\fa+{\rm m}_f\pm\epsilon_+)}{\sinh(2\fa)\sinh(2\fa\pm2\epsilon_+)}~,
\ee 
where $s={\rm sign}(q)$. \footnote{Again this required fixing the overall normalization of the path integral.} 
This means that we have 
\be
Z_{mono}^{(extra)}(1,0;q=\frac{s}{2})=\begin{cases}
e^{s( \sum_f m_f+\epsilon_+)}& \xi>0\\
e^{s(\sum_f m_f-\epsilon_+)}& \xi<0\end{cases}
\ee
This is exactly given by $I_{asymp}(q=\frac{s}{2})$  as shown in  Appendix \ref{app:A}. 

\subsubsection{$\langle L_{1,0}\rangle$ in the ${\rm N}_f=2$ Theory}

Now we can go to the ${\rm N}_f=2$ theory by decoupling the $3^{rd}$ fundamental hypermultiplet. This can be done in two ways (Re$[m_3]\to\pm\infty$) which can induce a Wilson line of charge $q=-1,0,1$. The result for $q=0$, which can be achieved by taking Re$[m_3]\to -{\rm sgn}(q_{{\rm N}_f=4})\times\infty$, is given in the previous section. In the case of $q=\pm1$, we have that $Z_{mono}(1,0)$ is given by
\begin{align}\begin{split}
&Z_{mono}({\rm N}_f=2)=\lim_{{\rm Re}[m_4],{\rm Re}[m_3]\to q\infty }e^{-s m_4-s  m_3}Z_{mono}({\rm N}_f=4)\\
&=-e^{-2q\fa\pm 2q\epsilon_+}\frac{\prod_f \sinh(\fa-{\rm m}_f\mp\epsilon_+)}{\sinh(2\fa)\sinh(2\fa\mp2\epsilon_+)}-e^{2q\fa\pm 2q\epsilon_+} \frac{\prod_f \sinh(\fa+{\rm m}_f\pm\epsilon_+)}{\sinh(2\fa)\sinh(2\fa\pm2\epsilon_+)}+e^{s\sum_f m_f\pm2s \epsilon_+}~,
\end{split}\end{align}
where $s={\rm sign}[q]$. 
Again, by introducing the Wilson line in the SQM, this changes the localization computation to give 

\be
Z_{mono}^{(Loc)}(1,0)=\half\oint_{JK(\xi)}\frac{d\varphi}{2\pi i} \frac{e^{2q \varphi}\sinh(2\epsilon_+)\prod_{f=1}^2\sinh(\varphi- m_f)}{\prod_\pm\sinh(\varphi\pm \fa+\epsilon_+)\sinh(-\varphi\pm \fa+\epsilon_+)}~.
\ee
As before, this can be evaluated as \footnote{Again this required fixing the overall normalization of the path integral.} 
\be
Z^{(JK)}_{mono}(1,0)=-e^{-2q\fa\pm 2q\epsilon_+}\frac{\prod_f \sinh(\fa-{\rm m}_f\mp\epsilon_+)}{\sinh(2\fa)\sinh(2\fa\mp2\epsilon_+)}-e^{2q\fa\pm 2q\epsilon_+} \frac{\prod_f \sinh(\fa+{\rm m}_f\pm\epsilon_+)}{\sinh(2\fa)\sinh(2\fa\pm2\epsilon_+)}~.
\ee
\be
Z_{mono}^{(extra)}=\begin{cases}
e^{s\sum_f m_f+2s \epsilon_+}& \xi>0\\
e^{s\sum_f m_f-2s \epsilon_+}& \xi<0\end{cases}\quad, \quad s={\rm sign}(q)~.
\ee
Note that as shown in the previous section, $I_{asymp}(q=0)=0$. 
This is exactly given by $I_{asymp}(q)$ as appropriate as shown in  Appendix \ref{app:A}. 

\section*{Acknowledgements}

The authors would like to thank Clay Cordova, Tudor Dimofte, Heeyeon Kim, Andrew Neitzke,  Wolfger Peelaers, Shu-Heng Shao, and  Piljin Yi  for englightening discussions and comments on the manuscript.  TDB and AD would like to thank the organizers of ``Simons Summer Workshop'' and GM would like to thank the Aspen Center for physics for hospitality while completing this work. 
The authors are supported by DOE grant DOE-SC0010008.

\appendix

\section{$U(1)$ $\CN=(0,4)$ SQM Analysis}
\label{app:A}

In this appendix we will present a general analysis of the $\CN=(0,4)$ SQM with a $U(1)$ gauge group coming from monopole bubbling in 4D $SU(2)$ $\CN=2$, asymptotically free, supersymmetric gauge theories. See \cite{Tong:2014yna} for a review of $\CN=(0,4)$ SUSY. 

The $\CN=(0,4)$ SQM we are considering consists of a $U(1)$ vector multiplet, two fundamental hypermultiplets and up to four	 $\CN=(0,2)$ Fermi-multiplets $\Psi^{(i)}$, that have been embedded into the $\CN=(0,4)$ theory by either embedding $\Psi_{\CN=(0,2)}^{(i)}\hookrightarrow \widetilde\Psi_{\CN=(0,4)}^{(i)}$ as short $\CN=(0,4)$ Fermi-multiplets or by combining $(\Psi_{\CN=(0,2)}^{(i)}\oplus \bar\Psi_{\CN=(0,2)}^{(i+2)})=\hat\Psi_{\CN=(0,4)}$ to make long $\CN=(0,4)$ Fermi-multiplets. In the case of a 4D  theory with ${\rm N}_f$ fundamental hypermultiplets we will have ${\rm N}_f$ short $\CN=(0,4)$ fundamental Fermi-multiplets. 
 In the case of the 4D $SU(2)$ $\CN=2^\ast$ theory, the multiplets form $\CN=(4,4)$ SUSY multiplets which are then mass deformed to the $\CN=(0,4)$ theory. This means that in studying the $\CN=2^\ast$ theory, we must include a massive $\CN=(0,4)$ adjoint (twisted) hypermultiplet $\Gamma$ and two long fundamental Fermi-multiplets in the SQM. 
 
 In general, these SQMs have a Lagrangian given by
\be
L=L_{univ}+L_{theory}~.
\ee
Here $L_{theory}$ is the part of the Lagrangian that is dependent on the details of the 4D $\CN=2$ theory, and $L_{univ}$ is the universal part of the Lagrangian that does not change with the matter content of the 4D Theory. Here we have 
\begin{align}\begin{split}
L_{univ}=L_{vec}+L_{hyp}+L_{FI}~,
\end{split}\end{align}
where 
\begin{align}\begin{split}
&L_{vec}=\frac{1}{e^2}\left[ \half(\partial_t \sigma)^2+i \bar\lambda_{A}\partial_t \lambda^{A}+\half D^2+|F|^2\right]~,\\
&L_{hyp}=\frac{1}{e^2}\left(|D_t\phi^{A,f}|^2-|\sigma \phi^{A,f}|^2-  \bar\phi_{A,f}(\sigma^r)_{~B}^{A} M_r \phi^{B,f}\right)\\&
+\frac{i}{2e^2}\left(\bar\psi_1^f (D_t+i \sigma)\psi_{1,f}+\psi_{1,f}(\tilde{D}_t-i\sigma)\bar\psi_1^f+\bar\psi^f_2(\tilde{D}_t-i \sigma)\psi_{2,f}+\psi_{2,f}(D_t+i \sigma)\bar\psi_2^f\right)\\
&+\frac{i}{\sqrt{2}e^2} (\barphi_{A,f}\lambda^{A}\psi_1^f-\bar\psi_{1,f}\bar\lambda_{A}\phi^{A,f}+\bar\phi_{A,f}\bar\lambda^{A}\bar\psi_2^f-\psi_{2,f}\lambda_{A}\phi^{A,f})~,\\
&L_{FI}=-\xi D~,\\
&M_r=(f~,~g~,~ D)\quad, \quad r=1,2,3\quad,\qquad f=\shalf(F+\bar{F})\quad, \quad g=\frac{1}{\sqrt{2}i}(F-\bar{F})~. 
\end{split}\end{align}
Here $A=1,2$ is an $SU(2)_R$ index, $f=1,2$ is an index for the global $SU(2)$ flavor symmetry, 
 $e$ is the gauge coupling, and $\sigma^{r=1,2,3}$ are the Pauli matrices acting on the $SU(2)_R$ indices. Here we use the convention
 \be
 (\phi^A)^\ast=\bar\phi_A\quad, \quad \phi_A=\epsilon_{AB}\phi^B\quad, \quad \epsilon_{21}=\epsilon^{12}=1~. 
 \ee

We additionally have that $L_{theory}$ can have contributions from terms 
\be
L_{theory}=L_{Fermi}+L_{adj\,hyp}~, 
\ee
which are of the form
\begin{align}\begin{split}
&L_{Fermi}=\frac{1}{e^2}\sum_j\left[\ihalf (\bar\eta_j(D_t+i \sigma)\eta_j+\eta_j(D_t-i \sigma)\bar\eta_j)+|G_j|^2+{\rm m}_j[\bar\eta_j,\eta_j]\right]~,
\\
&L_{adj\,hyper}=\frac{1}{e^2}\left[|\partial_t \rho^A|^2+\ihalf \sum_{I=1}^2(\bar\chi_I \partial_t \chi_I+\chi_I\partial_t \bar\chi_I)
\right]~,
\end{split}\end{align}
where in the case of a 4D theory with ${\rm N}_f$ fundamental hypermultiplets, we only include ${\rm N}_f$ fundamental Fermi multiplets and in the case of the $\CN=2^\ast$ theory we include both the adjoint hypermultiplet and 4 (short) fundamental Fermi multiplets. \footnote{Note that four short fundamental Fermi-multiplets is equivalent to two long fundamental Fermi-multiplets. }

In the following analysis we will decompose the  $\CN=(0,4)$ hypermultiplet that transforms in a quaternionic representation $\CR$ into two $\CN=(0,2)$ chiral multiplets transforming in conjugate representations  ${\rm R}\oplus \bar{\rm R}$:
\begin{align}\begin{split}
{\rm Fundamental~ Hypermultiplet}~\Phi&=(\phi^A,\psi_I)_\CR
=\Phi_1\oplus\bar{\tilde{\Phi}}_2=
(\phi,\psi)_{\rm R}\oplus (\tilde\phi,\tilde\psi)_{\bar{\rm R}}~,
\end{split}\end{align}
where $I=1,2$. 

In this notation,  the field content of this theory is given by 
\begin{center}
\begin{tabular}{|c|c|c|c|c|c|c|}
\hline
Lagrangian Term&Multiplet&Fields&$Q_{\rm Gauge}$&$Q_\fa$&$Q_\epsilon$&$F_j$\\\hline\hline
Universal
&$\CN=(0,4)$ Vector-&$\sigma$&0&0&0&0\\
&&$\lambda^1$&0&0&0&0\\
&&$\lambda^2$&0&0&4&0\\\hline Universal
&$\CN=(0,4)$ Fund. Hyper-& 
$\phi_1$&1&-1&1&0\\
&&$\tilde\phi_1$&-1&1&1&0\\
&&$\phi_2$&1&1&1&0\\
&&$\tildephi_2$&-1&-1&1&0\\
&&$\psi_1$&1&-1&1&0\\
&&$\tilde\psi_1$&-1&1&1&0\\
&&$\psi_2$&1&1&1&0\\
&&$\tilde\psi_2$&-1&-1&1&0\\\hline\hline
4D N$_f$ Fund. Hypermultiplets 
&$\CN=(0,4)$ Fund. Fermi-&$\eta_j$&1&0&0&-2\\\hline\hline
$\CN=2^\ast$ Theory
&$\CN=(0,4)$ Adjoint&$\rho$&0&0&1&1\\
&Twisted Chiral&$\tilde\rho$&0&0&-1&1\\
&&$\chi$&0&0&0&1\\
&&$\tilde\chi$&0&0&0&-1\\
\hline $\CN=2^\ast$ Theory
&$\CN=(0,4)$ Fund. Fermi&$\eta_1$&1&-1&1&0\\
&&$\tilde\eta_1$&-1&1&1&0\\
&&$\eta_2$&1&1&1&0\\
&&$\tilde\eta_2$&-1&-1&1&0\\\hline
\end{tabular}
\end{center}
This non-$SU(2)_R$ invariant notation is explicitly related to the $SU(2)_R$-invariant notation by 
\begin{align}\begin{split}
\phi^{A,f}=\left(\begin{array}{c}\phi^f\\\bar{\tilde{\phi}}^f\end{array}\right)
\quad,\quad \psi^f_I=\left(\begin{array}{c}\psi^f\\\bar{\tilde{\psi}}^f
\end{array}\right)\quad,\quad \rho^A=\left(\begin{array}{c}\rho\\\bar{\tilde{\rho}}\end{array}\right)\quad, \quad 
\chi_I=\left(\begin{array}{c}\chi\\\bar{\tilde{\chi}}\end{array}\right)~. 
\end{split}\end{align}
Here we will use the convention $\epsilon^{12}=\epsilon_{21}=1$ and will use a notation that is not flavor $SU(2)$-invariant by mapping 
\be
\phi^f,\tilde\phi^f,\psi^f,\tilde\psi^f\mapsto \phi_f,\tilde\phi_f,\psi_f,\tilde\psi_f~,
\ee
so that the $\{\phi_f,\tilde\phi_f,\psi_f,\tilde\psi_f\}$ should be understood as normal complex valued scalar and fermion fields. It should be understood that all flavor and $SU(2)_R$ indices are contracted properly in the upcoming analysis. 
Additionally, note that in these theories, there are only $J$-type  Fermi-multiplet interactions.  See \cite{Hori:2014tda,Tong:2014yna} for a more complete discussion about this Lagrangian and the corresponding field content.

These fields transform under  supersymmetry as
\begin{align}\begin{split}\label{SQMSUSY1}
&\delta v_t=-\shalf\epsilon^A \bar\lambda_A-\shalf \bar\epsilon_A \lambda^A\quad,\quad 
\delta \sigma=\shalf\epsilon^A \bar\lambda_A+\shalf \bar\epsilon_A \lambda^A~,
\\&
\delta \lambda^A=\frac{i}{\sqrt{2}}\epsilon^AD_t \sigma+\frac{1}{\sqrt{2}e^2}\epsilon^BM_r(\sigma^r)^A_{~B}~,
\end{split}\end{align}
and
\begin{align}\begin{split}\label{SQMSUSY2}
&\delta \phi^{A,f}=-i\left(\epsilon^A \psi^f+\bar\epsilon^A\bar\tildepsi^f\right)\quad, \quad\delta \psi^f=\bar\epsilon_A\left(D_t-\frac{i}{e^2} \sigma\right)\phi^{A,f}\quad, \quad \delta \bar\tildepsi^f=-\epsilon_A(D_t-\frac{i}{e^2} \sigma)\phi^{A,f}~,\\
&\delta\eta_j=i \epsilon_1 G_j+\bar\epsilon_1 \bar{F}_j\qquad\,~, \quad
\delta G_j=-\half\bar\epsilon_1 (D_t+i \sigma)\eta_j-\half\epsilon_1 (\tilde{D}_t-i \sigma)\bar\eta_j~,\\
&\delta \rho^A=-i\left(\epsilon^A \chi_1+\bar\epsilon^A\chi_2\right)\quad, \quad\delta \chi_1= \bar\epsilon_A\partial_t\rho^A\quad, \quad \delta \chi_2=-\epsilon_A\partial_t\rho^A~,\\
\end{split}\end{align}
The supercharges generating the supersymmetry transformations for the vector multiplet fields $(\sigma,\lambda^A)$ and hypermultiplet fields $(\phi,\psi),(\tilde\phi,\tilde\psi)$ are given by
\begin{align}\begin{split}\label{eq:supercharge}
Q_A&= \psi^f\left(\pi_{A,f}+\frac{i \sigma}{e^2} \bar\phi_{A,f}\right)+\tilde\psi_f\left(\bar\pi_A^f-\frac{i \sigma}{e^2}\phi_A^f\right)-\shalf \bar\lambda_{A}(-i p_\sigma)+\frac{1}{\sqrt{2}e^2}(\sigma^r)_A^{~B}\bar\lambda_BM_r~.
\end{split}\end{align}
Additionally, when considering the 4D $\CN=2^\ast$ theory, we must add another hypermultiplet $(\rho^A, \chi_I)$ which contributes 
\be
Q_A=...+ \chi_1\left(\pi_{\rho,A}+\frac{i m}{e^2} \bar\rho_A\right)+\bar\chi_2\left(\bar\pi_{\rho,A}-\frac{i m}{e^2} \rho_A\right)
\ee
where we have included possible mass terms.

Now consider adding the masses: \footnote{These masses come from turning on a flat gauge connection for a flavor symmetry $\mapsto a$, and for a $U(1)_R$ symmetry $\mapsto \epsilon$. The $U(1)_R$ symmetry comes from the diagonal combination of $SU(2)_R\times SU(2)_r$ where $SU(2)_R$ is an $R$-symmetry and $SU(2)_r$ is an outer autormophism symmetry. Turning on the $\epsilon$ mass corresponds to gauging the combination $Q_\epsilon=Q_R-Q_r$ where $Q_R,Q_r$ are the generator of the Cartan for $SU(2)_R,SU(2)_r$ respectively. }
\begin{align}\begin{split}
m_{\Phi_1}&=-{\rm Im}[\fa/\beta]+{\rm Im}[\epsilon_+/\beta]=-2a+2\epsilon\quad,\quad m_{\tilde\Phi_1}=-{\rm Im}[\fa/\beta]-{\rm Im}[\epsilon_+/\beta]=-2a-2\epsilon~,\\
m_{\Phi_2}&={\rm Im}[\fa/\beta]+{\rm Im}[\epsilon_+/\beta]=2a+2\epsilon\qquad~~,\quad m_{\tilde\Phi_2}={\rm Im}[\fa/\beta]-{\rm Im}[\epsilon_+/\beta]=2a-2\epsilon~,\\
m_{\lambda^2}&=4 {\rm Im}[\epsilon_+/\beta]=4\epsilon~,
\end{split}\end{align}
Additionally, in the case of 4D theories with matter, we will add the masses
\begin{align}\begin{split}
&{\rm m}_{\rho^1}={\rm Re}[m/\beta]+{\rm Re}[\epsilon_+/\beta]=2{\rm m}+2\epsilon\qquad~~,\quad
{\rm m}_{\rho^2}={\rm Re}[m/\beta]-{\rm Re}[\epsilon_+/\beta]=2{\rm m}-2\epsilon~,\\
&{\rm m}_{\hat\Psi^{(i)}}=\pm {\rm Re}[\fa/\beta]+{\rm Re}[m/\beta]=\pm2 a+2{\rm m}\quad, \quad {\rm m}_{\widetilde\Psi^{(f)}}=2Re[{ m}_f/\beta]=2 {\rm m}_f~,
\end{split}\end{align}
as appropriate. These masses break SUSY down to $\CN=2$ where $Q_1,\bar{Q}^1$ are the conserved supercharges. Since we know that the Witten index depends holomorphically on the masses \cite{Festuccia:2011ws}, we will take the mass parameters to be real and positive with $a>\epsilon$ for simplicity and analytically continue in the final answer. \footnote{The analysis changes slightly for the case of mass parameters and when $\epsilon>a$, but the answer will be independent of these choices. }



In terms of component fields, the universal Lagrangian is given by 
\begin{align}\begin{split}
&L_{univ}=\frac{1}{e^2}\left[ \half(\partial_t \sigma)^2+\ihalf\left( \bar\lambda_{1}\partial_t \lambda^{1}+\lambda^{1}\partial_t \bar\lambda_{1}\right)+ i \bar\lambda_2(\partial_t-4 i \epsilon)\lambda^2-\half D^2-|F|^2\right]\\
&+
\frac{1}{e^2}\left[ |D_t \phi_1|^2+|D_t \phi_2|^2+|\tilde{D}_t\tilde\phi_1|^2+|\tilde{D}_t \tildephi_2|^2\right]\\
&-\frac{1}{e^2}\left[(\sigma-a+\epsilon)^2|\phi_1|^2+(\sigma-a-\epsilon)^2|\tilde\phi_1|^2+(\sigma+a+\epsilon)^2|\phi_2|^2+(\sigma+a-\epsilon)^2|\tilde\phi_2|^2\right]\\
&+\frac{i}{2e^2}\Big(\bar\psi_1(D_t+i (\sigma-a+\epsilon))\psi_1+\psi_1(\tilde{D}_t-i  (\sigma-a+\epsilon))\bar\psi_1\\
&\quad+\bar\psi_2(D_t+i  (\sigma+a+\epsilon))\psi_2+\psi_2(\tilde{D}_t-i  (\sigma+a+\epsilon))\bar\psi_2\\
&\quad+\bar\tildepsi_1(\tilde{D}_t-i  (\sigma-a-\epsilon))\tildepsi_1+\tildepsi_1(D_t+i (\sigma-a-\epsilon))\bar\tildepsi_1\\&\quad
+\bar\tildepsi_2(\tilde{D}_t-i (\sigma+a-\epsilon))\tildepsi_2+\tildepsi_2(D_t+i (\sigma+a-\epsilon))\bar\tildepsi_2\Big)\\
&\frac{D}{e^2}(|\phi_1|^2+|\phi_2|^2-|\tilde\phi_1|^2-|\tilde\phi_2|^2-e^2\xi)+\frac{\bar{F}}{e^2}(\phi_1\tilde\phi_1+\phi_2\tilde\phi_2)+\frac{F}{e^2}(\phi_1\tildephi_1+\phi_2\tildephi_2)\\
&-\frac{i}{\sqrt{2}e^2} \sum_i\left(\bar\phi_i\lambda^1\psi_i+\tildephi_i\lambda^2\psi_i+\tildephi_i\bar\lambda_1\bar\tildepsi_i-\barphi_i \bar\lambda_2\bar\tildepsi_i-\bar\psi_i\bar\lambda_1\phi_i-\bar\psi_i\bar\lambda_2\bar\tildephi_i-\tildepsi_i\lambda^1\bar\tildephi_i+\tildepsi_i\lambda^2\phi_i\right)~,
\end{split}\end{align}
where $D_t =\partial_t+i v_t$ and $\tilde{D}_t=\partial_t-i v_t$. 

Similarly, the $L_{theory}$ will be of the form
\begin{align}\begin{split}
L_{theory}&=\frac{1}{e^2} \sum_{j=1}^{{\rm N}_f}\left[\ihalf\left(\bar\eta_j (D_t+i \sigma)\eta_j+\eta_j(\tilde{D}_t-i \sigma)\bar\eta_j\right)-|G_j|^2+{\rm m}_j[\bar\eta_j,\eta_j]\right]~,
\end{split}\end{align}
in the case of 4D $SU(2)$ with ${\rm N}_f$-fundamental hypermultiplets and 
\begin{align}\begin{split}
L_{theory}&=\frac{1}{e^2}\Bigg[|\partial_t \rho|^2+|\partial_t\tilde\rho|^2-({\rm m}+\epsilon)^2|\rho|^2-({\rm m}-\epsilon)^2|\tilde\rho|^2\\
&\qquad\qquad+\ihalf(\bar\chi \partial_t \chi+\chi\partial_t \bar\chi+\bar{\tilde{\chi}}\partial_t \tilde{\chi}+\tilde\chi\partial_t \bar{\tilde{\chi}})-\frac{\rm m}{2}([\bar\chi,\chi]-[\bar{\tilde{\chi}},\tilde\chi])\Bigg]\\
&+\frac{1}{e^2}\Bigg[\ihalf \Bigg(\bar\eta_1(D_t+i (\sigma-a+\epsilon))\eta_1+\eta_1(\tilde{D}_t-i (\sigma-a+\epsilon))\bar\eta_1\\
&\qquad+\bar{\tilde{\eta}}_1(\tilde{D}_t-i (\sigma-a-\epsilon))\tilde{\eta}_1+\tilde\eta_1(D_t+i (\sigma-a-\epsilon))\bar{\tilde{\eta}}_1\\
&\qquad+(\bar\eta_2(D_t+i (\sigma+a+\epsilon))\eta_2+\eta_2(\tilde{D}_t-i (\sigma+a+\epsilon))\bar\eta_2\\
&\qquad+\bar{\tilde{\eta}}_2(\tilde{D}_t-i (\sigma+a-\epsilon))\tilde{\eta}_2+\tilde\eta_2(D_t+i (\sigma+a-\epsilon))\bar{\tilde{\eta}}_2\Bigg)-|G_1|^2-|G_2|^2\Bigg]~,
\end{split}\end{align}
for the case of the 4D $SU(2)$ $\CN=2^\ast$ theory. 

By defining the conjugate momenta to the elementary fields
\begin{align}\begin{split}\label{eq:momentum}
&p_\sigma=\frac{1}{e^2}\partial_t \sigma\quad~, \quad p_{\lambda^A}=\frac{i}{e^2} \bar\lambda_A~,\quad p_{\psi i}=\frac{i}{e^2} \bar\psi_i~, \quad \tildep_{\psi i}=\frac{i}{e^2} \bar\tildepsi_i~,\\
&\pi_i=\frac{1}{e^2}\tilde{D}_t \barphi_i ~, \quad \tilde\pi_i=\frac{1}{e^2}D_t\bar{\tilde{\phi}}_i~,\quad p_{\eta_j}=\frac{i}{e^2}\bar\eta_j~,\\&
p_{\chi}=\frac{i}{e^2}\bar\chi~,\quad \tilde{p}_{\chi}=\frac{i}{e^2}\bar{\tilde{\chi}}~,\quad \pi_{\rho}=\frac{1}{e^2}\partial_t \bar\rho~, \quad \tilde\pi_{\rho}=\frac{1}{e^2}\partial_t\bar{\tilde{\rho}}~,
\end{split}\end{align}
we can compute the Hamiltonian and integrate out the auxiliary fields $(D,F,G)$:
\be
H=\frac{e^2}{2}p_\sigma^2-\frac{4\epsilon}{e^2}\bar\lambda_2\lambda^2+U+H_{matter}+H_I+v_t Q_{\rm Gauge}~,
\ee
where
\begin{align}\begin{split}
U&=\frac{1}{2e^2}\left(|\phi_1|^2+|\phi_2|^2-|\tildephi_1|^2-|\tildephi_2|^2-e^2\xi\right)^2
+\frac{1}{e^2}|\phi_1\tildephi_1+\phi_2\tildephi_2|^2
\end{split}\end{align}
and
\begin{align}\begin{split}
&H_{matter}=e^2\left[|\pi_1|^2+|\pi_2|^2+|\tilde\pi_1|^2+|\tilde\pi_2|^2\right]\\&
+\frac{1}{e^2}\left[(\sigma-a+\epsilon)^2|\phi_1|^2+(\sigma-a-\epsilon)^2|\tilde\phi_1|^2+(\sigma+a+\epsilon)^2|\phi_2|^2+(\sigma+a-\epsilon)^2|\tilde\phi_2|^2\right]\\&
+\frac{1}{2e^2}\left((\sigma-a+\epsilon)[\bar\psi_1,\psi_1]+(\sigma+a+\epsilon)[\bar\psi_2,\psi_2]-(\sigma-a-\epsilon)[\bar\tildepsi_1,\tildepsi_1]-(\sigma+a-\epsilon)[\bar\tildepsi_2,\tildepsi_2]\right)\\
&+H_{theory}
\end{split}\end{align}
and 
\begin{align}\begin{split}
&H_I=-\frac{i}{\sqrt{2}e^2}\sum_i\left(\bar\phi_i\lambda^1\psi_i+\tildephi_i\lambda^2\psi_i+\tildephi_i\bar\lambda_1\bar\tildepsi_i-\barphi_i \bar\lambda_2\bar\tildepsi_i-\bar\psi_i\bar\lambda_1\phi_i-\bar\psi_i\bar\lambda_2\bar\tildephi_i-\tildepsi_i\lambda^1\bar\tildephi_i+\tildepsi_i\lambda^2\phi_i\right)
\end{split}\end{align}
where
\begin{align}\begin{split}
&Q_{\rm Gauge}=Q_{theory}+\sum_i\left(i\phi_i \pi_i-i\barphi_i\bar\pi_i-i\tildephi_i\tilde\pi_i+i\bar\tildephi_i\bar{\tilde{\pi_i}}-\galf[\bar\psi_i,\psi_i]+\galf[\bar\tildepsi_i,\tildepsi_i]\right)
\end{split}\end{align}
Here, $H_{theory}$ is of the form
\be
H_{theory}=\frac{1}{2e^2} \sum_{j=1}^{{\rm N}_f}(\sigma-2{\rm m}_j)[\bar\eta_j ,\eta_j]~,
\ee
for the 4D theory with ${\rm N}_f$-fundamental hypermultiplets and 
\begin{align}\begin{split}
&H_{theory}=\frac{{\rm m}}{2e^2}[\bar\chi_I,\chi_I]+e^2|\pi_{\rho}|^2+e^2|\tilde\pi_\rho|^2+\frac{({\rm m}+\epsilon)^2}{e^2}|\rho|^2+\frac{({\rm m}-\epsilon)^2}{e^2}|\tilde\rho|^2\\&-\frac{ (\sigma-a+{\rm m})}{2e^2}[\bar\eta_1,\eta_1]+\frac{ (\sigma-a-{\rm m})}{2e^2}[\bar{\tilde{\eta}}_1,\tilde{\eta}_1]-\frac{ (\sigma+a+{\rm m})}{2e^2}[\bar\eta_2,\eta_2]+\frac{ (\sigma+a-{\rm m})}{2e^2}[\bar{\tilde{\eta}}_2,\tilde{\eta}_2]~,
\end{split}\end{align}
for the $\CN=2^\ast$ theory. 
Additionally,
\be
Q_{theory}=-\half\sum_{j=1}^{{\rm N}_f}[\bar\eta_j,\eta_j]~,
\ee
or 
\be
Q_{theory}=-\half([\bar\eta_1,\eta_1]-[\bar{\tilde{\eta}}_1,\tilde\eta_1]+[\bar\eta_2,\eta_2]-[\bar{\tilde{\eta}}_2,\tilde\eta_2])~,
\ee
for the ${\rm N}_f$-fundamental hypermultiplet and $\CN=2^\ast$ theory respectively. By Gauss's law we have that $Q_{\rm Gauge}$ must annihilate all physical states.

The classical vacuum equations for this theory are given by\footnote{There is an additional vacuum equation for the $\CN=2^\ast$ theory, however it has only trivial solutions: $\rho^A=0$.}
\begin{align}\begin{split}
&|\phi_1|^2+|\phi_2|^2-|\tilde\phi_1|^2-|\tilde\phi_2|^2-e^2\xi=0\quad,\quad\phi_1\tilde\phi_1+\phi_2\tilde\phi_2=0~,\\
&(\sigma-a+\epsilon)^2|\phi_1|^2+(\sigma-a-\epsilon)^2|\tilde\phi_1|^2+(\sigma+a+\epsilon)^2|\phi_2|^2+(\sigma+a-\epsilon)^2|\tilde\phi_2|^2=0~.
\end{split}\end{align}

The classical vacua of this theory are described by a Coulomb and Higgs branch which are defined by
\begin{align}\begin{split}\label{Cvaccua}
&\CM_C=\{\sigma\in \IR~,~\phi_i,\tilde\phi_i=0\}\cong \IR~,\\
&\CM_H=\left\{\begin{array}{l}\sigma=\pm a \pm' \epsilon~,~|\phi_1|^2+|\phi_2|^2-|\tilde\phi_1|^2-|\tilde\phi_2|^2=-e^2\xi~,~\phi_1\tilde\phi_1+\phi_2\tilde\phi_2=0\\
(\sigma-a+\epsilon)^2|\phi_1|^2+(\sigma-a-\epsilon)^2|\tilde\phi_1|^2+(\sigma+a+\epsilon)^2|\phi_2|^2+(\sigma+a-\epsilon)^2|\tilde\phi_2|^2=0
\end{array}\right\}\Big\slash U(1)~. 
\end{split}\end{align}
Note that in this case, the Higgs branch is given by a disjoint union of 4 points given by the 4-different choices of $(\pm,\pm')$ in \eqref{Cvaccua}. 

Now if we rescale the fields $\psi_i,\tildepsi_i,\eta_i,\lambda_i$:
\be
\psi_i,\tildepsi_i,\eta_i\to \frac{1}{e}\psi_i,\frac{1}{e}\tildepsi_i,\frac{1}{e}\eta_i\quad, \quad \lambda^A,\chi,\tilde\chi\to \frac{1}{e}\lambda^A, \frac{1}{e}\chi,\frac{1}{e}\tilde\chi~,
\ee
 such that the   commutation relations become
\begin{align}\begin{split}
&\{\bar\psi_i,\psi_j\}=\delta_{ij}\quad, \quad \{\bar\tildepsi_i,\tildepsi_j\}=\delta_{ij}\quad, \quad 
\{\bar\eta_i,\eta_j\}=\delta_{ij}\quad, \quad \{\bar\chi,\chi\}=\{\bar{\tilde{\chi}},\tilde\chi\}=1~,\\
&[\phi_i,\pi_j]=\I\delta_{ij}\quad~, \quad [\tilde\phi_i,\tilde\pi_j]=\I\delta_{ij}\quad~, \quad [\rho,\pi_{\rho}]=\I \qquad~, \quad [\tilde\rho,\tilde\pi_\rho]=\I~.\end{split}\end{align}
with all other commutation relations (or anticommutation as appropriate) are trivial, we can define the oscillators
\begin{align}\begin{split}\label{oscillators}
&a_i=\frac{1}{\sqrt{2}e}\left(\sqrt{\omega_i }\phi_i+\frac{i e^2 \bar\pi_i}{\sqrt{\omega_i}}\right)\quad, \quad \bara_i=\frac{1}{\sqrt{2}e}\left(\sqrt{\omega_i}\bar\phi_i+\frac{i e^2 \pi_i}{\sqrt{\omega_i}}\right)~,\\&
\tildea_i=\frac{1}{\sqrt{2}e}\left(\sqrt{\tilde{\omega_i}}\tilde\phi_i+\frac{i e^2\bar{\tilde{\pi_i}}}{\sqrt{\tilde{\omega_i}}}\right)\quad, \quad \tilde\bara_{i}=\frac{1}{\sqrt{2}e}\left(\sqrt{\tilde{\omega_i}}\bar\tildephi_i+\frac{i e^2 \tilde{\pi}_i}{\sqrt{\tilde{\omega_i}}}\right)~,\\
&a_{\rho}=\frac{1}{\sqrt{2}e}\left(\sqrt{\omega_\rho }\rho+\frac{i e^2 \bar\pi_\rho}{\sqrt{\omega_\rho}}\right)\quad, \quad a_{\barrho}=\frac{1}{\sqrt{2}e}\left(\sqrt{\omega_\rho}\bar\rho+\frac{i e^2 \pi_\rho}{\sqrt{\omega_\rho}}\right)~,\\&
\tildea_{\rho}=\frac{1}{\sqrt{2}e}\left(\sqrt{\tilde\omega_\rho}\tilde\rho+\frac{i e^2\bar{\tilde{\pi_\rho}}}{\sqrt{\tilde\omega_\rho}}\right)\quad, \quad \tildea_{\bar\rho}=\frac{1}{\sqrt{2}e}\left(\sqrt{\tilde\omega_\rho}\bar\tilderho_\rho+\frac{i e^2 \tilde{\pi}_\rho}{\sqrt{\tilde\omega_\rho}}\right)~,
\end{split}\end{align}
where 
\begin{align}\begin{split}\label{omegas}
&\omega_1=|\sigma-a+\epsilon|\quad, \quad \omega_2=|\sigma+a+\epsilon|\quad, \quad \tilde\omega_1=|\sigma-a-\epsilon|\quad, \quad \tilde\omega_2=|\sigma+a-\epsilon|~,\\
&\omega_\rho=|{\rm m}+\epsilon|\quad, \quad \tilde\omega_\rho=|{\rm m}-\epsilon|~,
\end{split}\end{align}
and all other (anti-) commutation relations have zero on the right hand side. 
Using this we can define a Fock space vacuum
\begin{align}\begin{split}\label{FockVac}
&a_i|0\rangle=\bara_i|0\rangle=\tildea_i|0\rangle=\tilde\bara_i|0\rangle=a_{\rho}|0\rangle=a_{\barrho}|0\rangle=\tildea_{\rho}|0\rangle=\tildea_{\barrho}|0\rangle=0~,\\
&\psi_i|0\rangle=\tildepsi_i|0\rangle=\lambda^A|0\rangle=\eta_j|0\rangle=\chi|0\rangle=\tilde\chi|0\rangle=0~.
\end{split}\end{align}

Using these creation and annihilation operators, $H_{matter}$ and $Q_{\rm Gauge}$ can be written
\begin{align}\begin{split}
H_{matter}&=\omega_1(a^\dagger_{1}a_{1}+\bara^\dagger_{ 1}\bara_{1}+1)+(\sigma-a+\epsilon)(\bar\psi_1\psi_1-\half)\\
&+\omega_2(a^\dagger_{ 2}a_{2}+\bara^\dagger_{2}\bara_{2}+1)+(\sigma+a+\epsilon)(\bar\psi_2\psi_2-\half)\\&
+\tilde\omega_1(\tildea^\dagger_{1}\tildea_{ 1}+\bar\tildea^\dagger_{ 1}\bar\tildea_{1}+1)-(\sigma-a-\epsilon)(\bar\tildepsi_1\tildepsi_1-\half)\\
&+
\tilde\omega_2(\tildea^\dagger_{ 2}\tildea_{ 2}+\bar\tildea^\dagger_{ 2}\bar\tildea_{2}+1)
-(\sigma+a+\epsilon)(\bar\tildepsi_2\tildepsi_2-\half)+H_{theory}~,
\end{split}\end{align}
and 
\begin{align}\begin{split}
Q_{\rm Gauge}=-\left[a_{1}^\dagger a_{1}+a_{2}^\dagger a_{2}-\bara_{1}^\dagger \bara_{1}-\bara_{2}^\dagger \bara_{2}-\tildea_{1}^\dagger\tildea_{1}-\tildea_{2}^\dagger \tildea_{2}+\bar\tildea_{1}^\dagger \bar\tildea_{1}+\bar\tildea_{2}^\dagger \bar\tildea_{2}\right]\\
-\half\left[\bar\psi_1\psi_1 +\bar\psi_2\psi_2-\bar\tildepsi_1\tildepsi_1-\bar\tildepsi_2\tildepsi_2\right]+Q_{theory}~,
\end{split}\end{align}
where
\be
H_{theory}= \half\sum_{j=1}^{{\rm N}_f}(\sigma-2{\rm m}_j)[\bar\eta_j ,\eta_j]~,
\ee
or
\begin{align}\begin{split}
&H_{theory}=\frac{\rm m}{2}([\bar\chi,\chi]-[\bar{\tilde{\chi}},\tilde\chi]) +\omega_\rho\left[a^\dagger_{\rho}a_{\rho}+a^\dagger_{\barrho}a_{\barrho}+1 \right]+\tilde\omega_\rho\left[a^\dagger_{\tilde\rho}a_{\tilde\rho}+a^\dagger_{\bar\tilderho}a_{\bar\tilderho}+1\right]\\&
+ \half(\sigma-a+{\rm m})[\bar\eta_1,\eta_1]- \half(\sigma-a-{\rm m})[\bar{\tilde{\eta}}_1,\tilde{\eta}_1] +\half(\sigma+a+{\rm m})[\bar\eta_2,\eta_2]- \half(\sigma+a-{\rm m})[\bar{\tilde{\eta}}_2,\tilde{\eta}_2]~,
\end{split}\end{align}
and
\be
Q_{theory}=-\half\sum_{j=1}^{{\rm N}_f}[\bar\eta_j,\eta_j]\quad{\rm or}\quad 
Q_{theory}=-\half([\bar\eta_1,\eta_1]-[\bar{\tilde{\eta}}_1,\tilde\eta_1]+[\bar\eta_2,\eta_2]-[\bar{\tilde{\eta}}_2,\tilde\eta_2])~,
\ee
for the ${\rm N}_f$-fundamental hypermultiplet and $\CN=2^\ast$ theory respectively. 

\subsection{Matter Ground States}

We we can determine the asymptotic ground states by applying the Born-Oppenheimer approximation. In this approximation, we divide our fields into ``slow'' and ``fast'' fields. We then decompose the wave function as
\be
|\Psi\rangle=|\psi_{\rm slow}\rangle\otimes|\psi_{\rm fast}\rangle~,
\ee
and solve for the ground state of the fast degrees of freedom in the background determined by the slow degrees of freedom. This is described by $|\psi_{\rm fast}\rangle$. Then we solve for the ground state of the slow degrees of freedom in the effective potential created by integrating out the fast degrees of freedom.

For our purposes, we want to study Coulomb branch states that stretch out into the asymptotic region of the Coulomb branch. Here, the fast degrees of freedom are described by the matter fields (the fundamental hypermultiplets and fundamental Fermi-multiplets) while the slow degrees of freedom are then described by the vector multiplet fields. \footnote{In the case of the 4D $\CN=2^\ast$ theory, the slow degrees of freedom include the adjoint valued twisted hypermultiplet. } Now determining the vacuum state of the fast fields requires minimizing $H_{matter}$ and $U$ subject to the the constraint that $Q_{\rm Gauge}|\Psi\rangle=0$.  
%
%
%
In order to construct a basis of states for the fast fields, let us define the operators
\be
a_3=a_\rho\quad,\quad \bara_3=a_{\bar\rho}\quad, \quad \tildea_3=\tilde{a}_\rho\quad, \quad \tilde\bara_3=\tildea_{\bar\rho}~,
\ee
and similarly 
\be
\omega_3=\omega_\rho\quad, \quad \tilde\omega_3=\tilde\omega_\rho~. 
\ee
Now, $H_{matter},Q_{\rm Gauge}$ can be written in terms of the in terms of the complex creation and annihilation operators $a_{ i},\bara_{i},a_{i}^\dagger, \bara_{i}^\dagger, \tildea_{i}, \tilde\bara_{i}, \tildea_{i}^\dagger, \tilde\bara_{i}^\dagger$ and the Fermionic creation and annihilation operators $\psi_i,\tildepsi_i,\eta_i,\bar\psi_i,\bar\tildepsi_i,\bar\eta_i,\chi_i,\bar\chi_i$. Now let us pick a basis of states
\be
|(n_i, \barn_i,\tilden_i,\bar\tilden_i,m_i,\tildem_i, f_j)\rangle=\prod_{i=1}^2 \left(a_{ i}^\dagger\right)^{n_i} \left(\bara_{i}^\dagger\right)^{\barn_i}\left(\tildea_{ i}^\dagger\right)^{\tilden_i}\left(\tilde\bara_{i}^\dagger\right)^{\bar\tilden_i}\bar\psi_i^{m_i}\bar\tildepsi_i^{\tildem_i} \bar\chi_1^{m_3}\bar\chi^{\tilde{m}_3}\prod_{j=1}^4 \bar\eta_j^{f_j}\big{|}0\big\rangle~,
\ee
in the case of fundamental 4D matter and 
\be
|(n_i, \barn_i,\tilden_i,\bar\tilden_i,m_i,\tildem_i, f_j)\rangle=\prod_{i=1}^3 \left(a_{ i}^\dagger\right)^{n_i} \left(\bara_{i}^\dagger\right)^{\barn_i}\left(\tildea_{ i}^\dagger\right)^{\tilden_i}\left(\tilde\bara_{i}^\dagger\right)^{\bar\tilden_i}\bar\psi_i^{m_i}\bar\tildepsi_i^{\tildem_i} \bar\chi_1^{m_3}\bar\chi^{\tilde{m}_3}\prod_{j=1}^2 \bar\eta_j^{f_j}\bar{\tilde{\eta}}^{\tilde{f}_j}\big{|}0\big\rangle~,
\ee
for the case of the $\CN=2^\ast$ theory.  Note that this means that the quantum numbers are constrained
 \be
 n_i,\barn_i,\tilden_i,\bar\tilden_i\in \IZ_+\quad,\quad m_i,\tildem_i,f_j,\tilde{f}_j=0,1~.
 \ee

These quantum numbers have the interpretation of the eigenvalue of the number operator associated to the given fields. In this case, we have that the eigenvalues of $H_{matter},Q_{\rm Gauge}$ of these states are given by
\begin{align}\begin{split}
E_{matter}&=\omega_1(n_1+\barn_1+1)+(\sigma-a+\epsilon)(m_1-\half)+\omega_2(n_2+\barn_2+1)+(\sigma+a+\epsilon)(m_2-\half)\\&
+\tilde\omega_1(\tilden_1+\bar\tilden_1+1)-(\sigma-a-\epsilon)(\tildem_1-\half)+
\tilde\omega_2(\tilden_2+\bar\tilden_2+1)
-(\sigma+a-\epsilon)(\tildem_2-\half)\\
&+E_{theory}~,\\
q_{matter}&=-(n_1-\barn_1+n_2-\barn_2-\tilden_1+\bar\tilden_1-\tilden_2+\bar\tilden_2-m_1-m_2+\tildem_1+\tildem_2)\\&
+q_{theory}
\end{split}\end{align}
Note that the $m_i,\tilde{m}_i$ are quantum numbers and not masses. 
Here 
\be
E_{theory}=\sum_{j=1}^{{\rm N}_f}(\sigma-2{\rm m}_j)(f_j-\half)~,
\ee
or
\begin{align}\begin{split}
&E_{theory}={\rm m}[m_3-\tildem_3]+\omega_3(n_3+\barn_3+1)+\tilde\omega_3(\tilden_3+\bar\tilden_3+1)- (\sigma-a+{\rm m})(f_1-\half)\\&+ (\sigma-a-{\rm m})(\tildef_1-\half)- (\sigma+a+{\rm m})(f_2-\half)+ (\sigma+a-{\rm m})(\tildef_2-\half)~,
\end{split}\end{align}
and
\be
Q_{theory}=-\sum_{j=1}^{{\rm N}_f}(f_j-\half)\quad{\rm or}\quad 
Q_{theory}=-(f_1+f_2-\tildef_1-\tildef_2)~,
\ee
for the ${\rm N}_f$-fundamental hypermultiplet and $\CN=2^\ast$ theory respectively. 

Here we also need to define the flavor charges
\begin{align}\begin{split}\label{Qfa}
&Q_\fa=Q_\fa^{(theory)}+\half[\bar\psi_1,\psi_1]-\half[\bar\tildepsi_1,\tilde\psi_1]-\half[\bar\psi_2,\psi_2]+\half[\bar\tildepsi_2,\tildepsi_2]\\&
-\frac{2}{e^2}\begin{cases}
\omega_1|\phi_1|^2+\tilde\omega_1|\tilde\phi_1|^2-\omega_2|\phi_2|^2-\tilde\omega_2|\tilde\phi_2|^2
&\sigma>a+\epsilon\\
-\omega_1|\phi_1|^2-\tilde\omega_1|\tilde\phi_1|^2+\omega_2|\phi_2|^2+\tilde\omega_2|\tilde\phi_2|^2
&\sigma<-a-\epsilon
\end{cases}
\end{split}\end{align} 
where $Q_\fa^{(theory)}=0$ or 
\be
Q_\fa^{(theory)}=\half\big([\bar\eta_1,\eta_1]-[\bar{\tilde{\eta}}_1,\tilde\eta_1]-[\bar\eta_2,\eta_2]+[\bar{\tilde{\eta}}_2,\tilde\eta_2]\big)~.
\ee
and 
\begin{align}\begin{split}\label{Qepsilon}
&Q_\epsilon=Q_\epsilon^{(theory)}+4\bar\lambda_2\lambda^2+\half[\bar\psi_1,\psi_1]+\half[\bar\tildepsi_1,\tilde\psi_1]+\half[\bar\psi_2,\psi_2]+\half[\bar\tildepsi_2,\tildepsi_2]\
\\&-\frac{2}{e^2}\sum_{i=1}^2\begin{cases}
\left(\omega_i |\phi_i|^2-\tilde\omega_i |\tilde\phi_i|^2
\right)
&\sigma>a+\epsilon\\
-\left(\omega_i |\phi_i|^2-\tilde\omega_i |\tilde\phi_i|^2
\right)&\sigma<-a-\epsilon
\end{cases}
\end{split}\end{align} 
and $Q_\epsilon^{(theory)}=0$ or 
\be
Q_\epsilon^{(theory)}= \frac{2}{e^2}\big(-\omega_\rho|\rho|^2+\tilde\omega_\rho|\tilde\rho|^2\big)+\half\sum_i \left([\bar\eta_i,\eta_i]+[\bar{\tilde{\eta}}_i,\tilde\eta_i]\right)\quad, \quad \pm \sigma>0~.
\ee
In our basis of states, these can be written as 
\begin{align}\begin{split}
&Q_\fa=Q_\fa^{(theory)}\\
&+\begin{cases}
n_1+\barn_1+\tilden_1+\bar\tilden_1-n_2-\barn_2-\tilden_2-\barn_2+m_1-\tildem_1-m_2+\tildem_2&\sigma>a+\epsilon\\
-n_1-\barn_1-\tilden_1-\bar\tilden_1+n_2+\barn_2+\tilden_2+\barn_2+m_1-\tildem_1-m_2+\tildem_2&\sigma<-a-\epsilon
\end{cases}
\end{split}\end{align}
with $Q_\fa^{(theory)}=0$ or 
\be
Q_\fa^{(theory)}=(f_1-\tildef_1-f_2+\tildef_2)~,
\ee
and 
\begin{align}\begin{split}
&Q_\epsilon=Q_\epsilon^{(theory)}+4\bar\lambda_2\lambda^2\\
&+\begin{cases}
-n_1-\barn_1+\tilden_1+\bar\tilden_1-n_2-\barn_2+\tilden_2+\bar\tilden_2+m_1+\tildem_1+m_2+\tildem_2&\sigma>a+\epsilon\\
n_1+\barn_1-\tilden_1-\bar\tilden_1+n_2+\barn_2-\tilden_2-\bar\tilden_2+m_1+\tildem_1+m_2+\tildem_2&\sigma<-a-\epsilon
\end{cases}
\end{split}\end{align}
where $Q_\epsilon^{(theory)}=0$ or 
\be
Q_\epsilon^{(theory)}=-\left(n_3+\barn_3-\tilden_3-\bar\tilden_3\right)+\left(f_1+\tildef_1+f_2+\tildef_2\right)~.
\ee

The constraint for a supersymmetric ground state is now 
\be\label{susygs}
(H_{matter}+\epsilon Q_\epsilon+a Q_\fa+\sum_f {\rm m}_f F_f)|\Psi\rangle=0\quad, \quad Q_{\rm Gauge}|\Psi\rangle=0
\ee

There are 5 distinct regions in $\sigma$ space in which we can impose these conditions. The physically relevant ones are those for which $|\sigma|>|a|+|\epsilon|$. Therefore we will restrict to the regions in which $\sigma>a+\epsilon$ and $\sigma<-a-\epsilon$ where we are assuming $a>\epsilon>0$. 

It is actually more convenient to solve the equations
\be
(H_{matter}+\epsilon Q_\epsilon+a Q_\fa+\sum_f {\rm m}_f F_f+\sigma Q_{\rm Gauge})|\Psi\rangle=0~,
\ee
and then solve $Q_{\rm Gauge}|\Psi\rangle=0$. 

As it turns out there are only solutions only for the case of ${\rm N}_f=4$ and the $\CN=2^\ast$ theory. For the general theory, the zero energy condition 
 \eqref{susygs}, can be written as
\be
0=\sum_{i=1}^2\begin{cases}
n_i+\bar\tilden_i+m_i+(1-\tildem_i)& \sigma>a+\epsilon\\
-\barn_i-\tilden_i-\tildem_i-(1-m_i)&\sigma<-a-\epsilon
\end{cases}
\ee
For ${\rm N}_f$ fundamental hypermultiplet theories, the gauge invariance condition can be written
\be
0=\begin{cases}
\barn_1+\barn_2+\tilden_1+\tilden_2+\left(2-\frac{{\rm N}_f}{2}\right)+\sum_{j=1}^{{\rm N}_f} (1-f_j)&\sigma>a+\epsilon\\
-n_1-n_2-\bar\tilden_1-\bar\tilden_2-\left(2-\frac{{\rm N}_f}{2}\right)-\sum_{j=1}^{{\rm N}_f} f_j&\sigma<-a-\epsilon
\end{cases}
\ee
while for the $\CN=2^\ast$ theory, it can be written as 
\be
0=\begin{cases}
\barn_1+\barn_2+\tilden_1+\tilden_2+2-(f_1+f_2-\tildef_1-\tildef_2)&\sigma>a+\epsilon\\
-n_1-n_2-\bar\tilden_1-\bar\tilden_2-2-(f_1+f_2-\tildef_1-\tildef_2)&\sigma<-a-\epsilon
\end{cases}
\ee

These equations clearly have no solution for ${\rm N}_f=0,1,2,3$. 

The ground state solutions for $\CN=2^\ast$ and the ${\rm N}_f=4$ theory are given by:
\begin{align}\begin{split}
&\sigma>a+\epsilon~~\text{ : }\barn_i,\tilden_i=0~,~ \{f_j=1~{\rm or}~f_{1,2}=1~,~\tildef_{1,2}=0\}~,\\
&\sigma<-a-\epsilon\text{ : }n_i,\bar\tilden_i=0~,~\{ f_j=0~{\rm or}~\tildef_{1,2}=1~,~f_{1,2}=0\}~. 
\end{split}\end{align}

Now we can solve for the matter ground states in the regions $\sigma>a+\epsilon$ and $\sigma<-a-\epsilon$. We will define the matter ground states in these regions as \footnote{
Note that we do not need to worry about the normalization of $|\pm\rangle$, so long as it is normalizable. The reason is that the only physically relevant thing is for the total wave function to have unit norm. }
\begin{align}\begin{split}
&|+\rangle=\Theta(\sigma-a-\epsilon)|(n_i,\barn_i,\tilden_i,\bar\tilden_i,m_i=0~,~\tildem_i=1~,~\{f_j=1~{\rm or}~f_{1,2}=1~,~\tildef_{1,2}=0\}\rangle~,\\
&|-\rangle=\Theta(-\sigma+a+\epsilon)|(n_i,\barn_i,\tilden_i,\bar\tilden_i,\tildem_i=0~,~m_i=1~,~\{ f_j=0~{\rm or}~\tildef_{1,2}=1~,~f_{1,2}=0\}\rangle~.
\end{split}\end{align}

\subsection{Asymptotic States}

%
%
Thus far we have computed $|\psi_{\rm fast}\rangle=|\pm\rangle$ for $\pm\sigma>a+\epsilon$. 
%
%
Now we must find the state $|\psi_{\rm slow}\rangle$ that is dependent on the adjoint valued fields only, such that the entire state $|\Psi\rangle=|\psi_{\rm slow}\rangle\otimes |\pm\rangle$ is annihilated by the conserved  supercharge operators $Q_1,\bar{Q}^1$. To this effect, we can apply the Born rule to get an effective supercharge 
\be
Q_{\rm eff,A}=-\frac{\langle \psi_{\rm fast}|Q_A|\psi_{\rm fast}\rangle}{\langle \psi_{\rm fast}|\psi_{\rm fast}\rangle}=-\langle Q_A\rangle~. 
\ee
Using the fact that $|\mp\rangle$ is in the harmonic oscillator ground state of all bosonic, hypermultiplet fields, \footnote{Due to the form of the oscillators \eqref{oscillators}:
\be
a\sim\frac{e}{\sqrt{2\omega}}\left(i \bar\pi +\frac{\omega}{e^2}\phi\right)=\frac{e}{\sqrt{2\omega}}\left(\partial_{\barphi} +\frac{\omega}{e^2}\phi\right)~,
\ee 
the wave function $|\pm\rangle$ is of the form 
\be
|\pm \rangle\sim e^{-\frac{\omega}{e^2} |\phi|^2}~.
\ee
Note that this implies 
\be
\langle \pm|\,|\phi|^2|\pm\rangle=\frac{e^2}{2\omega}~. 
\ee
} we find that  $\langle F \rangle=0$ and
 that the effective supercharges are of the form
\be
Q_{\rm eff,A}=\frac{e}{\sqrt{2}}\left(-i \langle p_{\sigma}\rangle\,\delta_A^{~B}-\frac{1}{e^2}\langle D\rangle(\sigma^3)^{~B}_A \right)\bar\lambda_B\quad, \quad \bar{Q}^A_{\rm eff}=\frac{e}{\sqrt{2}}\left(i \langle p_{\sigma}\rangle \,\delta_A^{~B}-\frac{1}{e^2}\langle D\rangle(\sigma^3)^{~B}_A \right)\bar\lambda_B~,
\ee
where
\begin{align}\begin{split}&
\langle p_\sigma\rangle =-i \partial_\sigma+\ihalf\sum_\pm\left(\frac{1}{\sigma\pm a+\epsilon}+\frac{1}{\sigma\pm a-\epsilon}\right)~,\\&
\langle D\rangle=-e^2\xi+\frac{e^2}{2}\sum_i\left(\frac{1}{\omega_i}-\frac{1}{\tilde\omega_i}\right)~.
\end{split}\end{align}
Explicitly, the complex supercharges are given by
\begin{align}\begin{split}
Q_{\rm eff,1}=\frac{e}{\sqrt{2}}\bar\lambda_1\begin{cases}
\left(-\partial_\sigma 
+\frac{1}{\sigma+a-\epsilon}+\frac{1}{\sigma-a-\epsilon}
+\xi\right)&\sigma>a+\epsilon\\
\left(-\partial_\sigma+
\frac{1}{\sigma+a+\epsilon}+\frac{1}{\sigma-a+\epsilon}
+\xi\right)&\sigma<-a-\epsilon
\end{cases}\\
Q_{\rm eff,2}=\frac{e}{\sqrt{2}}\bar\lambda_2\begin{cases}
\left(-\partial_\sigma
+\frac{1}{\sigma+a+\epsilon}+\frac{1}{\sigma-a+\epsilon}
-\xi\right)& \sigma>a+\epsilon\\
\left(-\partial_\sigma +\frac{1}{\sigma+a-\epsilon}+\frac{1}{\sigma-a-\epsilon}
-\xi\right)&\sigma<-a-\epsilon
\end{cases}
\end{split}\end{align}
with similar expressions for the complex conjugate supercharges. \footnote{Note that if we had normalized the matter wave functions $|\pm \rangle$ such that $\langle\pm|\pm \rangle=1$, then we would have $\langle p_\sigma\rangle=-i \partial_\sigma$. However, we have made this choice of normalization such that when restricted to the Coulomb branch, all of the $\sigma$ dependence is manifested in $|\psi_{adjoint}\rangle$. This will make the discussion of normalizability of the state along the Coulomb branch simpler.}

 Using these complex supercharges, we can construct the real supercharges
\begin{align}\begin{split}\label{realQ}
\CQ_1=Q_{\rm eff,1}+\bar{Q}^1_{\rm eff}\quad, \quad \CQ_2=-i(Q_{\rm eff,1}-\bar{Q}^1_{\rm eff})~,\\
\CQ_3=Q_{\rm eff,2}+\bar{Q}^2_{\rm eff}\quad, \quad \CQ_4=-i(Q_{\rm eff,2}-\bar{Q}^2_{\rm eff})~,
\end{split}\end{align}

Since we are deforming by a mass parameter $\epsilon$, SUSY is broken from $\CN=(0,4)\to \CN=(0,2)$ such that $\CQ_1,\CQ_2$ are the conserved real supercharges. 
Therefore, supersymmetric ground states are in the kernel of $\CQ_1,\CQ_2$ or equivalently in the kernel of $Q_{\rm eff,1}$ and its complex conjugate operator. 

Now let us consider the states that are killed by $Q_{\rm eff,1}$ and its complex conjugate on the semi-infinite interval $\sigma>a+\epsilon.$

Here the relevant supercharges are given by 
\begin{align}\begin{split}
Q_{\rm eff,1}&=\frac{e}{\sqrt{2}}\bar\lambda_1
\begin{cases}
\left(-\partial_\sigma +\frac{1}{\sigma+a-\epsilon}+\frac{1}{\sigma-a-\epsilon}+\xi\right)&\sigma>a+\epsilon\\
\left(-\partial_\sigma +\frac{1}{\sigma+a+\epsilon}+\frac{1}{\sigma-a+\epsilon}+\xi\right)&\sigma<-a-\epsilon\\
\end{cases}\\
\bar{Q}_{\rm eff}^1&=\frac{e}{\sqrt{2}}\lambda^1
\begin{cases}
\left(\partial_\sigma -\frac{1}{\sigma+a+\epsilon}-\frac{1}{\sigma-a+\epsilon}+\xi\right)&\sigma>a+\epsilon\\
\left(\partial_\sigma -\frac{1}{\sigma+a-\epsilon}-\frac{1}{\sigma-a-\epsilon}+\xi\right)&\sigma<-a-\epsilon
\end{cases} 
\end{split}\end{align}
Since the $\epsilon$-mass deformation breaks SUSY from $\CN=(0,4)\to \CN=(0,2)$ we have that only the $Q_{{\rm eff},1}$ supercharge is preserved. These supercharges  satisfy the supersymmetry algebra
\be
\{Q_{{\rm eff},1},\bar{Q}_{\rm eff}^1\}=H_{\rm eff}-Z~,
\ee
where
\be
Z=-\epsilon Q_{\epsilon}-\fa Q_{\fa}-\sum_j {\rm m}_f F_j~. 
\ee
This corresponds to the effective Hamiltonian
\be\label{Hamshort}
H_{\rm eff}=\frac{e^2\langle p_\sigma\rangle^2}{2}+\langle D\rangle^2+\partial_\sigma \langle D\rangle[\lambda^1,\bar\lambda_1 ]~.
\ee

The wave functions dependent on the Clifford algebra spanned by the $\{\bar\lambda_1,\bar\lambda_2\}$ that are killed by both $Q_{\rm eff,1}$ and $\bar{Q}_{\rm eff}^1$ span a 4-dimensional Fock space dependent on the vector multiplet zero fields
\begin{align}\begin{split}\label{Q1BPS}
|\psi_{SUSY}\rangle&=\begin{cases}\tilde\omega_1\tilde\omega_2e^{\xi\sigma}\left(\alpha_1|+\rangle+\beta_1\bar\lambda_2|+\rangle\right)
+\omega_1\omega_2e^{-\xi\sigma}\left(\alpha_2\bar\lambda_1|+\rangle+\beta_2\bar\lambda_1\bar\lambda_2|+\rangle\right)&\sigma>a+\epsilon\\
\omega_1\omega_2 e^{\xi\sigma}(\alpha_1|-\rangle+\beta_1\bar\lambda_2|-\rangle)+\tilde\omega_1\tilde\omega_2e^{-\xi\sigma}(\alpha_2\bar\lambda_1|-\rangle+\beta_2\bar\lambda_1\bar\lambda_2|-\rangle)&\sigma<-a-\epsilon
\end{cases}
\end{split}\end{align}
where the $\alpha_i,\beta_i$ are undetermined coefficients and $\omega_i,\tilde\omega_i$ are given by \eqref{omegas}.

There are some additional considerations for the case of the $\CN=2^\ast$ theory. The reason is that there is a decoupled $\CN=(0,4)$ adjoint valued hypermultiplet field that pairs with the $\CN=(0,4)$ vector multiplet to make a $\CN=(4,4)$ vector multiplet. Because of the representation theory of the supersymmetry algebra, the Witten index is identically zero. 

This can be seen as follows. Since the $\CN=(0,4)$ adjoint valued hypermultiplet is completely decoupled, the vector multiplet state splits
\be
|\psi_{\rm slow}\rangle=|\psi_{vector}\rangle\otimes|\psi_{hyper}\rangle~.
\ee
Since we can write the Hamiltonian for the adjoint hypermultiplet fields in terms of simple harmonic oscillators, we can pick a basis of states for the hypermultiplet wave functions
\be
|n_3,\barn_3,\tilden_3,\bar\tilden_3~,~m_3,\tildem_3\rangle=(a_3^\dagger)^{n_3}(\bara_3^\dagger)^{\barn_3^\dagger}(\tildea_3^\dagger)^{\tilden_3}(\bar\tildea_3^\dagger)^{\bar\tilden_3}\bar\chi^{m_3}\bar{\tilde{\chi}}^{\tildem_3}|0\rangle_{hyp}~,
\ee
where 
\be
a_3|0\rangle_{hyp}=\bara_3|0\rangle_{hyp}=\tildea_3|0\rangle_{hyp}=\bar\tildea_3|0\rangle_{hyp}=\chi|0\rangle_{hyp}=\tilde\chi|0\rangle_{hyp}=0~.
\ee 
Because the hypermultiplet fields are completely decoupled (up to flavor symmetries), there are no constraints on the values of the $n's$ and $m's$ except 
\be
n_3,\barn_3,\tilden_3,\bar\tilden_3\in \IZ_+\quad,\quad m_3,\tildem_3=0,1~.
\ee
These states have definite charge under $Q_{\rm m}$ and $Q_{\epsilon}$. The eigenvalues of these charge operators is given by 
\be
q_\epsilon=(n_3-\barn_3-\tilden_3+\bar\tilden_3)\quad, \quad q_{\rm m}=2m(m_3-\tildem_3)+m(n_3-\barn_3+\tilden_3-\bar\tilden_3)~. 
\ee
Now pick a state $|\psi_0\rangle=|n^\ast_3,\barn^\ast_3,\tilden^\ast_3,\bar\tilden^\ast_3~,~m^\ast_3,\tildem^\ast_3\rangle$. Now to this state we can identify another state with the same charges under $Q_{\rm m}$ and $Q_{\epsilon}$ with different fermion numbers. Specifically, we can make a shift depending on the value of $(m^\ast_3-\tildem^\ast_3)$
\be
(m^\ast_3-\tildem^\ast_3)\mapsto(m^\prime_3-\tildem^\prime_3)= (m^\ast_3-\tildem^\ast_3)+1\quad, \quad (\barn^\ast_3,\bar\tilden^\ast_3)\mapsto(n^\prime_3,\tilden^\prime_3)=(\barn^\ast_3+1,\bar\tilden^\ast_3+1)~,
\ee
or 
\be
(m^\ast_3-\tildem^\ast_3)\mapsto(m^\prime_3-\tildem^\prime_3)= (m^\ast_3-\tildem^\ast_3)-1\quad, \quad (n^\ast_3,\tilden^\ast_3)\mapsto(n^\prime_3,\tilden^\prime_3)=(n^\ast_3+1,\tilden^\ast_3+1)~,
\ee
depending on the value of $(m^\ast_3-\tildem^\ast_3)$ where we only shift one of the $m^\ast_3,\tildem^\ast_3$. The state $|\psi_0^\prime\rangle=|n^\prime_3,\barn^\prime_3,\tilden^\prime_3,\bar\tilden^\prime_3~,~m^\prime_3,\tildem^\prime_3\rangle$ will then have the same eigenvalues $q_\epsilon,q_{\rm m}$ with different Fermion number by $\pm 1$, hence canceling the contribution of  $|\psi_0\rangle$ to the Witten index. 
Therefore, there is no contribution to the Witten index from asymptotic Coulomb branch states in  bubbling SQM for the case of the 4D $\CN=2^\ast$ theory. Thus, from hereon out, we will only consider the bubbling SQM for the ${\rm N}_f=4$ theory. 

It is a subtle point to define the fermion number of these states. As explained in \cite{Gaiotto:2015aoa}, the bosonic Fermi vacuum should be defined relative to the lowest energy state of the fermions. The Fermi vacuum of the Fermi-multiplet and multiplet fermions is defined by their bare mass terms in the full Hamiltonian. However, since the fermions in these multiplets come in pairs, the fermion number $(-1)^F$ is only dependent on the vector multiplet fermions that do not come in a symmetric pair. 

In our Born-Oppenheimer approximation, the vector multiplet fermion $\lambda^2$ is given a bare mass while $\lambda^1$ is given a mass from 1-loop terms. The mass terms are given by 
\be
H_{mass}=-4\epsilon\bar\lambda_2\lambda^2\mp e^2\sum_i\left(\frac{1}{\omega_i^2}-\frac{1}{\tilde\omega_i^2}\right)[\lambda^1,\bar\lambda_1]\quad, \quad \pm \sigma>a+\epsilon~.
\ee
Since $\omega_i>\tilde\omega_i$ for $\sigma>a+\epsilon$ ($\omega_i<\tilde\omega_i$ for $\sigma<-a-\epsilon$) for $\epsilon>0$ and similarly $\omega_i<\tilde\omega_i$ for $\sigma>a+\epsilon$ ($\omega_i>\tilde\omega_i$ for $\sigma<-a-\epsilon$) for $\epsilon<0$, the physical, bosonic vacuum state is defined by 
\begin{align}\begin{split}
\bar\lambda_2|0\rangle_{phys}=\lambda^1|0\rangle_{phys}=0\quad,\quad \epsilon>0~,\\
\lambda^2|0\rangle_{phys}=\bar\lambda_1|0\rangle_{phys}=0\quad, \quad \epsilon<0~,
\end{split}\end{align}
which differs from our $\epsilon$-invariant choice of Fock vacuum is defined in \eqref{FockVac}:
\be
\lambda^A|0\rangle_{ours}=0~. 
\ee
 These two choices are related by 
\be
|0\rangle_{phys}=\begin{cases}
\bar\lambda_2|0\rangle_{ours}&\epsilon>0\\
\bar\lambda_1|0\rangle_{ours}&\epsilon<0
\end{cases}~. 
\ee
 Thus, the Fermion number of our vacuum states are given by
\be
(-1)^F|0\rangle=-|0\rangle\quad\Longrightarrow\quad(-1)^F|\pm\rangle=-|\pm\rangle~.
\ee

\subsection{Hermitian Supercharge Operators and Boundary Conditions}

Note that the real supercharge operators defined in \eqref{realQ} are not actually self-adjoint on the relevant semi-infinite interval because integration by parts picks up boundary terms. Therefore, we must restrict the Hilbert space of BPS states to those on which the above supercharges are self-adjoint. It will be sufficient to impose that $\CQ_1=Q_{\rm eff,1}+\bar{Q}_{\rm eff}^1$ is Hermitian. In the seminifinite interval $\pm \sigma>a+\epsilon$, this has the form 
\be
\CQ=\frac{e}{\sqrt{2}}
(\lambda^1-\bar\lambda_1)
D_\sigma+\frac{e}{\sqrt{2}}(\lambda^1+\bar\lambda_1)\left(\frac{1}{|\sigma+a+\epsilon|}+\frac{1}{|\sigma-a+\epsilon|}-\frac{1}{|\sigma+a-\epsilon|}-\frac{1}{|\sigma-a-\epsilon|}+\xi\right)~,
\ee
where $D_\sigma=\langle i p_\sigma\rangle$. 

Since this only has $\bar\lambda_1$ and $\lambda^1$ Clifford elements, it is natural to divide the Hilbert space as 
\be
\CH={\rm span}\{\mathds{1}, \bar\lambda_2\}\otimes \{|+\rangle~,~\bar\lambda_1|+\rangle\}~. 
\ee

Now consider a generic state that is annihilated by $\lambda^2$:
\be
|\Psi\rangle=f(\sigma)|+\rangle+g(\sigma)\bar\lambda_1|+\rangle~. 
\ee
In this subspace, the supercharge $\CQ_1$ (which we choose to be our localizing supercharge) is the form of a Dirac operator:
\be
\CQ=\frac{e}{\sqrt{2}}\left(\begin{array}{cc}0 &D_\sigma+A(\sigma)\\-D_\sigma+A(\sigma)&0\end{array}\right) \quad{\rm where }\quad |\Psi\rangle=\left(\begin{array}{c} f(\sigma)\\g(\sigma)\end{array}\right)~,
\ee
and
\be
A(\sigma)=\left(\frac{1}{|\sigma+a+\epsilon|}+\frac{1}{|\sigma-a+\epsilon|}-\frac{1}{|\sigma+a-\epsilon|}-\frac{1}{|\sigma-a-\epsilon|}+\xi\right)~.
\ee
On these states, we have that 
\be
\langle \Psi_1|\CQ_1 \Psi_2\rangle=\langle \CQ_1\Psi_1|\Psi_2\rangle-\Big[\bar{f}_1 g_2-\bar{g}_1 f_2\Big]_{\pm \sigma=a+\epsilon}~. 
\ee
And therefore, for the $\CQ_i$ to be self-adjoint, we must impose
\be\label{eq:BC}
\Big[\bar{f}_1 g_2-\bar{g}_1 f_2\Big]_{\pm\sigma=a+\epsilon}=0~. 
\ee
%
A similar argument holds for the pair of states 
\be
|\tilde\Psi\rangle=\tilde{f}(\sigma)\bar\lambda_2|+\rangle+\tilde{g}(\sigma)\bar\lambda_1\bar\lambda_2|+\rangle~. 
\ee
Now we see that there are more than 10 different restrictions we can impose on the Hilbert space such that \eqref{eq:BC} is satisfied. We will impose the same condition on the Hilbert space for $\sigma=a+\epsilon$ and $\sigma=-a-\epsilon$. These choices are given by a combination of restricting wave functions and completely eliminating  all wave functions in different factors of the Hilbert space under the decomposition
\be
\CH=\oplus_{n_1,n_2=0,1} \CH_{n_1,n_2}=\oplus_{n_1,n_2=0,1} {\rm span}_{L^2}\{\bar\lambda_1^{n_1}\bar\lambda_2^{n_2}|0\rangle\}~.
\ee
These different conditions that we can impose are:
\begin{itemize}
\item Type I: restricting the wave functions in a factor of $\CH_{n_1,n_2}$ such that $\langle \sigma|\psi\rangle\big{|}_{\sigma=\pm (a+\epsilon)}=0$ for $|\psi\rangle\in \CH_{n_1,n_2}$
\item Type II:  eliminating a factor of $\CH_{n_1,n_2}$ 
\end{itemize}
We will choose either purely Type I or Type II conditions. 


Given our assumptions that the boundary conditions are symmetric and purely Type I or Type II, there is a unique such choice such that $I_{asymp}=Z_{mono}^{(extra)}$. If we choose any other boundary condition, then we have that $I_{asymp}\neq Z_{mono}^{(extra)}$.  Therefore, we believe that the physics of relating $I_{\CH_0}^{(Loc)}$ with a counting of Higgs branch states suggests that we 
we should choose boundary conditions that restrict our wave functions to be of the form $|\Psi_f\rangle$:
\be\label{BPSplus}
\CH_{BPS}^{\sigma>a+\epsilon}={\rm span}\left\{N_1\,\omega_1\omega_2 e^{\xi \sigma}\bar\lambda_2|+\rangle~,~N_2\,\tilde\omega_1\tilde\omega_2e^{-\xi \sigma}\bar\lambda_1|+\rangle\right\}~. 
\ee
A similar computation shows that 
\be\label{BPSminus}
\CH_{BPS}^{\sigma<-a-\epsilon}={\rm span}\left\{N_1\,\tilde\omega_1\tilde\omega_2 e^{\xi \sigma}\bar\lambda_2|-\rangle~,~N_2\,\omega_1\omega_2e^{-\xi \sigma}\bar\lambda_1|-\rangle\right\}~. 
\ee
Thus, what we have really shown is that in the Born-Oppenheimer approximation, there is a suitable boundary condition so that $I_{asymp}=Z_{mono}^{(extra)}$. Clearly this aspect of our proposal needs to be improved. 

\subsection{Extra Contribution to the Witten Index}

Now we have found the BPS states for the semi-infinite intervals $\sigma>a+\epsilon$ and $\sigma<-a-\epsilon$. Interestingly, these states undergo wall crossing with the sign of $\xi$. Essentially, as is evident from equations \eqref{BPSplus}-\eqref{BPSminus}, as one approaches the wall of marginal stability at $\xi=0$, the states contributing to the Witten index go off to infinity as $1/\xi$ and become non-normalizable at $\xi=0$. Then as we again increase $|\xi|$ from 0, another state comes in from infinity. 


By using the results of \eqref{BPSplus} and \eqref{BPSminus}, we have that there are only 2 normalizable BPS states for a given choice of $\xi>0$ or $\xi<0$. The corresponding (unnormalized, but normalizable) states are given by:
\begin{align}\begin{split}\label{U1BPS}
|\Psi_1\rangle=\begin{cases}
\tilde\omega_1\tilde\omega_2e^{-\xi \sigma}\bar\lambda_1|+\rangle& \xi>0\\
\omega_1\omega_2 e^{\xi \sigma}\bar\lambda_2|+\rangle& \xi<0\end{cases}~,\\ 
|\Psi_2\rangle=\begin{cases}
\tilde\omega_1\tilde\omega_2 e^{\xi \sigma}\bar\lambda_2|-\rangle & \xi>0\\
\omega_1\omega_2e^{-\xi \sigma}\bar\lambda_1|-\rangle& \xi<0
\end{cases}~.
\end{split}\end{align}

Now we can ask how these contribute to the Witten index. Here the flavor charges associated to $\fa$, $\epsilon$ are given by equations \eqref{Qfa} and \eqref{Qepsilon}. For our cases, these reduce to  
\be
Q_\fa=0\quad,\quad Q_\epsilon=2-4\bar\lambda_2\lambda^2~.
\ee
and similarly 
\be
Q_{{\rm m}_f}=[\bar\eta_f,\eta_f]~.
\ee
This means that the flavor charges of the ground state are given by  \\
\begin{center}
\begin{tabular}{|l|l|l|l|}
\hline
&$Q_\fa$&$Q_\epsilon$&$F_j$\\\hline
$|+\rangle$&0&+2&+1\\\hline
$|-\rangle$&0&+2&-1\\
\hline
\end{tabular}
\end{center}

Then, using the fact that $\bar\lambda_2$ has charge $-4$ under $Q_\epsilon$ with all other charges annihilating $\bar\lambda_1,\bar\lambda_2$, we see that the  
charges evaluated on the different states are given by 
\begin{align}\begin{split}\label{statecharges}
\left(\begin{array}{c}
a  Q_a \\{\rm m}_f F_f\\\epsilon Q_\epsilon\\(-1)^F
\end{array}\right)\bar\lambda_1|+\rangle
=\left(\begin{array}{c}
0\\+{\rm m}_f\\ +2\epsilon\\+1
\end{array}\right)\bar\lambda_1|+\rangle
\quad,\quad \left(\begin{array}{c}
a  Q_a \\{\rm m}_f F_f\\\epsilon Q_\epsilon\\(-1)^F
\end{array}\right)\bar\lambda_2|-\rangle
=\left(\begin{array}{c}
0\\-{\rm m}_f\\-2\epsilon\\+1
\end{array}\right)\bar\lambda_2|-\rangle~, \\
\left(\begin{array}{c}
a  Q_a \\{\rm m}_f F_f\\\epsilon Q_\epsilon\\(-1)^F
\end{array}\right)\bar\lambda_2|+\rangle
=\left(\begin{array}{c}
0\\+{\rm m}_f\\ -2\epsilon\\+1
\end{array}\right)\bar\lambda_2|+\rangle
\quad,\quad \left(\begin{array}{c}
a  Q_a \\{\rm m}_f F_f\\\epsilon Q_\epsilon\\(-1)^F
\end{array}\right)\bar\lambda_1|-\rangle
=\left(\begin{array}{c}
0\\-{\rm m}_f\\+2\epsilon\\+1
\end{array}\right)\bar\lambda_1|-\rangle~.
\end{split}\end{align}

We are interested in the contribution of these states to the ground state index:
\be\label{SQMGSI}
I_{\CH_0}\Big{|}_{\CM_C}=I_{asymp}=\lim_{\beta\to \infty}I_{\CH_{\CM_C}}(-1)^F e^{-\frac{\beta}{2}\{\CQ,\CQ\}+\fa Q_a+ \epsilon_+ Q_\epsilon+\sum_f m_f F_f}~. 
\ee
By using the fact that BPS states are annihilated by $\CQ$ and the charges \eqref{statecharges}, the asymptotic Coulomb branch states as in \eqref{U1BPS} give a contribution to the Witten index:
\be
I_{asymp}=\begin{cases} e^{-\sum_f m_f-2\epsilon}+e^{\sum_f m_f +2\epsilon}=2 \cosh\left(\sum_f m_f+2\epsilon\right)&\xi>0\\e^{-\sum_f m_f+2\epsilon}+e^{\sum_f m_f -2\epsilon}=2 \cosh\left(\sum_f m_f-2\epsilon\right)&\xi<0\end{cases}
=Z_{mono}^{(extra)}. 
\ee
This is exactly the contribution $Z_{mono}^{(extra)}(1,0)$. 

\subsection{1D Wilson Lines}

We can additionally add supersymmetric Wilson lines to the SQM. These are labeled by a parameter $q$ that is quantized $q\in \IZ+\frac{{\rm N}_f}{2}$. This adds a term to the total Lagrangian:
\be
L_{Wilson}=- q(v_t+\sigma)~. 
\ee
Note that this is supersymmetric due to the fact that $\delta v_t=-\delta \sigma$. This only changes the above analysis by changing the gauge invariance condition (recall that we solved the condition $(H+Z+\sigma Q_{\rm Gauge})|\Psi\rangle=0$):
\be
(Q_{\rm Gauge}-q)|\Psi\rangle=0~. 
\ee
This only changes the choice of matter ground states. Let us consider the ${\rm N}_f$ fundamental hypermultiplet theory. Here the gauge invariance condition is given by equation 
\be
0=q-
\begin{cases}
\barn_1+\barn_2+\tilden_1+\tilden_2+\left(2-\frac{{\rm N}_f}{2}\right)+\sum_{j=1}^{{\rm N}_f} (1-f_j)&\sigma>a+\epsilon\\
-n_1-n_2-\bar\tilden_1-\bar\tilden_2-\left(2-\frac{{\rm N}_f}{2}\right)-\sum_{j=1}^{{\rm N}_f} f_j&\sigma<-a-\epsilon
\end{cases}
\ee
Therefore, for ${\rm N}_f<4$, we only have solutions for $q=\pm \frac{4-{\rm N}_f}{2}=\pm q_{crit}$ in the $\pm \sigma>a+\epsilon$ region.

Once we have the existence of the matter ground states, the analysis for the vector-multiplet part of the states carries over from the ${\rm N}_f=4$ theories. This leads to the contribution to the ground state index \eqref{SQMGSI} from the asymptotic states
\begin{align}\begin{split}
I_{asymp}=\begin{cases}
e^{\sum_f m_f+2\epsilon_+} & q=q_{crit}\\
e^{-\sum_f m_f-2\epsilon_+} & q=-q_{crit}\\
0&{\rm else}
\end{cases}
\end{split}\end{align}
for $\xi>0$ and 
\be
I_{asymp}=\begin{cases}
e^{\sum_f m_f-2\epsilon_+}& q=q_{crit}\\
e^{-\sum_f m_f+2\epsilon_+}& q=-q_{crit}\\
0&{\rm else}
\end{cases}
\ee
for $\xi<0$. 

Note that here the fermion number is always even due to the sign of the mass term of the Fermi-multiplets which is determined by the sign of $\sigma$. This relies on the fact that we are working in the limit where $\sigma>>{\rm m}_f$, $\forall f$. 

\section{$U(1)^3$ $\CN=(0,4)$ SQM Analysis}
\label{app:B}

In this appendix we will  analyze the ground states on the Coulomb branch of the $\CN=(0,4)$ SQM with gauge group $U(1)_1\times U(1)_2\times U(1)_3$ corresponding to the monopole bubbling term $Z_{mono}(\fa,m_f,\epsilon;2,1)$. Here we have three $U(1)$ vector multiplets $(\sigma_i,\lambda^A_i,M^r_i)$ where $i=1,2,3$, two fundamental hypermultiplets $(\phi^A_i,\psi_{i,I})$ where $i=1,2$, and two bifundamental hypermultiplets $(\underline{\phi}_i^A,\underline{\psi}_{i,I})$ where $i=1,2$. Additionally, dependent on the specific 4D theory, we have up to 4 fundamental (short) Fermi-multiplets $(\eta_i, F_i)$ and 3 adjoint valued hypermultiplets $(\rho^A_i, \chi_{i,I})$. The quivers for the bubbling SQMs are given by:

\begin{center}\begin{tikzpicture}[node distance=2cm,cnode/.style={circle,draw,thick,minimum size=10mm},snode/.style={rectangle,draw,thick,minimum size=10mm}]
\node[cnode] (1) {1};
\node[cnode] (2) [right of=1]{1};
\node[cnode] (3) [right of=2]{1};
\node[snode] (4) [below of=1,yshift=0.3cm]{1};
\node[snode] (6) [below of=3,yshift=0.3cm]{1};
\node[snode] (5) [above of=2,yshift=-0.2cm]{N$_f$};
\draw[-] (1) -- (2);
\draw[-] (3) -- (2);
\draw[dashed,thick] (2) -- (5);
\draw[-] (4) -- (1);
\draw[-] (6) -- (3);
\end{tikzpicture}
\qquad
\begin{tikzpicture}[node distance=2cm,cnode/.style={circle,draw,thick,minimum size=10mm},snode/.style={rectangle,draw,thick,minimum size=10mm}]
\node[cnode] (1) {1};
\node[cnode] (2) [right of=1]{1};
\node[cnode] (3) [right of=2]{1};
\node[snode] (4) [below of=1,yshift=0.3cm]{1};
\node[snode] (6) [below of=3,yshift=0.3cm]{1};
\draw[-] (1) -- (2);
\draw[-] (3) -- (2);
\draw[-] (4) -- (1);
\draw[-] (6) -- (3);
\end{tikzpicture}
\end{center}
in the case of the theory with ${\rm N}_f$ fundamental hypermultiplets (given by a $\CN=(0,4)$ quiver SQM) and the $\CN=2^\ast$ theory (given by a $\CN=(4,4)$ quiver SQM). 

The total Lagrangian again decomposes as
\be
L=L_{univ}+L_{theory}
\ee
where $L_{univ}$ is the universal term describing 4D SYM field content and $L_{theory}$ depends on the matter content of the 4D theory. The universal term decomposes as
\be
L_{univ}=L_{vec}+L_{hyp}+L_{bf}+L_{FI}~.
\ee
After introducing notation analagous to that of Appendix \ref{app:A}, the field content of this theory and their charges are given by 

\begin{center}
\begin{tabular}{|c|c|c|c|c|c|c|c|c|}
\hline
Lagrangian Term&$\CN=(0,4)$ Multiplet&Fields&$Q_{G}^{(1)}$&$Q_{G}^{(2)}$&$Q_{G}^{(3)}$&$Q_\fa$&$Q_\epsilon$&$F_j$\\\hline\hline
Universal
& Vector-&$\sigma_1$&0&0&0&0&0&0\\
&&$\lambda^1_1$&0&0&0&0&0&0\\
&&$\lambda^2_1$&0&0&0&0&4&0\\
&&$\sigma_2$&0&0&0&0&0&0\\
&&$\lambda^1_2$&0&0&0&0&0&0\\
&&$\lambda^2_2$&0&0&0&0&4&0\\
&&$\sigma_3$&0&0&0&0&0&0\\
&&$\lambda^1_3$&0&0&0&0&0&0\\
&&$\lambda^2_3$&0&0&0&0&4&0\\
\hline Universal
& Fund. Hyper-& 
$\phi_1$&1&0&0&-1&1&0\\
&&$\tilde\phi_1$&-1&0&0&1&1&0\\
&&$\phi_2$&0&0&1&1&1&0\\
&&$\tildephi_2$&0&0&-1&1&1&0\\
&&$\psi_1$&1&0&0&-1&1&0\\
&&$\tilde\psi_1$&-1&0&0&1&1&0\\
&&$\psi_2$&0&0&1&1&1&0\\
&&$\tilde\psi_2$&0&0&-1&1&1&0\\
\hline Universal
& Bifund. Hyper-& 
$\uphi_1$&-1&1&0&0&1&0\\
&&$\tilde\uphi_1$&1&-1&0&0&1&0\\
&&$\uphi_2$&0&-1&1&0&1&0\\
&&$\tilde\uphi_2$&0&1&-1&0&1&0\\
&&$\upsi_1$&-1&1&0&0&1&0\\
&&$\tilde\upsi_1$&1&-1&0&0&1&0\\
&&$\upsi_2$&0&-1&1&0&1&0\\
&&$\tilde\upsi_2$&0&1&-1&0&1&0\\
\hline\hline
N$_f$ Fund. Hyper- 
& Fund. Fermi-&$\eta_f$&0&1&0&0&0&-2\\\hline\hline
$\CN=2^\ast$ Theory
& Twisted Chiral&$\rho_1$&0&0&0&0&1&1\\
&&$\tilde\rho_1$&0&0&0&0&-1&1\\
&&$\chi_1$&0&0&0&0&0&1\\
&&$\tilde\chi_1$&0&0&0&0&0&-1\\
&&$\rho_2$&0&0&0&0&1&1\\
&&$\tilde\rho_2$&0&0&0&0&-1&1\\
&&$\chi_2$&0&0&0&0&0&1\\
&&$\tilde\chi_2$&0&0&0&0&0&-1\\
&&$\rho_3$&0&0&0&0&1&1\\
&&$\tilde\rho_3$&0&0&0&0&-1&1\\
&&$\chi_3$&0&0&0&0&0&1\\
&&$\tilde\chi_3$&0&0&0&0&0&-1\\
\hline $\CN=2^\ast$ Theory
& Fund. Fermi&$\eta_1$&1&0&0&1&-1&0\\
&&$\tilde\eta_1$&-1&0&0&1&1&0\\
&&$\eta_2$&0&0&1&1&1&0\\
&&$\tilde\eta_2$&0&0&-1&-1&1&0\\\hline
\end{tabular}
\end{center}
In terms of these component fields, the Lagrangian terms can be written as 
\begin{align}\begin{split}
L_{vec}+L_{FI}&=\sum_{i=1}^3\frac{1}{2e^2}\left[ (\partial_t \sigma_i)^2+i\left( \bar\lambda_{i,1}\partial_t \lambda^{1}_i+\lambda^{1}_i\partial_t \bar\lambda_{i,1}\right)+2i\bar\lambda_{i,2}(\partial_t-4i \epsilon)\lambda^2_i-(M_{i,r})^2-2e^2 \xi_i M_{i,3}\right]~,\\
e^2 L_{hyp}&=|D_{t,1}\phi^{A}_1|^2-|\sigma_1 \phi^{A}_1|^2+ \bar\phi_{1,A}(\sigma^r)_{~B}^{A} M_{1,r} \phi^B_1
	+\ihalf \Big(\bar\psi_1 (D_{t,1}+i\sigma_1)\psi_1\\&+
	\psi_1(\tilde{D}_{t,1}-i\sigma_1)\bar\psi_1+\bar\tildepsi_1(\tilde{D}_{t,1}-i\sigma_1)\tildepsi_1
	+\tildepsi_1(D_{t,1}+i\sigma_1)\bar\tildepsi_1\Big)\\
&+\frac{i}{\sqrt{2}} \Big(\barphi_{1,A}\lambda^{A}_1\psi_1-\bar\psi_1\bar\lambda_{1,A}\phi^{A}_1+\bar\phi_{1,A}\bar\lambda_1^{A}\bar\tildepsi_1-\tildepsi_1\lambda_{1,A}\phi_1^{A}\Big)\\&
+|D_{t,3}\phi^{A}_2|^2-|\sigma_3 \phi^{A}_1|^2+ \bar\phi_{2,A}(\sigma^r)_{~B}^{A} M_{3,r} \phi^B_2
+\ihalf \Big(\bar\psi_2 (D_{t,3}+i\sigma_3)\psi_2\\&
+\psi_2(\tilde{D}_{t,3}-i\sigma_3)\bar\psi_2+\bar\tildepsi_2(\tilde{D}_{t,3}-i\sigma_3)\tildepsi_2+\tildepsi_2(D_{t,3}+i\sigma_3)\bar\tildepsi_2\Big)\\
&+\frac{i}{\sqrt{2}} \Big(\barphi_{2,A}\lambda^{A}_3\psi_2-\bar\psi_2\bar\lambda_{3,A}\phi^{A}_2+\bar\phi_{2,A}\bar\lambda_3^{A}\bar\tildepsi_2-\tildepsi_2\lambda_{3,A}\phi_2^{A}\Big)
~,\\
e^2L_{bf}&=
|D_{t,21}\uphi^{A}_1|^2-|\sigma_{21} \uphi^{A}_1|^2+ \bar\uphi_{1,A}(\sigma^r)_{~B}^{A} M_{21,r} \uphi^B_1
+\ihalf \Big(\bar\psi_1 (D_{t,21}+i\sigma_{21})\psi_1\\&
+\psi_1(\tilde{D}_{t,21}-i\sigma_{21})\bar\psi_1+\bar\tildepsi_1(\tilde{D}_{t,21}-i\sigma_{21})\tildepsi_1+\tildepsi_1(D_{t,21}+i\sigma_{21})\bar\tildepsi_1\Big)\\
&+\frac{i}{\sqrt{2}} \Big(\underline{\barphi}_{1,A}\lambda^{A}_{21}\psi_1-\bar\psi_1\bar\lambda_{21,A}\uphi^{A}_1+\bar\uphi_{1,A} \bar\lambda_{21}^{A}\bar\tildepsi_1-\tildepsi_1\lambda_{21,A}\uphi_1^{A}\Big)
+|D_{t,32}\uphi^{A}_2|^2\\&-|\sigma_{32} \uphi^{A}_2|^2+ \bar\uphi_{2,A}(\sigma^r)_{~B}^{A} M_{32,r} \uphi^B_2
+\ihalf \Big(\bar\psi_2 (D_{t,32}+i\sigma_{32})\psi_2\\&
+\psi_2(\tilde{D}_{t,32}-i\sigma_{32})\bar\psi_2+\bar{\tilde{\upsi}}_2(\tilde{D}_{t,32}-i\sigma_{32})\tilde\upsi_2
+\tilde\upsi_2(D_{t,32}+i\sigma_{32})\bar{\tilde{\upsi}}_2\Big)\\
&+\frac{i}{\sqrt{2}} \Big(\underline{\barphi}_{2,A}\lambda^{A}_{32}\psi_2-\bar\psi_2\bar\lambda_{32,A}\uphi^{A}_2 +\bar\uphi_{2,A}\bar\lambda_{32}^{A}\bar{\tilde{\upsi}}_2-\tilde\upsi_2\lambda_{32,A}\uphi_2^{A}\Big)~.
\end{split}\end{align}
where we used the notation
\be
(D_{t,ij}\pm i\sigma_{ij})= \partial_t +2i(v_{i,t}-v_{j,t})\pm i (\sigma_i-\sigma_j)\quad,\quad \lambda^A_{ij}=\lambda^A_i-\lambda^A_j~. 
\ee

$L_{theory}$ can have contributions from terms of the form
\be
L_{theory}=L_{Fermi}+L_{adj\,hyp}~.
\ee
In components, this can be written
\begin{align}\begin{split}
e^2L_{Fermi}&=\sum_j\left[\ihalf \left(\bar\eta_i (D_{t,2}+i \sigma_2)\eta_i+\eta_i(\tilde{D}_{t,2}-i \sigma_2)\bar\eta_i\right)+|G_i|^2+{\rm m}_j[\bar\eta_j,\eta_j]\right]~,\\
L_{adj\,hyp}&=\sum_i\frac{1}{e^2}\Bigg[|\partial_t \rho_i^A|^2+({\rm m}+\epsilon)|\rho_i|^2+({\rm m}-\epsilon)^2|\tilde\rho_i|^2\\&\qquad\qquad\quad
+\sum_I \left(\ihalf(\bar\chi_{i,I} \partial_t \chi_{i,I}+\chi_{i,I}\partial_t \bar\chi_{i,I})-\frac{\rm m }{2}[\bar\chi_{i,I},\chi_{i,I}]\right)\Bigg]~,
\end{split}\end{align}
where $\chi_{i,I}=\left(
\chi_i~,~\bar{\tilde{\chi}}_i
\right).$ 

Now we can go to the Hamiltonian formalism and integrate out the auxiliary fields. We will again scale our  fermionic fields such that they obey  anti-commutation relations of the form
\be
\{\bar\psi_i,\psi_j\}=\delta_{ij}\quad, \quad\{\bar\eta_i,\eta_j\}=\delta_{ij}~.
\ee
Now the total Hamiltonian will be form the form
\be
H=H_{vm}+U+H_{matter}+H_I+\sum_i v_{t,i}Q_{\rm Gauge}^{(i)}~,
\ee
where 
\begin{align}\begin{split}
&H_{vm}=\frac{e_1^2p_{\sigma 1}^2}{2}+\frac{e_2^2p_{\sigma2}^2}{2}
+\frac{e_3^2p_{\sigma3}^2}{2}-4\epsilon\sum_i\bar\lambda_{i,2}\lambda^2_i+H_{adj\,hyp}~,\\
&H_{matter}=e^2\left[|\pi_1|^2+|\pi_2|^2+|\tilde\pi_1|^2+|\tilde\pi_2|^2+|\upi_1|^2+|\upi_2|^2+|\tilde\upi_1|^2+|\tilde\upi_2|^2\right]\\&
+\frac{1}{e^2}\left(\omega_1^2|\phi_1|^2+\omega_2^2|\phi_2|^2+\tilde\omega_1^2|\tilde\phi_1|^2+\tilde\omega_2^2|\tilde\phi_2|^2+\underline\omega_1^2|\uphi_1|^2+\underline\omega_2^2|\uphi_2|^2+\underline{\tilde{\omega}}_1^2|\tilde\uphi_1|^2+\underline{\tilde{\omega}}_2^2|\tilde\uphi_2|^2\right)\\
&+\frac{1}{2}\left((\sigma_1-a+\epsilon)[\bar\psi_1,\psi_1]-(\sigma_1-a-\epsilon)[\bar\tildepsi_1,\tildepsi_1]+(\sigma_3+a+\epsilon)[\bar\psi_2,\psi_2]-(\sigma_3+a+\epsilon)[\bar\tildepsi_2,\tildepsi_2]\right)
\\&
-\half\left((\sigma_{21}+\epsilon)[\bar\upsi_1,\upsi_1]-(\sigma_{21}-\epsilon)[\bar\utpsi_1,\utpsi_1]+(\sigma_{21}+\epsilon)[\bar\upsi_2,\upsi_2]-(\sigma_{32}-\epsilon)[\bar\utpsi_2,\utpsi_2]\right) +H_{Fermi}~,
\end{split}\end{align}
where 
\begin{align}\begin{split}
&\omega_1=|\sigma_1-a+\epsilon|\quad, \quad \tilde\omega_1=|\sigma_1-a-\epsilon|\quad, \quad \omega_2=|\sigma_3+a+\epsilon|\quad, \quad \tilde\omega_2=|\sigma_3+a-\epsilon|~,\\
&\underline\omega_1=|\sigma_{21}+\epsilon|\quad, \quad \tilde{\underline{\omega}}_1=|\sigma_{21}-\epsilon|\quad,\quad\underline\omega_2=|\sigma_{32}+\epsilon|\quad, \quad\tilde{\underline{\omega}}_2=|\sigma_{32}-\epsilon|~, \\& 
\omega_{i+2}=|{\rm m}+\epsilon|\quad, \quad \tilde\omega_{i+2}|{\rm m}-\epsilon|\qquad, \quad {\rm for}~ a,\epsilon>0~.
\end{split}\end{align}
and 
\begin{align}\begin{split}\label{HFermiNf}
H_{Fermi}=\half\sum_{j=1}^{{\rm N}_f}(\sigma_2-2{\rm m}_f)[\bar\eta_j,\eta_j]~,
\end{split}\end{align}
or
\begin{align}\begin{split}\label{HFermiN2star}
H_{Fermi}=&(\sigma_1-a+m))[\bar\eta_1,\eta_1]-(\sigma_1-a-m))[\bar{\tilde{\eta}}_1,\tilde\eta_1]\\&
+(\sigma_3+a+m))[\bar\eta_2,\eta_2]-(\sigma_3+a-m))[\bar{\tilde{\eta}}_2,\tilde\eta_2]~,
\end{split}\end{align}
for the 4D theories with ${\rm N}_f$-hypermultiplets or for the $\CN=2^\ast$ theory respectively and $H_{adj\, hyp}$ is only included for the $\CN=2^\ast$ theory and is given by
\be
H_{adj\,hyp}=\sum_{i=1}^3 \left[e^2|\pi_{i+2}|^2+e^2|\tilde\pi_{i+2}|^2+\frac{\omega_{i+2}^2}{e^2}|\rho_{i}|^2+\frac{\tilde\omega_{i+2}^2}{e^2}|\tilde\rho_i|^2+\frac{\rm m}{2}\sum_I[\bar\chi_{i,I},\chi_{i,I}]\right]~. 
\ee
Additionally, 
\begin{align}\begin{split}
&U=\frac{1}{2e^2}\left(|\phi_1|^2-|\tilde\phi_1|^2-|\uphi_1|^2+|\tilde\uphi_1|^2-e^2\xi_1\right)^2+\frac{1}{2e^2}\left(|\uphi_1|^2-|\tilde\uphi_1|^2-|\uphi_2|^2+|\tilde\uphi_2|^2-e^2\xi_2\right)^2\\
&+\frac{1}{2e^2}\left(|\uphi_2|^2-|\tilde\uphi_2|^2+|\phi_2|^2-|\tilde\phi_2|^2-e^2\xi_3\right)^2\\&
+\frac{1}{e^2}|\phi_1\tilde\phi_1-\uphi_1\tilde\uphi_1|^2+\frac{1}{e^2}|\uphi_1\tilde\uphi_1-\uphi_2\tilde\uphi_2|^2+\frac{1}{e^2}|\uphi_2\tilde\uphi_2+\phi_2\tilde\phi_2|^2\\&
H_I=\frac{i}{\sqrt{2}}  \left(\bar\phi_1\lambda_1^1\psi_1+\tildephi_1\lambda_1^2\psi_1+\tildephi_1\bar\lambda_{1,1}\bar\tildepsi_1-\barphi_1 \bar\lambda_{1,2}\bar\tildepsi_1-\bar\psi_1\bar\lambda_{1,1}\phi_1-\bar\psi_1\bar\lambda_{1,2}\bar\tildephi_1-\tildepsi_1\lambda_1^1\bar\tildephi_1+\tildepsi_1\lambda_1^2\phi_1\right)\\
&
+\frac{i}{\sqrt{2}}  \left(\bar\phi_2\lambda_3^1\psi_2+\tildephi_2\lambda_3^2\psi_2+\tildephi_2\bar\lambda_{3,1}\bar\tildepsi_2-\barphi_2 \bar\lambda_{3,2}\bar\tildepsi_2-\bar\psi_2\bar\lambda_{3,1}\phi_2-\bar\psi_2\bar\lambda_{3,2}\bar\tildephi_2-\tildepsi_2\lambda_3^1\bar\tildephi_2+\tildepsi_2\lambda_3^2\phi_2\right)\\
&
+\frac{i}{\sqrt{2}} \left(\bar\uphi_1\lambda_{21}^1\upsi_1+\utphi_1\lambda_{21}^2\upsi_1+\utphi_1\bar\lambda_{21,1}\bar\utpsi_1-\bar\uphi_1 \bar\lambda_{21,2}\bar\utpsi_1-\bar\upsi_1\bar\lambda_{21,1}\uphi_1-\bar\upsi_1\bar\lambda_{21,2}\bar\utphi_1-\utpsi_1\lambda_{21}^1\bar\utphi_1+\utpsi_1\lambda_{21}^2\uphi_1\right)\\
&
+\frac{i}{\sqrt{2}}  \left(\bar\uphi_2\lambda_{32}^1\upsi_2+\utphi_2\lambda_{32}^2\upsi_2+\utphi_2\bar\lambda_{32,1}\bar\utpsi_2-\bar\uphi_2 \bar\lambda_{32,2}\bar\utpsi_2-\bar\upsi_2\bar\lambda_{32,1}\uphi_2-\bar\upsi_2\bar\lambda_{32,2}\bar\utphi_2-\utpsi_2\lambda_{32}^1\bar\utphi_2+\utpsi_2\lambda_{32}^2\uphi_2\right)~.
\end{split}\end{align}
Now by identifying $\phi_{i+2}=\rho_i$, we can define the operators
\begin{align}\begin{split}
&a_i=\frac{1}{\sqrt{2}e}\left(\omega_i \phi_i+\frac{i e^2 \bar\pi_i}{\omega_i}\right)\quad, \quad \bara_i=\frac{1}{\sqrt{2}e}\left(\omega_i\bar\phi_i+\frac{i e^2 \pi_i}{\omega_i}\right)~,\\&
\tildea_{ i}=\frac{1}{\sqrt{2}e}\left(\tilde\omega_i\tilde\phi_i+\frac{i e^2\bar{\tilde{\pi}}_i}{\omega_i}\right)\quad, \quad \bar\tildea_{ i}=\frac{1}{\sqrt{2}e}\left(\tilde\omega_i\bar\tildephi_i+\frac{i e^2 \tilde{\pi}_i}{\omega_i}\right)~,
\end{split}\end{align}
for $i=1,...,5$ and 
\begin{align}\begin{split}
&\ua_{i}=\frac{1}{\sqrt{2}e}\left({\underline{\omega}}_i \uphi_i+\frac{i e^2 \bar\upi_i}{{\underline{\omega}}_i}\right)\quad, \quad \bar\ua_{ i}=\frac{1}{\sqrt{2}e}\left({\underline{\omega}}_i\bar\uphi_i+\frac{i e^2 \upi_i}{{\underline{\omega}}_i}\right)~,\\&
\tildeua_{ i}=\frac{1}{\sqrt{2}e}\left(\tilde{\underline{\omega}}_i\tilde\uphi_i+\frac{i e^2\bar{\tilde{\upi}}_i}{{\underline{\omega}}_i}\right)\quad, \quad \bar\tildeua_{ i}=\frac{1}{\sqrt{2}e}\left(\tilde{\underline{\omega}}_i\bar\tildephi_i+\frac{i e^2 \tilde{\upi}_i}{{\underline{\omega}}_i}\right)~,
\end{split}\end{align}
for $i=1,2$. 

Now we can write the matter Hamiltonian as
\begin{align}\begin{split}
H_{matter}&=\omega_1(a_{1}^\dagger a_{1}+ \bar{a}_{1}^\dagger \bar{a}_{1}+1)+(\sigma_1-a+\epsilon)(\bar\psi_1\psi_1-\half)\\
&+\tilde\omega_1(\tildea_{1}^\dagger \tildea_{1}+\bar{\tilde{a}}_{1}^\dagger \bar{\tilde{a}}_{1}+1)-(\sigma_1-a-\epsilon)(\bar\tildepsi_1\tildepsi_1-\half)\\&
+\omega_2(a_{2}^\dagger a_{2}+ \bar{a}_{2}^\dagger \bar{a}_{2}+1)+(\sigma_3+a+\epsilon)(\bar\psi_2\psi_2-\half)\\&
+\tilde\omega_2(\tildea_{2}^\dagger \tildea_{2}+\bar{\tilde{a}}_{2}^\dagger \bar{\tilde{a}}_{2}+1)-(\sigma_3+a-\epsilon)(\bar\tildepsi_2\tildepsi_2-\half)\\&
+\underline\omega_1(\underline{a}_{1}^\dagger \underline{a}_{1}+\underline{\bar{a}}_{1}^\dagger \underline{\bar{a}}_{1}+1)+(\sigma_2-\sigma_1+\epsilon)\big(\bar\upsi_1\upsi_1-\half\big)\\&
+\tilde{\underline{\omega}}_1(\tilde{a}_{1}^\dagger\tilde{a}_{1}+\bar{\tilde{a}}_{1}^\dagger \bar{\tilde{a}}_{1}+1)-(\sigma_2-\sigma_1-\epsilon)\big(\bar\utpsi_1\utpsi_1-\half)\\&
+\underline{\omega}_2(\underline{a}_{2}^\dagger \underline{a}_{2}+\underline{\bar{a}}_{2}^\dagger \underline{\bar{a}}_{2}+1)+(\sigma_3-\sigma_2+\epsilon)\big(\bar\upsi_2\upsi_2-\half)\\&
+\tilde{\underline{\omega}}_2(\tilde{a}_{2}^\dagger\tilde{a}_{2}+\bar{\tilde{a}}_{2}^\dagger \bar{\tilde{a}}_{2}+1)-(\sigma_3-\sigma_2-\epsilon)\big(\bar\utpsi_2\utpsi_2-\half)\\&
+H_{Fermi}~.
\end{split}\end{align}
here again $H_{Fermi}$ is given by \eqref{HFermiNf} or \eqref{HFermiN2star} for the case of the corresponding 4D theory having  ${\rm N}_f$ fundamental hypermultiplets or being the $\CN=2^\ast$ theory respectively. 

These operators also allow us to write  $Q_{\rm Gauge}^{(i)}$ simply as
\begin{align}\begin{split}
Q_{\rm Gauge}^{(1)}&=-\left(a_{1}^\dagger a_{1}-\bar{a}_{1}^\dagger \bar{a}_{1}-\tildea_{1}^\dagger \tildea_{1}+\bar{\tilde{a}}_{1}^\dagger \bar{\tilde{a}}_{1}-\underline{a}_{1}^\dagger \underline{a}_{1}+\underline{\bar{a}}_{1}^\dagger \underline{\bar{a}}_{1}+\tilde{a}_{1}^\dagger\tilde{a}_{1}-\bar{\tilde{a}}_{1}^\dagger\bar{\tilde{a}}_{1}\right)\\&\qquad
-\left(\bar\psi_1\psi_1-\bar\tildepsi_1\tildepsi_1-\bar\upsi_1\upsi_1+\bar\utpsi_1\utpsi_1\right)+Q_{theory}^{(1)}~,\\
Q_{\rm Gauge}^{(2)}&=-\left(\underline{a}_{1}^\dagger \underline{a}_{1}-\underline{\bar{a}}_{1}^\dagger \underline{\bar{a}}_{1}-\tilde{a}_{1}^\dagger \tilde{a}_{1}+\bar{\tilde{a}}_{1}^\dagger \bar{\tilde{a}}_{1}\right)+\left(\underline{a}_{2}^\dagger \underline{a}_{2}-\underline{\bar{a}}_{2}^\dagger \underline{\bar{a}}_{2}-\tilde{a}_{2}^\dagger \tilde{a}_{2}+\bar{\tilde{a}}_{2}^\dagger \bar{\tilde{a}}_{2}\right)\\&\qquad
-(\bar\upsi_1\upsi_1-\bar\utpsi_1\utpsi_1)+(\bar\upsi_2\upsi_2-\bar\utpsi_2\utpsi_2)+Q_{theory}^{(2)}~,\\
Q_{\rm Gauge}^{(3)}&=-\left(a_{2}^\dagger a_{2}-\bar{a}_{2}^\dagger \bar{a}_{2}-\tildea_{2}^\dagger \tildea_{2}+\bar{\tilde{a}}_{2}^\dagger \bar{\tilde{a}}_{2}\right)-\left(\underline{a}_{2}^\dagger \underline{a}_{2}-\underline{\bar{a}}_{2}^\dagger \underline{\bar{a}}_{2}-\tilde{a}_{2}^\dagger \tilde{a}_{2}+\bar{\tilde{a}}_{2}^\dagger \bar{\tilde{a}}_{2}\right)\\&\qquad
-(\bar\psi_2\psi_2-\bar\tildepsi_2\tildepsi_2)-(\bar\upsi_2\upsi_2-\bar\utpsi_2\utpsi_2)+Q_{theory}^{(3)}~. 
\end{split}\end{align}
where 
\begin{align}\begin{split}
Q_{theory}^{(2)}=-\half\sum_{j=1}^{{\rm N}_f}[\bar\eta_j,\eta_j]\quad, \quad Q_{theory}^{(1)}=Q_{theory}^{(3)}=0
\end{split}\end{align}
for the theory with ${\rm N}_f$ fundamental hypermultiplets and 
\begin{align}\begin{split}
Q_{theory}^{(1)}=-\half[\bar\eta_1,\eta_1]+ \half[\bar{\tilde{\eta}}_1,\tilde\eta_1]\quad, \quad
Q_{theory}^{(3)}=-\half[\bar\eta_2,\eta_2]+ \half[\bar{\tilde{\eta}}_2,\tilde\eta_2]\quad, \quad Q_{theory}^{(2)}=0~,
\end{split}\end{align}
for the case of the $\CN=2^\ast$ theory.

Let us again pick a basis of states for our Hilbert space
\begin{align}\begin{split}
&\Bigg|(n_i, \barn_i,\tilden_i,\bar\tilden_i,m_i,\tildem_i)~;~(\un_i,\bar\un_i,\utn_i,\bar\utn_i)~;~ (f_j)\Bigg\rangle=\\&
\qquad\left(\prod_{i=1}^5 \left(a_{ i}^\dagger\right)^{n_i} \left(\bara_{i}^\dagger\right)^{\barn_i}\left(\tildea_{ i}^\dagger\right)^{\tilden_i}\left(\bar\tildea_{ i}^\dagger\right)^{\bar\tilden_i}\bar\psi_i^{m_i}\bar\tildepsi_i^{\tildem_i} \right)\times\\
&\qquad
\left(\prod_{i=1}^2 \left(\ua_{ i}^\dagger\right)^{n_i} \left(\bar\ua_{i}^\dagger\right)^{\barn_i}\left(\tilde\ua_{ i}^\dagger\right)^{\tilden_i}\left(\bar{\tilde{\ua}}_{ i}^\dagger\right)^{\bar\tilden_i}\bar\upsi_i^{\um_i}\bar{\tilde{\upsi}}_i^{\tilde\um_i} \right)\times\left(
\prod_{j=1}^4 \bar\eta_j^{f_j}\right)\Big{|}0\Big{\rangle}~,
\end{split}\end{align}
for the case of 4D fundamental matter or 
\begin{align}\begin{split}
&\Bigg|(n_i, \barn_i,\tilden_i,\bar\tilden_i,m_i,\tildem_i)~;~(\un_i,\bar\un_i,\utn_i,\bar\utn_i)~;~ (f_j)\Bigg\rangle=\\&
\qquad\left(\prod_{i=1}^5 \left(a_{ i}^\dagger\right)^{n_i} \left(\bara_{i}^\dagger\right)^{\barn_i}\left(\tildea_{ i}^\dagger\right)^{\tilden_i}\left(\bar\tildea_{ i}^\dagger\right)^{\bar\tilden_i}\bar\psi_i^{m_i}\bar\tildepsi_i^{\tildem_i} \right)\times\\
&\qquad
\left(\prod_{i=1}^2 \left(\ua_{ i}^\dagger\right)^{n_i} \left(\bar\ua_{i}^\dagger\right)^{\barn_i}\left(\tilde\ua_{ i}^\dagger\right)^{\tilden_i}\left(\bar{\tilde{\ua}}_{ i}^\dagger\right)^{\bar\tilden_i}\bar\upsi_i^{\um_i}\bar{\tilde{\upsi}}_i^{\tilde\um_i} \right)\times\left(
\prod_{j=1}^2 \bar\eta_j^{f_j}\bar{\tilde{\eta}}^{\tilde{f}_j}\right)\Big{|}0\Big{\rangle}~,
\end{split}\end{align}
for the case of $\CN=2^\ast$ theory where we have identified $\chi_{1,i}=\psi_{i+2}$ and $\chi_{2,i}=\tilde\psi_{i+2}$ and the vacuum state is defined as 
\begin{align}\begin{split}
&a_i|0\rangle=\bara_i|0\rangle=\tildea_i|0\rangle=\bar\tildea_i|0\rangle=\ua_i|0\rangle=\bar\ua_i|0\rangle=\tilde\ua_i|0\rangle=\bar{\tilde{\ua}}_i|0\rangle=0~,\\
&\psi_i|0\rangle=\tilde\psi_i|0\rangle=\upsi_i|0\rangle=\tilde\upsi_i|0\rangle=\eta_j|0\rangle=\tilde\eta_j|0\rangle=0~.
\end{split}\end{align}
 Thus, the quantum numbers are constrained
 \be
 n_i,\barn_i,\tilden_i,\bar\tilden_i, \un_i,\bar\un_i,\tilde\un_i,\bar{\tilde{\un}}_i\in \IZ_+\quad,\quad m_i,\tildem_i,f_j=0,1~.
 \ee

Now as before we want to solve for gauge invariant BPS states. These satisfy
\be
(H-Z)|\Psi\rangle=0\quad, \quad Q_{\rm Gauge}|\Psi\rangle=0~,
\ee
where 
\be
Z=-aQ_\fa-\epsilon Q_\epsilon-{\rm m}F_{\rm m}-\sum_{j =1}^{{\rm N}_f} {\rm m}_j F_j~,
\ee
where $F_{\rm m}$ and $F_j$ are the flavor charges associated with the 4D adjoint hypermultiplet and 4D fundamental hypermultiplets respectively.

As in the $U(1)$ case, there are unique matter ground states for the regions\footnote{Recall that we are assuming $a,\epsilon>0$.}
\be
S_+:=\{\sigma_1>a+\epsilon~,~\sigma_3>-a+\epsilon~,~\sigma_2>\sigma_1+\epsilon~,~\sigma_3>\sigma_2+\epsilon\}~. 
\ee
and
\be
S_-:=\{\sigma_1<a-\epsilon~,~\sigma_3<-a-\epsilon~,~\sigma_2< \sigma_1-\epsilon,\sigma_3<\sigma_2-\epsilon\}~,
\ee
which have the quantum numbers
\begin{align}\begin{split}
S_+~&:~N_i,\bar{N}_i,\tilde{N}_i,\bar{\tilde{N}}_i,\tilde{m}_i,\underline{\tilde{m}}_1,\underline{m}_2=0~,~m_i,\underline{m}_1,\underline{\tilde{m}}_2=1~,~f_j=0~,\\
S_-~&:~N_i,\bar{N}_i,\tilde{N}_i,\bar{\tilde{N}}_i,m_i,\underline{m}_1,\tilde{\underline{m}}_2=0~,~\tilde{m}_i,\tilde{\underline{m}}_1,\underline{m}_2=1~,~f_j=1~,
\end{split}\end{align}
where here we use the notation $\{N_i,\bar{N}_i,\tilde{N}_i,\bar{\tilde{N}}_i,M_i,\tilde{M}_i\}_{i=1}^4$ to collectively refer to the quantum numbers of all hypermultiplets where $i=1,2$ correspond to the fundamental hypermultiplets and $i=3,4$ correspond to the $1^{st}$ and $2^{nd}$ bi-fundamental hypermultiplets respectively. 

We will denote the matter ground state wave functions in these regions as
\begin{align}\begin{split}
|\Psi_+\rangle=\delta_{S_+}|N_i,\bar{N}_i,\tilde{N}_i,\bar{\tilde{N}}_i,\tilde{m}_i,\underline{\tilde{m}}_1,\underline{m}_2=0~,~m_i,\underline{m}_1,\underline{\tilde{m}}_2=1~,~f_j=0\rangle~,\\
|\Psi_-\rangle=\delta_{S_-}|N_i,\bar{N}_i,\tilde{N}_i,\bar{\tilde{N}}_i,m_i,\underline{m}_1,\tilde{\underline{m}}_2=0~,~\tilde{m}_i,\tilde{\underline{m}}_1,\underline{m}_2=1~,~f_j=1\rangle~,
\end{split}\end{align}
where $\delta_S$ is the indicator function for the set $S$.

\subsection{Effective Hamiltonian}

In analogy with the procedure in Appendix \ref{app:A} we can compute the effective Hamiltonian by integrating out the fundamental hypermultiplet and Fermi-multiplet matter. In this SQM, the supercharge is of the form:
\begin{align}\begin{split}
Q_A&=Q_{{\rm matter},A}-Q_{{\rm vec},A}~,\\
Q_{{\rm vec},A}&=\sum_{i=1}^3\frac{e}{\sqrt{2}}\left(-i p_{\sigma i}\bar\lambda_{i,A}+M_r(\sigma^r)^{~B}_A\bar\lambda_B\right)~,
\end{split}\end{align}
and $Q_{{\rm matter},A}$ is analogous to the first terms of \eqref{eq:supercharge} which annihilate the harmonic oscillator wave functions of the matter fields. Now the effective supercharge is of the form
\be
Q_{{\rm eff},A}=\langle Q_{{\rm vec},A}\rangle=\sum_{i=1}^3 \frac{e}{\sqrt{2}}\left(-i \langle p_{\sigma i}\rangle \bar\lambda_{i,A}-\frac{1}{e^2}\langle D_i\rangle (\sigma^3)_A^{~B}\bar\lambda_B-\frac{\sqrt{2}}{e^2}\langle F_i\rangle(\sigma^+)_A^{~B}\bar\lambda_B-\frac{\sqrt{2}}{e^2}\langle \bar{F_i}\rangle (\sigma^-)_A^{~B}\bar\lambda_B\right)~,
\ee
where
\begin{align}\begin{split}\label{AuxL21}
F_{1}&=\left(\phi_1\tilde\phi_1-\uphi_1\tilde\uphi_1\right)~,\quad
F_{2}=\left(\uphi_1\tilde\uphi_1-\uphi_2\tilde\uphi_2\right)~,\quad
F_{3}=\left(\uphi_2\tilde\uphi_2+\phi_2\tilde\phi_2\right)~,\\
D_1&=\left(|\phi_1|^2-|\tilde\phi_1|^2-|\uphi_1|^2+|\tilde\uphi_1|^2-e^2\xi_1\right)~,\\
D_2&=\left(|\uphi_1|^2-|\tilde\uphi_1|^2-|\uphi_2|^2+|\tilde\uphi_2|^2-e^2\xi_2\right)~,\\
D_3&=\left(|\uphi_2|^2-|\tilde\uphi_2|^2+|\phi_2|^2-|\tilde\phi_2|^2-e^2 \xi_3\right)~.
\end{split}\end{align}

Now by using the form of \eqref{AuxL21}, we see that $\langle F_i\rangle=\langle \bar{F}_i\rangle=0$, $\forall i$. Again, due to having a non-zero $\epsilon,a$, we have broken SUSY to $\CN=(0,2)$, preserving the supercharges $Q_{\rm eff 1},\bar{Q}^1_{\rm eff}$.

We can now compute the effective Hamiltonian by squaring the supercharges
\be
H_{\rm eff}=\{\bar{Q}^1_{\rm eff},Q_{\rm eff,1}\}-Z~. 
\ee
Using the fact that Gauss's law imposes $Q_{\rm Gauge}^{(i)}=0$, $\forall i$, we have that only flavor charges contribute to the central charge. This gives rise to the central charge:
\be
Z=4\epsilon \sum_i\bar\lambda_{i,2} \lambda_i^2-6\epsilon- \sum_{f=1}^4 {\rm m}_f [\bar\eta_f,\eta_f]~. 
\ee
This gives us the full effective Hamiltonian:
\begin{align}\begin{split}
H_{\rm eff}
&= \sum_i \frac{e^2 \langle p_{\sigma i}\rangle ^2}{2}+\frac{1}{2e^2}\langle D_i\rangle^2-\frac{1}{2}[\bar\lambda_{i,1},\lambda_i^1]\partial_{\sigma_i}\langle D_i\rangle-4\epsilon \sum_i\bar\lambda_{i,2} \lambda_i^2+6\epsilon+ 2\sum_{f=1}^4{\rm m}_f  [\bar\eta_f,\eta_f]~,
\end{split}\end{align}
where
\begin{align}\begin{split}
\langle D_i\rangle
&=-e^2 \xi+\frac{e^2}{2}\begin{cases}
\frac{1}{\omega_1}-\frac{1}{\tilde\omega_1}-\frac{1}{\umega_1}+\frac{1}{\tilde\umega_1}&i=1\\
\frac{1}{\umega_1}-\frac{1}{\tilde\umega_1}-\frac{1}{\umega_2}+\frac{1}{\tilde\umega_2}&i=2\\
\frac{1}{\umega_2}-\frac{1}{\tilde\umega_2}+\frac{1}{\omega_2}-\frac{1}{\tilde\omega_2}&i=3
\end{cases}
\end{split}\end{align}
and
\begin{align}\begin{split}
i\langle p_{\sigma i}\rangle
&=\partial_{\sigma i}-\half\begin{cases}
\frac{1}{\sigma_1-a+\epsilon}-\frac{1}{\sigma_1-a-\epsilon}-\frac{1}{\sigma_{21}+\epsilon}-\frac{1}{\sigma_{21}-\epsilon}&i=1\\
\frac{1}{\sigma_{21}+\epsilon}+\frac{1}{\sigma_{21}-\epsilon}-\frac{1}{\sigma_{32}+\epsilon}-\frac{1}{\sigma_{32}-\epsilon}
&i=2\\
\frac{1}{\sigma_3+a+\epsilon}+\frac{1}{\sigma_3+a-\epsilon}+\frac{1}{\sigma_{32}+\epsilon}+\frac{1}{\sigma_{32}-\epsilon}
&i=3
\end{cases}
\end{split}\end{align}
where the $s_i={\rm sign}({\rm arg}(\omega_i))$, $\tilde{s}_i={\rm sign}({\rm arg}(\tilde\omega_i))$ where $\omega_i,\tilde\omega_i$ are treated as the absolute value function of its argument.  This gives rise to the effective supercharges:
\begin{align}\begin{split}
&Q_{\rm eff,1}=
\frac{e}{\sqrt{2}}\lambda^1_{1}\left(-\partial_{\sigma 1}-\frac{1}{\tilde\omega_1}+\frac{1}{\tilde{\underline{\omega}}_1}+\xi_1\right)
+\frac{e}{\sqrt{2}}\lambda^1_{2}\left(-\partial_{\sigma 2}-
\frac{1}{\tilde{\underline{\omega}}_1}+\frac{1}{\tilde{\underline{\omega}}_2}+\xi_2\right)\\
&+\frac{e}{\sqrt{2}}\lambda^1_{3}\left(\partial_{\sigma 3}-
\frac{1}{\tilde{\underline{\omega}}_2}-\frac{1}{\tilde\omega_2}+\xi_3\right)\\
&\bar{Q}_{\rm eff}^1=\frac{e}{\sqrt{2}}\lambda^1_{1}\left(-\partial_{\sigma 1}+\frac{1}{\omega_1}-\frac{1}{\underline\omega_1}+\xi_1\right)+
+\frac{e}{\sqrt{2}}\lambda^1_{2}\left(\partial_{\sigma 2}+
\frac{1}{\underline\omega_1}-\frac{1}{\underline\omega_2}+\xi_2\right)\\
&+\frac{e}{\sqrt{2}}\bar\lambda_{3}\left(\partial_{\sigma 3}+
\frac{1}{\underline\omega_2}+\frac{1}{\omega_2}+\xi_3\right)~,
\end{split}\end{align}
for $S_+$ and:
\begin{align}\begin{split}
&Q_{\rm eff,1}=\frac{e}{\sqrt{2}}\bar\lambda_{1,1}\left(-\partial_{\sigma 1}+\frac{1}{\omega_1}-\frac{1}{\underline\omega_1}+\xi_1\right)
+\frac{e}{\sqrt{2}}\bar\lambda_{2,1}\left(-\partial_{\sigma 2}+
\frac{1}{\underline\omega_1}-\frac{1}{\underline\omega_2}+\xi_2\right)\\
&+\frac{e}{\sqrt{2}}\bar\lambda_{3,1}\left(-\partial_{\sigma 3}+
\frac{1}{\underline\omega_2}+\frac{1}{\omega_2}\right)~,\\
&\bar{Q}_{\rm eff}^1=
\frac{e}{\sqrt{2}}\lambda^1_{1}\left(\partial_{\sigma 1}-\frac{1}{\tilde\omega_1}+\frac{1}{\tilde{\underline{\omega}}_1}+\xi_1\right)
+\frac{e}{\sqrt{2}}\lambda^1_{2}\left(\partial_{\sigma 2}-
\frac{1}{\tilde{\underline{\omega}}_1}+\frac{1}{\tilde{\underline{\omega}}_2}+\xi_2\right)\\
&+\frac{e}{\sqrt{2}}\lambda^1_{3}\left(\partial_{\sigma 3}-
\frac{1}{\tilde{\underline{\omega}}_2}-\frac{1}{\tilde\omega_2}+\xi_3\right)~,
\end{split}\end{align}
for $S_-$. 

\subsection{Ground States}
\label{sec:BGS}

Unfortunately, solving for the ground states of this system is significantly more complicated than the last section. We have to balance an unknown choice of boundary conditions, Born-Oppenheimer approximation, and solving a system of partial differential equations. 

Recall that in the Born-Oppenheimer approximation, we can only truly make sense of the quantum physics away from the boundaries. Thus, we are working in the limit 
\be
\epsilon/\sigma_1~,~\epsilon/\sigma_3~,~\epsilon/\sigma_{21}~,~\epsilon/\sigma_{32} <<1~.
\ee
Therefore, we will solve for ground states that are to first order in these parameters. 

In order to study this differential operator, we will introduce a basis for the Clifford algebra:
\be
\left(\begin{array}{c}
f_1\\f_2\\f_3\\f_4\\f_5\\f_6\\f_7\\f_8
\end{array}\right)_\pm=\begin{array}{l}f_1|\pm\rangle +f_2\bar\lambda_{1,1}|\pm\rangle +f_3 \bar\lambda_{2,1}|\pm \rangle +f_4\bar\lambda_{1,1}\bar\lambda_{2,1}|\pm \rangle+f_5\bar\lambda_{3,1}|\pm\rangle \\
+f_6\bar\lambda_{1,1}\bar\lambda_{3,1}|\pm \rangle+f_7 \bar\lambda_{2,1}\bar\lambda_{3,1}|\pm \rangle +f_8\bar\lambda_{1,1}\bar\lambda_{2,1}\bar\lambda_{3,1}|\pm \rangle
\end{array}~. 
\ee
In this basis, the Dirac operator $\CQ_1=Q_{\rm eff,1}+\bar{Q}_{\rm eff}^1$, can be written as
\be
\CQ_1=\left(\begin{array}{cccccccc}
0&-\CD_3+ D_3&-\CD_2+ D_2&0&-\CD_1+ D_1 &0&0&0\\
\CD_3+ D_3&0&0&-\CD_2+ D_2&0&-\CD_1+ D_1 &0&0\\
\CD_2+D_2&0&0&-\CD_3+D_3&0&0&-\CD_1+D_1&0\\
0&\CD_2+D_2&\CD_3+D_3&0&0&0&0&-\CD_1+D_1\\
\CD_1+D_1&0&0&0&0&-\CD_3+D_3&-\CD_2+D_2&0\\
0&\CD_1+D_1&0&0&\CD_3+D_3&0&0&-\CD_2+D_2\\
0&0&\CD_1+D_1&0&\CD_2+D_2&0&0&-\CD_3+D_3\\
0&0&0&\CD_1+D_1&0&\CD_2+D_2&\CD_3+D_3&0\\
\end{array}\right)
\ee
where 
\be
\CD_i=-i \langle p_{\sigma_i}\rangle\quad, \quad D_i=\langle D_i\rangle~. 
\ee
Now by taking wave functions that are functionally of the form 
\be
|\psi\rangle=\prod_i \omega_i \tilde\omega_i \underline\omega_i \tilde{\underline{\omega}}_i|\chi\rangle~, 
\ee
we can simplify the Dirac operator to 
\be
\hat\CQ_1=\left(\begin{array}{cccccccc}
0&-\partial_3+ D_3&-\partial_2+ D_2&0&-\partial_1+ D_1 &0&0&0\\
\partial_3+ D_3&0&0&-\partial_2+ D_2&0&-\partial_1+ D_1 &0&0\\
\partial_2+D_2&0&0&-\partial_3+D_3&0&0&-\partial_1+D_1&0\\
0&\partial_2+D_2&\partial_3+D_3&0&0&0&0&-\partial_1+D_1\\
\partial_1+D_1&0&0&0&0&-\partial_3+D_3&-\partial_2+D_2&0\\
0&\partial_1+D_1&0&0&\partial_3+D_3&0&0&-\partial_2+D_2\\
0&0&\partial_1+D_1&0&\partial_2+D_2&0&0&-\partial_3+D_3\\
0&0&0&\partial_1+D_1&0&\partial_2+D_2&\partial_3+D_3&0\\
\end{array}\right)
\ee

\subsubsection{Ground States in $S_+$}
Now we can try to solve the equations 
\be
\hat\CQ_1|\chi\rangle=0~. 
\ee
Let us consider states in $S_+$ for which $\xi_i>0$, $\forall i$. First let us restrict to $\xi_2<\xi_1,\xi_3$. In this case the only states that are normalizable have exponential dependence that  goes as $e^{-\xi_i \sigma_i}$. Therefore let us consider states on which  $\partial_i+D_i$ vanishes:
\be
|\chi\rangle=
\frac{(\sigma_1-a+\epsilon)(\sigma_3+a+\epsilon)}{(\sigma_1-a-\epsilon)(\sigma_3+a-\epsilon)}\frac{(\sigma_{21}-\epsilon)(\sigma_{32}+\epsilon)}{(\sigma_{21}+\epsilon)(\sigma_{32}-\epsilon)} e^{-\xi_1\sigma_1-\xi_2\sigma_2-\xi_3\sigma_3}|\hat\chi\rangle~. 
\ee
Now $\hat\CQ_1$ acting on $|\hat\chi\rangle$ is of the form
\be
\tilde\CQ_1=\left(\begin{array}{cccccccc}
0&2 D_3&2D_2&0&2D_1 &0&0&0\\
0&0&0&2D_2&0&2D_1 &0&0\\
0&0&0&2D_3&0&0&2D_1&0\\
0&0&0&0&0&0&0&2D_1\\
0&0&0&0&0&2D_3&2D_2&0\\
0&0&0&0&0&0&0&2D_2\\
0&0&0&0&0&0&0&2D_3\\
0&0&0&0&0&0&0&0\\
\end{array}\right)~.
\ee
Now we are reduced to finding zero-eigenvectors of this matrix. 

Recall that 
\be
 D_i=2\begin{cases}
\frac{1}{\omega_1}-\frac{1}{\tilde\omega_1}-\frac{1}{\umega_1}+\frac{1}{\tilde\umega_1}+\frac{\xi_1}{2}&i=1\\
\frac{1}{\umega_1}-\frac{1}{\tilde\umega_1}-\frac{1}{\umega_2}+\frac{1}{\tilde\umega_2}+\frac{\xi_2}{2}&i=2\\
\frac{1}{\umega_2}-\frac{1}{\tilde\umega_2}+\frac{1}{\omega_2}-\frac{1}{\tilde\omega_2}+\frac{\xi_3}{2}&i=3
\end{cases}~.
\ee
Using the properties of the $\omega_i$'s, we have that 
\be
\frac{1}{\omega_i}-\frac{1}{\tilde\omega_i}\sim O(\epsilon/\sigma^2)\sim 0~. 
\ee
Therefore, in our approximation, we only need to cancel the $\xi_i$'s which are not parametrically small and hence we can effectively replace $D_i$ by $\xi_i$.

We see that $(1,0,0,0,0,0,0,0)$ is clearly a 0-eigenvector and hence is a normalizable SUSY ground state. Now by rescaling our basis of eigenvectors by factors of $\xi_i$, we can see that there are additional approximate 0-eigenvectors such that the full space of ground states is given by
\begin{align}\begin{split}
&{\rm span}_{\IC}\big\{|v^{(+)}_1\rangle,|v^{(+)}_2\rangle,|v^{(+)}_3\rangle\big\}\\
&~={\rm span}_{\IC}\big\{(1,0,0,0,0,0,0,0)^{\rm tr}~,~(0,1,-1,0,0,0,0,0)^{\rm tr}~,~ (0,1,0,0,-1,0,0,0)^{\rm tr}\big\}~. 
\end{split}\end{align}

\subsubsection{Ground States in $S_-$}

We can similarly perform the same analysis in the negative wedge. Here the analysis changes by looking for states that are annihilated by $\partial_i-D_i$. These states are of the form
\be
|\chi\rangle= 
\frac{(\sigma_1-a-\epsilon)(\sigma_3+a-\epsilon)}{(\sigma_1-a+\epsilon)(\sigma_3+a+\epsilon)}\frac{(\sigma_{21}+\epsilon)(\sigma_{32}-\epsilon)}{(\sigma_{21}-\epsilon)(\sigma_{32}+\epsilon)}
e^{\xi_1\sigma_1+\xi_2\sigma_2+\xi_3\sigma_3}|\hat\chi\rangle~. 
\ee
Acting on these states, the supercharge operator $\CQ_1$ is of the form 
\be
\hat\CQ_1=\left(\begin{array}{cccccccc}
0&0&0&0&0 &0&0&0\\
2D_3&0&0&0&0&0 &0&0\\
2D_2&0&0&0&0&0&0&0\\
0&2D_2&2D_3&0&0&0&0&0\\
2D_1&0&0&0&0&0&0&0\\
0&2D_1&0&0&2D_3&0&0&0\\
0&0&2D_1&0&2D_2&0&0&0\\
0&0&0&2D_1&0&2D_2&2D_3&0\\
\end{array}\right)
\ee

Again we find 3 approximate BPS states
\begin{align}\begin{split}
&
\big\{|v^{(-)}_1\rangle,|v^{(-)}_2\rangle,|v^{(-)}_3\rangle\big\}\\
&~=
\big\{(0,0,0,0,0,0,0,1)^{\rm tr}~,~(0,0,0,0,0,-1,1,0)^{\rm tr}~,~ (0,0,0,-1,0,0,1,0)^{\rm tr}\big\}~. 
\end{split}\end{align}

\subsubsection{Hermiticity}

Again we have to impose boundary conditions so that the supercharges are Hermitian. The minimal boundary conditions to impose hermiticity allows us to keep all BPS states in both sectors. We will make a choice that is symmetric between exchange of $\bar\lambda_{I,1}$ and $\bar\lambda_{I,2}$ in analogy with the $U(1)$ case, and that is symmetric under $\bar\lambda_{1,I}$ and $\bar\lambda_{3,I}$. 

So let us define the (unnormalized) states
\begin{align}\begin{split}&
|\psi_1^{(+)}\rangle=\left(\sum_i \bar\lambda_{i,2}\right)|v_1^{(+)}\rangle\quad, \quad |\psi_2^{(+)}\rangle=|v_2^{(+)}\rangle+|v_3^{(+)}\rangle~,\\
&|\psi_1^{(-)}\rangle=\prod_i \bar\lambda_{i,2}|v_1^{(-)}\rangle\quad, \quad |\psi_2^{(-)}\rangle=\bar\lambda_{1,2}\bar\lambda_{2,2}|v_2^{(-)}\rangle+\bar\lambda_{2,2}\bar\lambda_{3,2}|v_3^{(-)}\rangle~.
\end{split}\end{align}
Then it is consistent to pick boundary conditions such that the supersymmetric ground states are given by 
\be
\{|\psi_1^{(+)}\rangle\,,\, |\psi_2^{(+)}\rangle \,,\, |\tilde{\psi}_1^{(+)}\rangle\,,\, |\tilde{\psi}_2^{(+)}\}_{S_+}\cup\{|\psi_1^{(-)}\rangle\,,\, |\psi_2^{(-)}\rangle \,,\, |\tilde{\psi}_1^{(-)}\rangle\,,\, |\tilde{\psi}_2^{(-)}\}_{S_-}~,
\ee
where 
\be
|\tilde{\psi}_i^{(\pm)}\rangle=\tilde{P}|\psi_i^{(\pm)}~,~{\rm replace}~\xi_i\leftrightarrow -\xi_i\rangle\quad,\quad \tilde{P}=\prod_{i,A}\left(\lambda^A_{i}+\bar\lambda_{A,i}\right)~.
\ee

Under this choice of boundary conditions, the normalizable asymptotic Coulomb branch states for $\xi_i>0$ and $\xi_i<0$ are given by
\begin{align}\begin{split}
\{|\Psi_{BPS}\rangle\}=\begin{cases}
\{|\psi_1^{(\pm)}~,~|\psi_2^{(\pm)}\rangle \}&\xi_i>0~,~\forall i\\
\{|\tilde{\psi}_1^{(\pm)}~,~|\tilde{\psi}_2^{(\pm)}\rangle \}&\xi_i<0~,~\forall i
\end{cases}
\end{split}\end{align}

\subsection{Mixed Branch}
\label{app:mixed}

As it turns out, there are no mixed branch states in this theory. The reason is the following. The localization principal states that only finite enegy states that survive in the limit $e^2\to 0$ contribute to the Witten index. Due to the form of the potential \eqref{massdefvac}, we 
 must simultaneously solve the mass equations 
\begin{align}\begin{split}&
0=(\sigma_1-a+\epsilon)^2|\phi_1|^2=(\sigma_1-a-\epsilon)^2|\tilde\phi_1|^2=(\sigma_{21}+\epsilon)^2|\uphi_1|^2=(\sigma_{21}-\epsilon)^2|\tilde\uphi_1|^2~,\\
&
0=(\sigma_3+a+\epsilon)^2|\phi_2|^2=(\sigma_3+a-\epsilon)^2|\tilde\phi_2|^2=(\sigma_{32}+\epsilon)^2|\uphi_2|^2=(\sigma_{32}-\epsilon)^2|\tilde\uphi_2|^2~,
\end{split}\end{align}
the F-term equations
\begin{align}\begin{split}&
0=\phi_1\tilde\phi_1-\uphi_1\tilde\uphi_1~,\\&
0=\uphi_1\tilde\uphi_1-\uphi_2\tilde\uphi_2~,\\&
0=\uphi_2\tilde\uphi_2+\phi_2\tilde\phi_2~,
\end{split}\end{align}
and the D-term equations 
\begin{align}\begin{split}
&0=|\phi_1|^2-|\tilde\phi_1|^2-|\uphi_1|^2+|\tilde\uphi_1|^2-e^2\xi_1~,\\&
0=|\uphi_1|^2-|\tilde\uphi_1|^2-|\uphi_2|^2+|\tilde\uphi_2|^2-e^2\xi_2~,\\&
0=|\uphi_2|^2-|\tilde\uphi_2|^2+|\phi_2|^2-|\tilde\phi_2|^2-e^2\xi_3~,
\end{split}\end{align}
to order $O(e)$. 

Let us consider the case where $\xi_i>0$ or $\xi_i<0$, $\forall i$. 
Here there are only solutions to the D-term equations when there are light fundamental hypermultiplet fields with non-zero expectation value due to the repeated appearance of bifundamental hypermultiplet fields. Therefore the mixed branches are 
\begin{align}\begin{split}
{\rm I_+~:~}& \sigma_1=a-\epsilon~,~|\sigma_2|,|\sigma_3|>>0~,~\\
{\rm II_+~:~}& \sigma_1=\sigma_2-\epsilon=a-\epsilon~,~|\sigma_3|>>0~,~\\
{\rm III_+~:~}& \sigma_3=-a-\epsilon~,~|\sigma_2|,|\sigma_1|>>0~, \\
{\rm IV_+~:~}& \sigma_3=\sigma_2-\epsilon=-a-\epsilon~,~|\sigma_1|>>0~,\\
{\rm V_+~:~}& \sigma_1=a-\epsilon~,~ \sigma_3=-a-\epsilon~,~|\sigma_2|>>0~,
\end{split}\end{align}
for $\xi_i>0$ and the mixed branches 
\begin{align}\begin{split}
{\rm I_-~:~}& \sigma_1=a+\epsilon~,~|\sigma_2|,|\sigma_3|>>0~,~\\
{\rm II_-~:~}& \sigma_1=\sigma_2+\epsilon=a+\epsilon~,~|\sigma_3|>>0~,~\\
{\rm III_-~:~}& \sigma_3=-a+\epsilon~,~|\sigma_2|,|\sigma_1|>>0~, \\
{\rm IV_-~:~}& \sigma_3=\sigma_2+\epsilon=-a+\epsilon~,~|\sigma_1|>>0~,\\
{\rm V_+~:~}& \sigma_1=a+\epsilon~,~ \sigma_3=-a+\epsilon~,~|\sigma_2|>>0~,
\end{split}\end{align}
for $\xi_i<0$. We conjecture that there are no BPS states localized on these vacuum branches. \footnote{See upcoming dissertation of the first author for more details.}

\subsection{Contribution to the Witten Index}

It follows from our conjecture in Section \ref{app:mixed} that in the $U(1)^3$ bubbling SQM,  
only Coulomb branch states contribute to the non-compact index $I_{asymp}$. 
These states give rise to the results
\begin{align}\begin{split}
I_{asymp}(\xi_i>0)=
e^{\sum_f m_f+ 6\epsilon_+}+e^{-\sum_f m_f- 6 \epsilon_+}+e^{\sum_f m_f+ 2\epsilon_+}+e^{-\sum_f m_f- 2 \epsilon_+}\\=2\cosh\left(\sum_f m_f+6\epsilon_+\right)+2\cosh\left(\sum_f m_f+2\epsilon_+\right)~,
\end{split}\end{align}
or
\begin{align}\begin{split}
I_{asymp}(\xi_i<0)=
e^{\sum_f m_f- 6\epsilon_+}+e^{-\sum_f m_f+ 6 \epsilon_+}+e^{\sum_f m_f- 2\epsilon_+}+e^{-\sum_f m_f+ 2 \epsilon_+}\\=2\cosh\left(\sum_f m_f-6\epsilon_+\right)+2\cosh\left(\sum_f m_f-2\epsilon_+\right)~.
\end{split}\end{align}

\section{Behavior of Localized Path Integral at Infinity}

\label{app:C}

In this appendix we will consider the behavior of the integrand of the localized path integral \eqref{Zint}, $Z_{int}(\varphi)$,  at $\varphi\to \partial \ft_\IC/\Lambda_{cr}$. Let us take $G=\prod_{i=1}^{n-1} U(k^{(i)})$ to be the gauge group of the SQM such that the corresponding Lie algebra $\fg$ decomposes as $\fg=\bigoplus_{i=1}^{n-1}\fg^{(i)}=\bigoplus_{i=1}^{n-1} \fu(k^{(i)})$. Consider taking the limit 
\be
\tau\to \infty \quad {\rm  where }\quad \varphi=\tau \uu\quad, \quad \uu\in \ft
\ee
where $\ft$ is the Lie algebra of $\fg$ which itself decomposes as $\ft=\bigoplus_{i=1}^{n-1}\fu(k^{(i)})=\bigoplus_{i=1}^{n-1}\ft^{(i)}$. The element $\uu$  can be written with respect to this decomposition as 
\be
\uu=\bigoplus_{i=1}^{n-1} \uu^{(i)} \quad, \quad \uu^{(i)}=\sum_{a=1}^{k^{(i)}}\uu_a^{(i)} e_a^{(i)}\quad, \quad \ft^{(i)}\subset \fu(k^{(i)})\quad, \quad \ft^{(i)}={\rm span}_{\IR}\{e_a^{(i)}\}_{a=1}^{k^{(i)}}~,
\ee
and as a matrix $e_a^{(i)}=\delta_{a,a}$. The matter content of a generic bubbling SQM transforms under the representations
\begin{align}\begin{split}
{\rm bifundamental~ hyper:}&~\bigoplus_{i=1}^{n-1}\left[{\rm k}^{(i)}\otimes {\rm k}^{(i+1)}\right]\oplus\left[\overline{\rm k}^{(i)}\otimes \overline{\rm k}^{(i+1)}\right]~,\\
{\rm fundamental ~hyper:}&~\bigoplus_{i=1}^{n-1}\left[\delta_{s(i),1}{\rm k}^{(i)}\oplus \overline{\rm k}^{(i)}\right]\oplus 2\left[\delta_{s(i_m),2}
{\rm k}^{(i_m)}\oplus \overline{\rm k}^{(i_m)} \right]~,\\
{\rm fundamental ~Fermi:} & ~{\rm N}_f {\rm k}^{(i_m)}~,
\end{split}\end{align}
where 
\be
s(i)=2k^{(i)}-k^{(i+1)}-k^{(i-1)}\quad, \quad i_m=\half n-1~. 
\ee
Using this, we can compute the limiting form of the different terms in $Z_{det}$ as $\tau\to \infty$. Using \eqref{veccont}, we can see that
\begin{align}\begin{split}
|Z_{vec}|\underset{\substack{\tau\to \infty\\\varphi=\tau \uu}}{\sim} \prod_{i=1}^{n-1} {\rm exp}\left\{2\tau\sum_{\alpha\in \Delta^{(i)}_{adj}}|\alpha(\uu^{(i)})|\right\}=\prod_{i=1}^{n-1} {\rm exp}\left\{4\tau \sum_{\alpha\in \Delta^{(i)+}_{adj}}\alpha(\uu^{(i)})\right\}=\prod_{i=1}^{n-1} e^{4 \tau \rho^{(i)}\cdot \uu}~,
\end{split}\end{align}
where $\Delta^{(i)+}_{adj}$ are the set of positive weights of the adjoint representation with respect to the splitting of the weight lattice where $\uu$ is in the fundamental chamber and 
\be
\rho=\half \sum_{\alpha\in \Delta^+(\fg)}\alpha=\sum_{i=1}^{n-1} \rho^{(i)}\quad, \quad \rho^{(i)}=\half\sum_{\alpha\in \Delta^{+}(\fg^{(i)})}\alpha~,
\ee
is the Weyl element of $\fg$ and $\fg^{(i)}$ respectively.  Then using the form of 
\be
\rho^{(i)}=\half\sum_{a=1}^{k^{(i)}}(k^{(i)}-2a+1)e^{(i)}_a~, 
\ee
we can rewrite the limiting form as
\be
|Z_{vec}|\underset{\substack{\tau\to \infty\\\varphi=\tau \uu}}{\sim}\prod_{i=1}^{n-1}e^{4\tau \sum_{a=1}^{k^{(i)}}(k^{(i)}-2a+1)\uu_a^{(i)}}~. 
\ee
The contribution from the bifundamental hypermultiplets \eqref{hypermultIntegrand} has the limiting form 
\be
|Z_{hyper;bf}|\underset{\substack{\tau\to \infty\\\varphi=\tau \uu}}{\sim}\prod_{i=1}^{n-1} e^{-2 \tau (k^{(i-1)}+k^{(i+1)})\sum_{a=1}^{k^{(i)}}|\uu_a^{(i)}|}~.
\ee
Similarly, the fundamental hypermultiplets \eqref{hypermultIntegrand} and \eqref{fermicont} contributions have the limiting forms
\begin{align}\begin{split}
|Z_{hyper;f}|&\underset{\substack{\tau\to \infty\\\varphi=\tau \uu}}{\sim}\prod_{i=1}^{n-1}e^{-2 \tau\left(\delta_{s(i),1}+2\delta_{s(i),2}\right)\sum_{a=1}^{k^{(i)}}|\uu_a^{(i)}|}~,\\
|Z_{Fermi;f}|&\underset{\substack{\tau\to \infty\\\varphi=\tau \uu}}{\sim}e^{\tau {\rm N}_f \sum_{a=1}^{k^{(i_m)}} |\uu_a^{(i_m)}|}~. 
\end{split}\end{align}
Putting these factors all together, we find that 
\begin{align}\begin{split}
|Z_{int}|\underset{\substack{\tau\to \infty\\\varphi=\tau \uu}}{\sim}\prod_{i=1}^{n-1}{\rm exp}\Bigg\{&4\tau \sum_{a=1}^{k^{(i)}}(k^{(i)}-2a+1)\uu_a^{(i)}+\tau {\rm N}_f \sum_{a=1}^{k^{(i_m)}} |\uu_a^{(i_m)}|\\&
-2 \tau \big(k^{(i-1)}+k^{(i+1)}+\delta_{s(i),1}+2\delta_{s(i),2}\big)\sum_{a=1}^{k^{(i)}}|\uu_a^{(i)}|
\Bigg\}~.
\end{split}\end{align}
This is bounded from above by 
\begin{align}\begin{split}
|Z_{int}|\underset{\substack{\tau\to \infty\\\varphi=\tau \uu}}{\lesssim}\prod_{i=1}^{n-1}{\rm exp}\Bigg\{&4\tau \sum_{a=1}^{k^{(i)}}(k^{(i)}-1)|\uu_a^{(i)}|+\tau {\rm N}_f \sum_{a=1}^{k^{(i_m)}} |\uu_a^{(i_m)}|\\&
-2 \tau (k^{(i-1)}+k^{(i+1)}+\delta_{s(i),1}+2\delta_{s(i),2})\sum_{a=1}^{k^{(i)}}|\uu_a^{(i)}|
\Bigg\}~,
\end{split}\end{align}
which can further be simplified to 
\be
|Z_{int}|\underset{\substack{\tau\to \infty\\\varphi=\tau \uu}}{\lesssim}\prod_{i=1}^{n-1}{\rm exp}\Bigg\{2\tau\left(s(i)-2-2\delta_{s(i),1}-4\delta_{s(i),2}+\frac{{\rm N}_f}{2}\delta_{i,i_m}\right)\sum_{a=1}^{k^{(i)}}|\uu_a^{(i)}|\Bigg\}~.
\ee
Using the fact that $s(i_m)=0$ or 2 and the fact that ${\rm N}_f\leq4$, we see that 
the exponential factors can at most completely cancel as $\tau\to \infty$. 
In this case, the behavior of the 1-loop determinant at infinity will be polynomially suppressed by the Yukawa terms for the hypermultiplet fields to order 
 $O(\prod_i\tau^{-3k^{(i)} })$. Therefore,  since the measure goes as $\prod_i \tau^{2k^{(i)}-1}$, we have that the product of the integrand and measure will vanish as $O(\prod_i \tau^{-k^{(i)}-1})$ and the integral is convergent. 

\section{A Useful Integral}
\label{app:D}

Often in the text we make use of a non-standard integral identity which we will now precisely derive. Consider the integral 
\be
F(a,b,\eta)= 
\int_{\IR + \I \eta}  \frac{d D}{D} e^{- a D^2 + i b D }
\ee
where
\be\label{eq:2}
a > 0  \qquad  b\in \IC  \qquad \eta \in \IR^* 
\ee
We claim this integral is just 
\be
F(a,b,\eta) = +\I \pi {\rm erf}\left( \frac{b}{2 \sqrt{a}} \right) -  \pi \I  {\rm sign}(\eta)
\ee
where we choose the positive square root of $a$ 
%
%
and
\be
{\rm erf}(x)=\frac{2}{\sqrt{\pi}}\int_{0}^x dy e^{-y^2}~. 
\ee

Proof:   $F(a,b,\eta)$ is an entire function of $b$. 
Moreover, it satisfies the 
differential equation 
\be
\frac{\p F}{\p b} = \I \sqrt{\frac{\pi}{a}} \exp[ - \frac{b^2}{4a} ] 
\ee
so 
\be
\begin{split}
F(a,b,\eta) & =  \int_0^b  \I \sqrt{\frac{\pi}{a}} \exp[ - \frac{s^2}{4a} ]  ds  + F(a,0,\eta) \\
& = F(a,0,\eta) + 
\I \pi\, {\rm erf}\left( \frac{b}{2 \sqrt{a}} \right) \\
\end{split}
\ee
It thus remains to determine
\be
F(a,0,\eta)=H(\eta/\sqrt{a})~,
\ee
where
\be
H(\eta/\sqrt{a}):= \int_{\IR + \I \eta} \frac{dD}{D} e^{-D^2}~.
\ee

Now by contour integration arguments $H(\eta)$ only depends on the sign of $\eta$. Let $H_+$ be the 
value for $\eta>0$ and $H_-$ the value for $\eta<0$. We can take the limit as $\eta \to 0^+$
and use 
\be
\frac{1}{D+\I\eta} \rightarrow P \left(\frac{1}{D}\right) - \I \pi \delta(D)
\ee
for $D$ real, where $P$ is the principal part. But 
\be
P \int \frac{dD}{D} e^{-D^2} := \lim_{\epsilon\to 0^+} \left[
\int_{-\infty}^{-\epsilon} \frac{dD}{D} e^{-D^2} + \int_{\epsilon}^{\infty} \frac{dD}{D} e^{-D^2} \right] = 0 
\ee
Moreover, $H_-^* = H_+$, so 
\be
H(\eta) = \begin{cases}  - \I \pi  & \eta > 0 \\   + \I \pi  & \eta < 0  \\  \end{cases}
\ee

\section{Bubbling Contribution in the $SU(2)\times SU(2)$ SCFT}

\label{app:E}

Consider the case of a superconformal $\CN=2$ quiver gauge theory with $G=SU(2)_1\times SU(2)_2$ with fundamental matter:
\begin{center}
\begin{tikzpicture}[node distance=2cm,
cnode/.style={circle,draw,thick,minimum size=9mm},snode/.style={rectangle,draw,thick,minimum size=9mm}]
\node[snode] (1) {2};
\node[cnode] (2) [right of=1] {SU(2)};
\node[cnode] (3) [right of=2] {SU(2)};
\node[snode] (4) [right of=3] {2};
\draw[-] (1) -- (2);
\draw[-] (2) -- (3);
\draw[-] (3) -- (4);
\end{tikzpicture}
\end{center}
Now consider the bubbling sector where 
\be
P=\bigoplus_{i=1}^2 P_i\quad, \quad \vv=\bigoplus_{i=1}^2 \vv_i\quad,\quad (P_i,\vv_i)=\big({\rm diag}(1,-1),{\rm diag}(0,0)\big)~.
\ee
In this case, the $\CN=(0,4)$ bubbling SQM is of the form
\begin{center}\begin{tikzpicture}[node distance=2cm,cnode/.style={circle,draw,thick,minimum size=10mm},snode/.style={rectangle,draw,thick,minimum size=10mm}]
\node[cnode] (1) {1};
\node[cnode] (2) [right of=1,xshift=1cm]{1};
\node[snode] (3) [below of=1]{2};
\node[snode] (4) [below of=2]{2};
\node[snode] (5) [above of=1]{2};
\node[snode] (6) [above of=2]{2};
\draw[double,double distance=6pt,thick] (1) -- (2);
\draw[double,double distance=6pt,thick] (1) -- (3);
\draw[double,double distance=6pt,thick] (2) -- (4);
\draw[dashed,thick] (1) -- (5);
\draw[dashed,thick] (2) -- (6);
\draw[dashed,thick] (3) -- (2);
\draw[dashed,thick] (4) -- (1);
\path[dashed,every loop/.style={looseness=5}] (1)
         edge  [in=210,out=150,loop] node {} (); 
\path[dashed,every loop/.style={looseness=5}] (2)
         edge  [in=30,out=330,loop] node {} (); 
\end{tikzpicture}\end{center}
The localization contribution to $Z_{mono}\big((1,0)\oplus(1,0)\big)$  is then given by the contour integral
\begin{align}
\begin{split}
Z_{mono}^{(Loc)}=\sinh^2(2\epsilon_+)&\oint_{JK(\xi_i,\xi_2)}\frac{d\varphi_1d\varphi_2}{(2\pi i)^2}\frac{\prod_{f=1}^2\sinh(\varphi_1-m_f)\sinh(\varphi_2-m_{f+2})}{\prod_{i=1}^2\prod_{\pm}\sinh(\pm (\varphi_i-\fa_i)+\epsilon_+)\sinh(\pm (\varphi_i+\fa_i)+\epsilon_+)}\\&
\times \frac{%
\prod_{\pm}\sinh(-\varphi_1\pm \fa_2+m+\epsilon_+)\sinh(\varphi_2\pm \fa_1+m+\epsilon_+)
}{4\sinh(\varphi_2-\varphi_1)\sinh(\varphi_1-\varphi_2+2\epsilon_+)}
~.
\end{split}\end{align}
Let us choose $\xi_1,\xi_2>0$. In this case there are 8 poles contributing this to this path integral:
\begin{align}\begin{split}
{\rm I:}&~\varphi_1=\fa_1-\epsilon_+\quad~~, \quad \varphi_2=\fa_2-\epsilon_+~,\\
{\rm II:}&~\varphi_1=\fa_1-\epsilon_+\quad~~ ,\quad \varphi_2=-\fa_2-\epsilon_+~,\\
{\rm III:}&~\varphi_1=-\fa_1-\epsilon_+\quad, \quad \varphi_2=\fa_2-\epsilon_+~,\\
{\rm IV:}&~\varphi_1=-\fa_1-\epsilon_+\quad, \quad \varphi_2=-\fa_2-\epsilon_+~,\\
{\rm V:}&~\varphi_1=\fa_1-\epsilon_+\quad~~\,, \quad \varphi_2=\fa_1-\epsilon_+~,\\
{\rm VI:}&~\varphi_1=-\fa_1-\epsilon_+\quad, \quad \varphi_2=-\fa_1-\epsilon_+~,\\
{\rm VII:}&~\varphi_1=\fa_2-3\epsilon_+\quad~, \quad \varphi_2=\fa_2-\epsilon~,\\
{\rm VIII:}&~\varphi_1=-\fa_2-3\epsilon_+~\, ,\quad \varphi_2=-\fa_2-\epsilon~.
\end{split}\end{align}
Using this set of poles as defined via the Jeffrey-Kirwan residue prescription, we find that the localization computation of $Z_{mono}^{(Loc)}((1,0)\oplus(1,0))$ is given by 

\begin{align}\begin{split}
&Z_{mono}^{(Loc)}((1,0)\oplus(1,0))=-\frac{\prod_{f=1}^2\sinh(\fa_1-m_f-\epsilon_+)\sinh(\fa_2-m_{f+2}-\epsilon_+)}{\sinh(2\fa_1)\sinh(2\fa_1-2\epsilon_+)\sinh(2\fa_2)\sinh(2\fa_2-2\epsilon_+)}\\&\quad\times
 \frac{
\sinh(\fa_2-\fa_1+m+2\epsilon_+)\sinh(\fa_1+\fa_2-m-2\epsilon_+)\sinh(\fa_2-\fa_1+m)\sinh(\fa_2+\fa_1+m)
}{\sinh(\fa_2-\fa_1)\sinh(\fa_1-\fa_2+2\epsilon_+)}
\\&-
\frac{\prod_{f=1}^2\sinh(\fa_1-m_f-\epsilon_+)\sinh(\fa_2+m_{f+2}+\epsilon_+)}{\sinh(2\fa_1)\sinh(2\fa_1-2\epsilon_+)\sinh(2\fa_2)\sinh(2\fa_2+2\epsilon_+)}\\&\quad\times
 \frac{
\sinh(\fa_2+\fa_1-m-2\epsilon_+)\sinh(\fa_1-\fa_2-m-2\epsilon_+)\sinh(\fa_2+\fa_1-m)\sinh(\fa_2-\fa_1-m)
}{\sinh(\fa_2+\fa_1)\sinh(\fa_1+\fa_2+2\epsilon_+)}
\\&-
\frac{\prod_{f=1}^2\sinh(\fa_1+m_f+\epsilon_+)\sinh(\fa_2-m_{f+2}-\epsilon_+)}{\sinh(2\fa_1)\sinh(2\fa_1+2\epsilon_+)\sinh(2\fa_2)\sinh(2\fa_2-2\epsilon_+)}\\&\quad\times
 \frac{
\sinh(\fa_2+\fa_1+m+2\epsilon_+)\sinh(\fa_1-\fa_2+m+2\epsilon_+)\sinh(\fa_2+\fa_1+m)\sinh(\fa_2-\fa_1+m)
}{\sinh(\fa_2+\fa_1)\sinh(\fa_1+\fa_2-2\epsilon_+)}
\\&+
\frac{\prod_{f=1}^2\sinh(\fa_1+m_f+\epsilon_+)\sinh(\fa_2+m_{f+2}+\epsilon_+)}{\sinh(2\fa_1)\sinh(2\fa_1+2\epsilon_+)\sinh(2\fa_2)\sinh(2\fa_2+2\epsilon_+)}\\&\quad\times
 \frac{
\sinh(\fa_2-\fa_1-m-2\epsilon_+)\sinh(\fa_1+\fa_2+m+2\epsilon_+)\sinh(\fa_2-\fa_1-m)\sinh(\fa_2+\fa_1-m)
}{\sinh(\fa_2-\fa_1)\sinh(\fa_1-\fa_2-2\epsilon_+)}
\\&-
\sinh(m)\frac{\prod_{f=1}^2 \sinh(\fa_1-m_f-\epsilon_+)\sinh(\fa_1-m_{f+2}-\epsilon_+)}{\sinh(2\fa_1)\sinh(2\fa_1-2\epsilon_+)}\\&\quad\times
\sinh(2\fa_1+m)\prod_\pm\frac{\sinh(\fa_1\pm \fa_2-m-2\epsilon_+)}{\sinh(\fa_1\pm\fa_2)\sinh(\fa_1\pm \fa_2-2\epsilon_+)}\\&+
\sinh(m)\frac{\prod_{f=1}^2 \sinh(\fa_1+m_f+\epsilon_+)\sinh(\fa_1+m_{f+2}+\epsilon_+)}{\sinh(2\fa_1)\sinh(2\fa_1+2\epsilon_+)}\\&\quad\times
\sinh(2\fa_1-m)\prod_\pm\frac{\sinh(\fa_1\pm \fa_2+m+2\epsilon_+)}{\sinh(\fa_1\pm \fa_2+2\epsilon_+)\sinh(\fa_1\pm \fa_2)}
\\&+
\sinh(m+4\epsilon_+)
\frac{\prod_{f=1}^2\sinh(\fa_2-m_f-3\epsilon_+)\sinh(\fa_2-m_{f+2}-\epsilon_+)}{\sinh(2\fa_2)\sinh(2\fa_2-2\epsilon_+)}\\&\quad\times
\sinh(2\fa_2-m-4\epsilon_+)\prod_\pm \frac{\sinh(\fa_2\mp \fa_1+m)}{\sinh(\fa_2\pm \fa_1-2\epsilon_+)\sinh(\fa_2\pm \fa_1-4\epsilon_+)}
\\&-
\sinh(m+4\epsilon_+)
\frac{\prod_{f=1}^2\sinh(\fa_2+m_f+3\epsilon_+)\sinh(\fa_2+m_{f+2}+\epsilon_+)}
{\sinh(2\fa_2)\sinh(2\fa_2+2\epsilon_+)}\\&\quad\times 
\sinh(2\fa_2+m+4\epsilon_+)\prod_\pm \frac{\sinh(\fa_2\pm \fa_1-m)}{\sinh(\fa_2\pm \fa_1+2\epsilon_+)\sinh(\fa_2\pm \fa_1+4\epsilon_+)}
~.
\end{split}\end{align}
One can check that the localization result for $Z_{mono}((1,0)\oplus(1,0))$ from residues associated to these poles is not invariant under the Weyl group of the flavor symmetry groups which is generated by the elements $W=\langle a_1,a_2, b_1,b_2\rangle $ that act on the masses in the previous formula as 
\begin{align}\begin{split}
&a_1:(m_1,m_2,m_3,m_4)\mapsto (m_2,m_1,m_3,m_4)~,\\
&a_2:(m_1,m_2,m_3,m_4)\mapsto (m_1,m_2,m_4,m_3)~,\\
&b_1:(m_1,m_2,m_3,m_4)\mapsto (-m_2,-m_1,m_3,m_4)~,\\
&b_2:(m_1,m_2,m_3,m_4)\mapsto  (m_1,m_2,-m_4,-m_3)~.
\end{split}\end{align}  

\end{document}